\documentclass[journal, twoside]{IEEEtran}
\usepackage{amstext,amssymb}
\usepackage{latexsym,epsfig}
 \newtheorem{theorem}{Theorem}[section]
 \newtheorem{lemma}[theorem]{Lemma}

 \newtheorem{corollary}[theorem]{Corollary}

\newcommand{\Z}{{\mathbb{Z}}}

\newcommand{\R}{{\mathbb{R}}}
\newcommand{\C}{{\mathbb{\CC}}}

\newcommand{\CC}{{\mathcal{C}}}
\newcommand{\D}{{\mathcal{D}}}

\newcommand{\I}{{\mathcal{I}}}
\newcommand{\J}{{\mathcal{J}}}

\renewcommand{\P}{{\mathcal{P}}}

\newcommand{\PP}{{\mathcal{P}}}

\newcommand{\V}{{\mathcal{V}}}
\newcommand{\W}{{\mathcal{W}}}
\newcommand{\X}{{\mathcal{X}}}

\newcommand{\Gh}{G\hat{\ }}
\newcommand{\Ghh}{G\hat{\ }\hat{\ }}
\newcommand{\Hh}{H\hat{\ }}

\newcommand{\ab}{\mathbf a}

\newcommand{\cb}{\mathbf c}
\newcommand{\db}{\mathbf d}

\newcommand{\gb}{\mathbf g}
\newcommand{\hb}{\mathbf h}

\newcommand{\qb}{\mathbf q}

\renewcommand{\sb}{\mathbf s}

\newcommand{\wb}{\mathbf w}
\newcommand{\xb}{\mathbf x}

\newcommand{\zerob}{\mathbf 0}

\newcommand{\sigmab}{\mbox{\boldmath$\sigma$}}

\newcommand{\ie}{{\em i.e., }}
\newcommand{\eg}{{\em e.g., }}

\newcommand{\inner}[2]{\langle{#1},{#2}\rangle}

\newcommand{\lra}{\leftrightarrow}
\newcommand{\qed}{\hspace*{1cm}\hspace*{\fill}\openbox}
\newcommand{\half}{\frac{1}{2}}

\newcommand{\mod}{\mathrm{~mod~}}  %(math mode only)
\newcommand{\im}{\mathrm{im~}}  %(math mode only)
\newcommand{\remove}[1]{}

\newcommand{\h}{\hat{\ }}
\newcommand{\cl}{^\mathrm{cl}}

\newcommand{\openbox}{\leavevmode
  \hbox to.77778em{%
  \hfil\vrule
  \vbox to.675em{\hrule width.6em\vfil\hrule}%
  \vrule\hfil}}
\newcommand{\proofname}{Proof}

\begin{document}
\title{The Dynamics of Group Codes: \\ Dual Abelian Group Codes and
Systems}
\author{G. David Forney, Jr.,~\IEEEmembership{Fellow,~IEEE,}
        and~Mitchell D. Trott,~\IEEEmembership{Member,~IEEE}% 
\thanks{Manuscript received xxxxxxx yy, 2000; revised zzzzzzz ww, 2004.
        Parts of the material of this paper were
presented at the 1993 IEEE International Symposium on Information Theory,
San Antonio, TX, Jan.\ 1993;  the 1993 IEEE Workshop on Coding, System
Theory and Symbolic Dynamics, Mansfield, MA, Oct.\ 1993;  the 1994 IEEE
Workshop on Information Theory, Moscow, June 1994,  and the 34th
Conference on Decision and Control, New Orleans, LA, Dec.\ 1995.}%
\thanks{G. D. Forney is with the Laboratory for Information and Decision Systems,
Massachusetts Institute of Technology, Cambridge, MA 02139 USA.}%
\thanks{M. D. Trott is with the Hewlett-Packard Laboratories, Palo Alto, CA 94304
USA.}%
\thanks{Communicated by F. R. Kschischang, Associate Editor for Coding Theory.}%
\thanks{Digital Object Identifier 10.1109/TIT.2004.xxxxxx}}%
% note the % following the last \IEEEmembership and also the first \thanks - 
% instance, "A\textbf{} \textbf{}B" will typeset as "A B" not "AB". If you want
% "AB" then you have to do: "A\textbf{}\textbf{}B"
\markboth{IEEE Transactions on Information Theory,~Vol.~50,
No.~11,~November~2004}{Forney and Trott: Duals of Abelian Group Codes and
Systems}
% *** Note that you probably will NOT want to include the author's name in ***
% *** the headers of peer review papers.                                   ***

% If you want to put a publisher's ID mark on the page
% (can leave text blank if you just want to see how the
% text height on the first page will be reduced by IEEE)
\pubid{0000--0000/00\$00.00~\copyright~2004 IEEE}

%\specialpapernotice{(Invited Paper)}

\maketitle

\begin{abstract}
Fundamental results concerning the dynamics of abelian group codes
(behaviors) and their duals are developed.  Duals of sequence spaces over
locally compact abelian groups may be defined via Pontryagin duality; 
dual group codes are orthogonal subgroups of dual sequence spaces.  The
dual of a complete code or system is finite, and the dual of a Laurent
code or system is (anti-)Laurent.  If $\C$ and $\C^\perp$ are dual codes, 
then the state spaces of $\C$ act as the character groups of the state
spaces of $\C^\perp$.  The controllability properties of $\C$ are the
observability properties of $\C^\perp$.  In particular, $\C$ is (strongly)
controllable if and only if  $\C^\perp$ is (strongly) observable, and the
controller memory of $\C$ is the observer memory of $\C^\perp$.  The
controller granules of $\C$ act as the character groups of the observer
granules of $\C^\perp$.  Examples of minimal observer-form encoder and
syndrome-former constructions are given.  Finally, every observer granule
of $\C$ is an ``end-around" controller granule of $\C$.
\end{abstract}


% Will be ignored for other modes.
%\IEEEpeerreviewmaketitle

\section{Introduction}

\PARstart{A}{group code} is a set of sequences that has a group property
under a  componentwise group operation \cite{FT93, LM96}.  For example, if $G$ is
any group and $G^\Z$  is the direct product group whose elements are the
bi-infinite sequences with components in $G$, then any subgroup $\C$ of
$G^\Z$  is a group code.

	A group code may be regarded as the behavior of a behavioral group 
system, in the sense of Willems \cite{W86, W89, W91, W97}.  It has
been shown in \cite{FT93, LFMT94, LM96} that many of the fundamental
properties of linear codes and systems depend only on their group
structure.  Most importantly, a group code or system has
naturally-defined minimal state spaces. 

	In this paper we study dual group codes and systems.  Our motivation is 
the importance of duality in the study of linear codes and systems.  
(For brevity, we will usually say ``code" rather than ``code or
system/behavior.")

	Our first problem is to define the dual $\C^\perp$ of a group code $\C$. 
For this purpose we use Pontryagin duality, a rather general notion of
duality that applies to abelian topological groups.  A closed abelian
group code $\C$ in a sequence space $\W$ may then be characterized as the
set of all sequences in $\W$ that are orthogonal to all sequences in the
dual code $\C^\perp$--Ñ \ie $\C$ may be characterized by a set of
constraints (``checks").

	An immediate consequence of this definition is that the dual of a 
complete code, namely a closed subgroup of a complete sequence space
such as $G^\Z$, is a finite code, namely a code all of whose sequences
are finite.  On the other hand, the dual of a Laurent code is
(anti-)Laurent.

	We derive fundamental duality relations between the dynamics of $\C$ and 
the dynamics of $\C^\perp$.  For example, the state spaces of $\C$ act as
the character groups of the state spaces of $\C^\perp$, and the
observability properties of $\C$ are the controllability properties of
$\C^\perp$.
(Here observability is defined as in \cite{LFMT94} as a property of a
code, not of a state space representation as in
\cite{W89}.)

	More precisely, we decompose the dynamics of a group code into observer 
granules, in a decomposition dual to the controller granule decomposition
of \cite{FT93}.

	Our original goal was to construct a minimal observer-form 
encoder and a minimal syndrome-former/state observer for $\C$ based on
its observability structure.  This is straightforward in many
particular cases, but surprisingly difficult in general. 
Fagnani and Zampieri \cite{FZ99} have succeeded in providing such
constructions for group codes over general finite nonabelian groups in a
purely algebraic setting.  Therefore we merely
present some general principles and examples of minimal observer-form
encoder and syndrome-former/state observer constructions.

	Finally, we show algebraically that every observer granule is isomorphic 
to an ``end-around" controller granule.  As corollaries, we obtain purely
algebraic proofs of many of our results.

	We should say that our restriction to abelian groups does not appear to 
us to be essential, except to allow the use of Pontryagin duality.  More
general notions of duality of nonabelian groups exist (see, \eg
\cite{ES92}), but are beyond us.  Most of the results of this paper do not
appear to depend on the abelian property.  (We show that
$\C$ has abelian dynamics if and only if $\C$ is normal in its output
sequence space;  however, normality appears to us to be no more
fundamental than abelianness.)  It is striking that the
syndrome-former construction of  \cite{FZ99}, like the
minimal encoder construction of  \cite{FT93}, applies to codes over
(finite) nonabelian groups and makes no use of duality, although it
employs the observability structure that we develop here.

% needed in second column of first page if using \pubid
\pubidadjcol
\pagebreak

	Section 2 briefly introduces Pontryagin duality.  Section 3 discusses dual
sequence spaces of several important types, namely
complete, finite and Laurent.  Section 4 discusses dual group codes,
proves that projections and subcodes are duals, gives dual
definitions of wide-sense controllability and observability, and presents
some examples of dual group codes.  Section 5 develops various results
about dual state spaces.  Section 6 is concerned with dual notions of
finite memory, including strong controllability and
observability.  Section 7 develops observability
decompositions into granules dual to the controllability decompositions of
\cite{FT93, LM96}.  Section 8 gives examples of the
construction of minimal observer-form encoders, state observers and
syndrome-formers.  Section 9 presents the end-around theorem and some
corollaries.  Section 10 is a brief conclusion.

\section{Pontryagin duality}

	Our treatment is based on Pontryagin duality, which applies to 
topological groups.  Pontryagin's original treatise \cite{P46} remains an
excellent reference.  For a more modern exposition, see any book on
Fourier (harmonic) analysis on groups;  \eg Rudin \cite{R90} or Hewitt and
Ross \cite{HR79}.

	A topological group is a group that is also a topological space, such 
that the group and topological properties are consistent.  We do not
expect the reader to have much background in topology.  We are not much
interested in the topology of individual symbol alphabets;  we usually
think of them as being finite or at least discrete and/or compact,
although we make more general statements when they appear to be warranted. 
However, topology does turn out to be important when considering codes
whose sequences are defined on infinite index sets, even with finite
symbol groups.  For an introduction to topology, see, \eg \cite{KF70} or
\cite{S75}.

	All topological groups in this paper will be assumed to be metric spaces;
\ie to have a topology induced by a distance function.  Group
homomorphisms will be assumed to be continuous, and group isomorphisms
will be assumed also to be homeomorphisms.  A subgroup of a topological
group is itself a topological group under the induced subspace topology,
but is considered to be a topological subgroup only if it is closed.

	In this section we review the two basic dualities of Pontryagin duality 
theory:  character group duality and orthogonal subgroup duality. 
Sequence space duality is defined in terms of the former, and code/system
duality in terms of the latter.  We also introduce some additional
fundamental duality principles:  direct product/direct sum duality,
sum/intersection duality, quotient group duality, and adjoint duality.

\subsection{Character group duality}

	A \emph{character} of a (topological) group $G$ is a (continuous)
homomorphism
$$
		h\colon G \to \R/\Z 
$$
from $G$ into the additive circle group (``1-torus") $\R/\Z$ (or
equivalently into the complex unit circle under multiplication, to which
$\R/\Z$ is isomorphic).

	The \textbf{character group} of $G$, denoted by
$\Gh$, is the set of all characters of $G$, with group operation defined
by 
$$
		(h_1 \circ h_2)(g) = h_1(g) + h_2(g).  
$$
Obviously $h_1 \circ h_2 = h_2 \circ h_1$, so $\Gh$ is abelian, and we
may use additive notation;  \ie the sum of two characters $h_1, h_2$ is
$h_1 + h_2$.  The identity of $\Gh$ is the zero (or principal) character
$0$, defined by
$0(g) = 0$ for all $g \in G$.  The inverse of $h \in \Gh$ is the
character $-h$ defined by $(-h)(g) = -h(g)$.  The characters of a group
$G$ are by definition unique, in the sense that no two characters $h_1,
h_2$ have equal values $h_1(g), h_2(g)$ for all $g \in G$.

	When $G$ is locally compact abelian (LCA), the fundamental
Pontryagin duality theorem holds:
\begin{theorem}[Pontryagin duality] Given an LCA group $G$, 
\begin{itemize}
	\item[(a)]  its character group $\Gh$ is LCA;
	\item[(b)]  the character group of $\Gh$ is naturally isomorphic to $G$: 
$\Ghh \cong G$.
\end{itemize}
\end{theorem}

	The natural isomorphism of this theorem associates $g \in G$ with the 
character $\phi_g \in \Ghh$ defined by $\phi_g(h) = h(g)$ for all $h \in
\Gh$.  The theorem says that the character group of $\Gh$ is precisely
the set of all such characters:  $\Ghh = \{\phi_g:  g \in G\}$.  In this
sense, we may say that $G$ \emph{acts as} the character group of $\Gh$,
and write
$\Ghh = G$ and $g(h) = h(g)$.

	Characters thus define a generalized inner product, called a
\emph{pairing},  from 
$\Gh \times G$ into $\R/\Z$, which we write as follows:
$$
		\inner{h}{g} = h(g) = g(h).
$$
A pairing satisfies the ``bihomomorphic" relationships
\begin{eqnarray*}
		\inner{0}{g} = \inner{h}{0} & = & 0; \\
		\inner{h_1 + h_2}{g} & = & \inner{h_1}{g} + \inner{h_2}{g}; \\
		\inner{h}{g_1 + g_2} & = & \inner{h}{g_1} + \inner{h}{g_2}.
\end{eqnarray*}
We say that $h \in \Gh$ and $g \in G$ are \textbf{orthogonal} if
$\inner{h}{g} = 0$.

	The \emph{character table} of $G$ (or of $\Gh$) is the ``matrix"
$$
		\inner{\Gh}{G} = \{\inner{h}{g} \mid h \in \Gh, g \in G\}.
$$
The ``rows" and ``columns" of this matrix are the ``vectors"
\begin{eqnarray*}
		\inner{h}{G} & = & \{\inner{h}{g} \mid g \in G\}; \\
		\inner{\Gh}{g} & = & \{\inner{h}{g} \mid h \in \Gh\},
\end{eqnarray*}
which explicitly specify the characters $h\colon  G \to \R/\Z \in \Gh$ and 
$g\colon \Gh \to \R/\Z \in G$, respectively.  The rows are distinct and
form a group under row addition that is naturally isomorphic to $\Gh$; 
similarly, the columns are distinct and form a group that is naturally
isomorphic to $G$.

	The elementary LCA groups in Pontryagin duality theory are the real 
numbers $\R$, the integers $\Z$, the circle group $\R/\Z$, and the finite
cyclic groups $\Z_m = \Z/m\Z$, which may be identified with the finite
subgroups
$(m^{-1}\Z)/\Z$ of $\R/\Z$.  The following table gives the corresponding
character groups and pairings:
$$
\begin{array}{l|l|l}
G & \Gh & \inner{h}{g} \\
\hline
\R	 & \R	& hg \mod \Z \mbox{~(in~} \R/\Z) \\
	\Z	& \R/\Z &	hg \mbox{~(in~} \R/\Z) \\
	\Z_m &	\Z_m &	hg \mbox{~(in~} \Z_m)
\end{array}
$$
	Note that in the cases of $\R$ and $\Z_m$, the character group $\Gh$ is
isomorphic to $G$;  however, in these cases we caution that the
isomorphism is not a ``natural" one.  Moreover, the case of $\Z$ and
$\R/\Z$ shows that $G$ and $\Gh$ need not even have the same cardinality.

	The fact that $\Z\h = \R/\Z$ illustrates an important general
result:  the character group of a discrete group is compact and
\emph{vice versa}
\cite{R90}.  Since a finite group with the discrete topology is both
discrete and compact, the character group of a finite group is finite; \eg
$(\Z_m)\h \cong
\Z_m$.

\subsection{Finite direct product duality}

	Let $\I$ denote a discrete index set, which throughout this section will
be finite.  We will often think of $\I$ as an ordered time axis, such as a
finite subinterval of $\Z$.  A set indexed by $\I$ such as  $\wb = \{w_k
\in G_k, k \in \I\}$ will correspondingly be called a \emph{sequence}.

 Given a finite set of LCA \emph{symbol groups}  $\{G_k, k \in \I\}$
indexed by
$\I$, their \textbf{direct product} is defined as the Cartesian product
set of all sequences $\wb = \{w_k \in G_k, k \in \I\}$, denoted by
$$
		\W = \prod_{k\in\I} G_k.
$$  
The group operation of $\W$ is defined
componentwise, using the symbol group operations.
If all $G_k$ are equal to a common group $G$, then we write $\W = G^\I$. 
If $|\I| = n$, then we may alternatively write $\W = G^n$.

The finite direct product $\W$ is
equipped with the natural product topology \cite{R90}.
  	If all $G_k$ are compact (resp.\ locally compact), then the
finite direct product
$\W = \prod_k G_k$ is compact (resp.\ locally compact) \cite{R90}.   
If all $G_k$ are discrete (resp.\ finite), then $\W$ is discrete (resp.
finite).

As expected, the character group of a finite direct product group is the
direct product of the symbol character groups:
\begin{theorem}[Finite direct product duality]  The character group
of a finite direct 
 product $\W = \prod_{k\in\I} G_k$ of LCA groups is the
finite direct product
$$\W\h = \prod_{k\in\I} G_k\h,$$ with pairing $\inner{h}{g}$
defined by the componentwise sum
$$
		\inner{h}{g} = \sum_{k\in\I} \inner{h_k}{g_k}, \quad h \in \W\h,
g \in \W.
$$
\end{theorem}

Note that  $\sum_{k\in\I} \inner{h_k}{g_k}$ is
well defined since $\I$ is finite.

	It follows that the character group of $G = \R^n$ is $\Gh = \R^n$, and
that the pairing $\inner{h}{g}$ between vectors $g \in \R^n, h \in \R^n$
is the ordinary inner (dot) product $h \cdot g$, mod $\Z$.

	Similarly, since every finite abelian group may be decomposed into a 
finite direct product of finite cyclic groups, it follows that every
finite abelian group $G$ is isomorphic to its
character group $\Gh$.  Moreover, if $m$ is the exponent of $G$ (the least
integer such that $mg = 0$ for all $g \in G$), then $G$ may be written as
a subgroup of $(\Z_m)^n$ for some $n$.  The character group of $(\Z_m)^n$
may be identified with $(\Z_m)^n$, and pairings may then be defined
in the usual manner as inner products over the ring $\Z_m$.

\subsection{Orthogonal subgroup duality}

	We now consider a second kind of duality, which will be the basis of our
definition of dual codes and systems.

	Let $G$ be an LCA group with character group $\Gh$, and let $S$ be a
subset of $G$.  The \textbf{orthogonal subgroup} to $S \subseteq G$ (the
\emph{annihilator} of $S$) is the set of all elements of $\Gh$ that are
orthogonal to all elements of $S$:
$$
		S^\perp = \{a \in \Gh \mid  \inner{a}{s} = 0 \mbox{~for all~} s \in S\}.
$$

	The orthogonal subgroup to $G$ itself is $G^\perp = \{0\}$, since the
zero character is the  unique character in $\Gh$ that is orthogonal to all
of $G$.  Similarly, $\{0\}^\perp = \Gh$.

	In topological groups, the group \emph{generated by} a subset $S \subseteq
G$ is defined as the smallest closed subgroup of $G$ that contains $S$,
called the
\emph{closure} $S\cl$ of $S$. $S$ is \emph{closed} if $S = S\cl$.  Thus in
topological groups the notion of closure involves both algebraic and
topological closure. 

	Orthogonal subgroups and closed subgroups are intimately linked by the 
following duality theorem \cite{N70}:
\begin{theorem}[Orthogonal subgroup duality]  If $G$ is an LCA group,
and $S$ is a subset of $G$,  then 
\begin{itemize}
	\item[(a)]  the orthogonal subgroup $S^\perp$ to $S$ is a closed subgroup
of
$\Gh$;
	\item[(b)]  the orthogonal subgroup $S^{\perp\perp}$ to $S^\perp$ is the
closure
$S\cl$ of $S$ in $G$.
\end{itemize}
\end{theorem}

It follows that  $S$ is a closed subgroup of $G$ if and only if
$S^{\perp\perp} = S$.  
Also, $S^{\perp\perp\perp} = S^\perp$.

	We shall say that two orthogonal closed subgroups $H
\subseteq G$ and $H^\perp \subseteq \Gh$ are \emph{dual subgroups}.  We
caution the reader that when we say that a group $H^\perp$ is the
orthogonal group to $H$, we do not imply that $H$ is closed, so that
$H^{\perp\perp} = H$.  However, if we say that two groups are dual or
orthogonal groups, then we imply mutual orthogonality, and thus that
both groups are closed.

This notion of duality is consistent with the usual definitions of duality
in a variety of contexts:
\begin{itemize}
\item  If $G = \R^n$ and $H$ is a \emph{subspace} of $G$ as a vector space
over
$\R$, then $H^\perp$ is the \emph{orthogonal subspace} to $H$ in $\Gh =
\R^n$. 
\emph{Proof}:  for $\gb
\in G$ and $\ab \in \Gh$, the pairing $\inner{\ab}{\gb}$ is the ordinary
dot product $\ab \cdot \gb$, mod $\Z$.  But a subspace $H$ of $G$ is
scale-invariant;  \ie $\hb
\in H$ implies $\alpha \hb \in H$ for all $\alpha \in \R$.  Now $\ab \cdot
\alpha\hb \equiv 0 \mod \Z$ for all $\alpha \in \R$ if and only if $\ab
\cdot \hb = 0$.  Thus 
$$
		H^\perp = \{\ab \in \Gh \mid \ab \cdot \hb = 0 \mbox{~for all~} \hb \in
H\}, 
$$
which is the usual definition of the orthogonal subspace to $H$.
	 \item   If $G = \R^n$ and $H$ is a \emph{lattice} in $\R^n$ (a discrete
subgroup of $\R^n$), then $H^\perp$ is the \emph{dual lattice} in $\Gh =
\R^n$.  \emph{Proof}:  Since
$\inner{\ab}{\gb} = \ab \cdot \gb \mod \Z$, 
$$
		H^\perp = \{\ab \in \Gh \mid  \ab \cdot \hb \equiv 0 \mod \Z 
\mbox{~for all~} \hb \in H\}, 
$$
which is the usual definition of the dual lattice to $H$.
	 \item   If $G = (\Z_m)^n$ and $H$ is a subgroup (a \emph{linear block
code} of length $n$ over $\Z_m$), then $H^\perp$ is the \emph{dual linear
block code} in $\Gh = (\Z_m)^n$.  \emph{Proof}:  Here the pairing
$\inner{\ab}{\gb}$ is the usual inner product over the ring $\Z_m$.
\end{itemize}

	It is important to distinguish character group duality from 
orthogonal subgroup duality.  The character group $\Gh$ is often called
the ``dual group" to $G$ in the mathematical literature.  However, these
examples show that the terms ``dual code" and ``dual lattice" are to be
understood in the orthogonal subgroup sense.  We use both types of
duality in this paper;  for example, we use the term ``dual sequence
space" in the character group sense, whereas we use the terms ``dual code"
and ``dual system" in the orthogonal subgroup sense.  We caution the
reader to keep this distinction in mind, and to refer to the notation
if in doubt.

\subsection{Sum/intersection duality}

	Let $G$ be a topological group, and let $\{S_j \subseteq G, j \in \J\}$
be a collection of subsets of $G$ indexed by an index set $\J$, possibly
infinite.  For topological groups, the group \emph{generated by} the
collection, called the \emph{sum} of the subsets $\{S_j\}$ and denoted  by
$\sum_{j\in\J} S_j$, is defined as the closure $S\cl$ of the set $S$ of
all finite sums $\sum_{j\in\J} s_j$, where $s_j$ denotes an element of
$S_j$.  Thus the sum (the group generated by the $S_j$) is closed both
algebraically and topologically.

	Let $\{S_j^\perp \subseteq \Gh, j \in \J\}$ be the collection of
orthogonal subgroups to the subsets $\{S_j, j \in \J\}$.  The intersection
$\bigcap_{j \in \J} S_j^\perp$ of this set of closed subgroups is a closed
subgroup of $\Gh$.  Moreover, by orthogonal subgroup duality, it is the
orthogonal group to the sum
$\sum_{j \in \J} S_j$:
\begin{theorem}[Sum/intersection duality]  
$$(\sum_{j \in \J} S_j)^\perp =
\bigcap_{j \in \J} S_j^\perp; \qquad \sum_{j \in \J} S_j = (\bigcap_{j \in
\J} S_j^\perp)^\perp.
$$
\end{theorem}

\emph{Proof}.  Let $S$ be the set of all finite sums $\sum_j s_j$ for all
$s_j \in S_j$.  Then $S^\perp = \bigcap_j S_j^\perp$, since  $h \in \Gh$
is orthogonal to $S$ if and only if $h$ is in all orthogonal subgroups
$S_j^\perp$.  But by definition
$\sum_j S_j = S\cl$, and by orthogonal subgroup duality $S\cl =
S^{\perp\perp} = (\bigcap_j S_j^\perp)^\perp$.	\qed

	This theorem applies particularly when the subsets $S_j$ consist of
single elements $s_j \in G$, called \emph{generators}.  The orthogonal
subgroup to $S_j$ is then the set of elements $a \in \Gh$ that
pass the test $\inner{a}{s_j} = 0$, called a \emph{check} (or
constraint).  This theorem then says that the
orthogonal subgroup to the subgroup generated by the generators $s_j, j
\in \J$, is the set of
$a \in \Gh$ that satisfy all checks $\inner{a}{s_j} = 0, j \in \J$.

\subsection{Quotient group duality}

	Let $H$ and $H^\perp$ be dual (closed) subgroups in $G$ and $\Gh$.  Every
character $g$ in the  character group $G$ of $\Gh$ is evidently a character
of $H^\perp$.  However, since for a given $h \in H^\perp$
$$
\inner{h}{g} = \inner{h}{g'} \Leftrightarrow \inner{h}{g - g'} = 0,
$$
two characters $g, g' \in G$ of $H^\perp$ are identical if and only if $g -
g' \in H$, the orthogonal subgroup to $H^\perp$.  Thus the characters of
$H^\perp$ naturally correspond one-to-one to the cosets $H + r$ of $H$ in
$G$, which form the quotient group $G/H$.  Indeed, it is easy to verify
that the correspondence $(H^\perp)\h
\lra G/H$ is an isomorphism.  In this sense, the quotient group $G/H$ acts
as the character group of $H^\perp$, with pairing defined by $\inner{h}{H
+ r} = \inner{h}{r}$, just as $G$ acts as the character group of $\Gh$. 
Correspondingly, $H^\perp$ acts as the character group of $G/H$ with the
same pairing \cite{V76}.  
\begin{theorem}[Subgroup/quotient group duality]  If $H$ and $H^\perp$
are dual closed subgroups in $G$ and $\Gh$, then $G/H$ acts as the
character group of $H^\perp$ and \emph{vice versa}: 
\end{theorem} 
$$
(H^\perp)\h = G/H; \qquad (G/H)\h = H^\perp.
$$

	For example, if $H$ is a subspace of $G = \R^n$, and $H^\perp$ is
its orthogonal subspace, then this theorem implies that
$\dim H^\perp = \dim G - \dim H$.

We note that each element of a group $G$ with a subgroup $H$ may be
written uniquely as $g = r + h$, where $r$ is a representative of the coset
$H + g \in G/H$ and $h \in H$.  There is thus a one-to-one correspondence
between $G$ and the Cartesian product $H \times G/H$, which may be viewed
as a decomposition of $G$ into two components, $H$ and $G/H$. However, the
two components play different roles.  In general, $G/H$ is not a subgroup
of
$G$; moreover, $G$ may have no subgroup isomorphic to
$G/H$.  For example, $\R$ has no subgroup isomorphic to $\R/\Z$.  Note
that although the character group $\Gh$ may similarly be thought of as
being composed of $\Hh$ and $(G/H)\h$, the two components exhange roles: 
$(G/H)\h = H^\perp$ is by definition a subgroup of $\Gh$, whereas $\Hh$ is
the quotient $\Gh/H^\perp$, which in general is not a subgroup of $\Gh$.

	This result may be straightforwardly extended to the quotients of a 
finite chain $J \subseteq H \subseteq G$ of closed subgroups of $G$. 
Since $h \in H^\perp$ implies $h \in J^\perp$, the orthogonal subgroup
chain runs in the reverse order:  $H^\perp \subseteq J^\perp \subseteq
\Gh$.  For $g \in H, h \in J^\perp$, the value of the pairing
$\inner{h}{g}$ depends only on the cosets $J + g, H^\perp + h$ of $J$ and
$H^\perp$ in $H$ and $J^\perp$, respectively.  Therefore $H/J$ and
$J^\perp/H^\perp$ act as dual character groups, with pairing defined by
$\inner{H^\perp + h}{J + g} = \inner{h}{g}$.  In summary:
\begin{theorem}[Quotient group duality]  If $J \subseteq H \subseteq G$,
then the dual  quotient group $J^\perp/H^\perp$ to $H/J$ acts as the
character group of $H/J$:  $(H/J)\h = J^\perp/H^\perp$.	
\end{theorem}

Quotient groups such as $H/J$ and $J^\perp/H^\perp$ will be called
\textbf{dual quotient groups}. 

	The \emph{dual diagrams} below illustrate two chains of subgroups, with their
quotients.   The right chain is obtained by inverting the left chain,
replacing subgroups by their orthogonal subgroups, and replacing quotient
groups by their character groups.
$$
\begin{array}{clccccl}
G &&&&& \{0\}^\perp = \Gh & \\
| & G/H &&&& | & \Gh/J^\perp = J\h \\
H &&&&& J^\perp & \\
| & H/J &&&& | & J^\perp/H^\perp = (H/J)\h \\
J &&&&& H^\perp & \\
| & J &&&& | & H^\perp = (G/H)\h \\
\{0\} &&&&& G^\perp = \{0\} & 
\end{array}
$$
	The following dual diagrams illustrate the chain of elementary groups
$
		\{0\} \subseteq m\Z \subseteq \Z \subseteq \R, 
$
whose quotients are $m\Z \cong \Z, \Z/m\Z = \Z_m$, and $\R/\Z$, and its
dual chain 
$$
		\{0\} \subseteq \Z^\perp = \Z \subseteq (m\Z)^\perp = m^{-1}\Z
\subseteq \R\h = \R, 
$$
whose quotients are congruent to $\Z \cong (\R/\Z)\h,
\Z_m \cong (\Z_m)\h$, and $\R/\Z \cong (m\Z)\h$,
respectively.  Indeed, the dual chain is just the primal chain scaled by
$m^{-1}$.

$$
\begin{array}{clcl}
\R && \R\h = \R & \\
| & \R/\Z & | & \R/(m^{-1}\Z) \cong \R/\Z \\
\Z && (m\Z)^\perp = m^{-1}\Z & \\
| & \Z/m\Z = \Z_m & | & (m^{-1}\Z)/\Z \cong \Z_m \\
m\Z && \Z^\perp = \Z & \\
| & m\Z \cong \Z & | & \Z \\
\{0\} && \{0\} & 
\end{array}
$$

\subsection{Adjoint duality}  \label{ad}

Quotient group duality is a special case of a general duality principle
for adjoint homomorphisms.

Let $\phi\colon  G \to U$ be a homomorphism of an LCA group $G$ to another 
LCA group $U$.  The \emph{adjoint homomorphism} $$\phi^*\colon  U\h \to G\h$$ 
is the unique homomorphism such that $\inner{v}{\phi(g)} =
\inner{\phi^*(v)}{g}$ for all $g \in G, v \in U\h$, where $G\h$ and $U\h$
are the character groups of $G$ and $U$, respectively.  Explicitly, the
adjoint character
$\phi^*(v)$ is the unique character in $\Gh$ whose values are given by
$\phi^*(v)(g) = \inner{v}{\phi(g)}$.  Evidently the adjoint of $\phi^*$ is
$\phi$;  \ie $\phi^{**} = \phi$.

For example, let $H$ be a closed subgroup of $G$, and let $\phi\colon  G 
\to G/H$ be the natural map defined by $\phi(g) = H + g$.  Since $H^\perp$
acts as the character group of $G/H$, with 
$\inner{v}{H + g} = \inner{v}{g}$ for $g \in G, v \in H^\perp \subseteq
\Gh$, the adjoint $\phi^*\colon H^\perp \to \Gh$ is the inclusion
of $H^\perp$ into $\Gh$.

The fundamental adjoint duality theorem is as follows:
\begin{theorem}[Adjoint duality] \label{ad}
Given  adjoint homomorphisms $\phi\colon  G \to U$, $\phi^*\colon   U\h
\to G\h$, the kernel of $\phi$ is the orthogonal subgroup in $\Gh$ to the
image of $\phi^*$.
\end{theorem}

\emph{Proof}.
We show that $g \in (\im \phi^*)^\perp$ if and only if $g \in \ker \phi$.
Let $g \in \ker \phi$;  \ie $\phi(g) = 0$.  Then
$\inner{\phi^*(v)}{g} = \inner{v}{\phi(g)} = 0$; \ie every $g \in \ker
\phi$ is orthogonal to $\phi^*(v) \in \Gh$ for all $v \in U\h$. 
Conversely, if $g$ is not in $\ker \phi$, then $\phi(g) \neq 0$, so
$\inner{\phi^*(v)}{g} = \inner{v}{\phi(g)} \neq 0$ for some $\phi^*(v) \in
G\h$, because $0 \in U$ is the unique character $u \in U = U\h\h$ such
that $\inner{v}{u} = 0$ for all $v \in U\h$.
\qed

Note that whereas the kernel of $\phi$ is necessarily closed, the image of
$\phi^*$ may not be closed;  the orthogonal subgroup to $\ker \phi$ is
therefore the closure of $\im \phi^*$.

In our example, the kernel $H$ of the natural map $\phi\colon  G \to
G/H$ is indeed the orthogonal subgroup in $G$ to the image $H^\perp$ of the
inclusion $\phi^*\colon H^\perp \to \Gh$.  Also, the kernel of $\phi^*$
is $\{0\} \subseteq H^\perp$ and the image of $\phi$ is the trivially
orthogonal subgroup $G/H = (H^\perp)\h$ in $(H^\perp)\h$.

 The decomposition of $G$ into $H$ and $G/H$
is sometimes illustrated by the following
\emph{short exact sequence}:
$$
\{0\} \to H \to G \to G/H \to \{0\},
$$
where the first two maps are inclusions and the second two are natural
maps.  (``Exact" means that the image of each map is the kernel
of the next.)  The adjoint short exact sequence 
$$
\{0\} \to (G/H)\h = H^\perp \to G\h \to  H\h = G\h/H^\perp \to \{0\},
$$
 illustrates the exchange of roles upon which we previously remarked.

A subgroup chain such as $\{0\} \subseteq J \subseteq H \subseteq G$
implies a chain of inclusion maps, \eg $$\{0\} \to J \to H \to G.$$  The
adjoint chain runs in the opposite direction, $$G\h \to H\h \to J\h \to
\{0\},$$ and consists of a chain of natural maps with kernels $H^\perp =
(G/H)\h, J^\perp/H^\perp = (H/J)\h$, and $\Gh/J^\perp = J\h$, illustrating
the same decomposition of $\Gh$ as in the first dual diagram above.

\section{Dual sequence spaces}

	A group code or system (behavior) $\C$ is a subgroup of a sequence space
$\W$.  In this section we define complete, Laurent and finite topological
sequence spaces, and determine their character groups (dual sequence
spaces) $\W\h$.  We briefly discuss more general memoryless
sequence spaces.

\subsection{Complete and finite sequence spaces}

	We now let the discrete index $\I$ be possibly
countably infinite: \eg $\I = \Z$.  In general, $\I$ need not be ordered; 
for example, we could consider an $n$-dimensional index set such as $\I =
\Z^n$.  However, for simplicity we will assume $\I \subseteq \Z$ from now
on.  We will continue to call a set indexed by
$\I$ a \emph{sequence}.

 Given a set of LCA symbol groups  $\{G_k, k \in \I\}$ indexed by $\I$,
their \textbf{direct product} is again defined as the Cartesian product
set of all sequences $\wb = \{w_k \in G_k, k \in \I\}$, now denoted by
$$
		\W^c = \prod_{k\in\I} G_k.
$$ 
We call a direct product $\W^c$ a \textbf{complete sequence space}.
Its group operation is still defined
componentwise.  We continue to write $\W^c = G^\I$ if all symbol groups are
equal to $G$. 

The complete sequence space $\W^c$ is equipped with the natural product
topology \cite{R90}.     	If all symbol groups
$G_k$ are compact, then under the product topology $\W^c$ is
compact.  However, even when all symbol groups are locally compact, $\W^c$
need not be locally compact \cite{R90}.  

In topology, ``completeness" is a property of metric spaces (every Cauchy
sequence converges).  A
\emph{metric space} is a topological space whose topology is induced by a
\emph{distance function} $d(\cdot, \cdot)$ that satisfies the distance
axioms: strict positivity, symmetry, and the triangle inequality.  

For example, if $\I \subseteq \Z$ and all $G_k$ are discrete, then the
product topology is induced by the distance metric $$d(\wb, \wb') =
2^{-\l(\wb, \wb')},$$ where
$\l(\wb, \wb')$ is the least absolute value $|k|$ of an index $k \in \I$
such that $w_k \neq w_k'$.  In other words, two sequences are regarded as
``close" if they agree over a large central interval.  In this case the
product topology is also called the \emph{topology of pointwise
convergence}, because a series $\{\wb^{n}, n \in \mathbb{N}\}$ converges
to $\wb$ if and only if, for all $k \in \I$, $w^{n}_k = w_k$ for all
sufficiently large $n$.

In general, a topological direct product $\W^c = \prod_\I G_k$ is complete
if and only if all $G_k$ are complete \cite[II.3.5]{S75}.  
%If $G_k$ is LCA, then $G_k$ is complete if it is discrete or has a topology
%with a countable basis for its open sets \cite[II.3.9]{S75}.
We will therefore assume from now on that all symbol groups $G_k$ are
complete metric spaces.  Moreover, a countable direct product $\W^c$ of
complete metric spaces is metrizable (can be endowed with a metric under
which it is a metric space) \cite[II.3.8]{S75}.

In a complete metric space, a subspace is complete if and only if it is
closed \cite[II.3.2]{S75}.  Since all sequence spaces we consider will be
complete metric spaces, we will generally use the term
``closed" rather than ``complete" for subspaces.  We will reserve the
term  ``complete" to mean ``closed in the product topology;" \ie as a
subspace of a complete sequence space $\W^c$.  

In behavioral system theory, a behavior $\C \subseteq \W^c$ is called
``complete" if whenever a sequence $\wb \in \W^c$ satisfies all finite
$\C$-checks, then $\wb \in \C$.  As we will discuss in Section 4.6, this
notion of completeness usually coincides with the topological definition,
but may need to be generalized.

	On the other hand, the \textbf{direct sum} of the symbol groups $\{G_k, k
\in \I\}$ is defined as the subset of $\W^c$ comprising the sequences 
$\wb = \{w_k\}$ in which only finitely many symbol values $w_k$ are
nonzero (sometimes called the set of ``Laurent polynomials" in system
theory), denoted by
$$
		\W_f = \bigoplus_{k\in\I} G_k.
$$ 
We will call a direct sum $\W_f$ a \textbf{finite sequence space}.
Sums are still defined componentwise, and  $\W_f$ is
evidently closed under finite sums.  If all
symbol groups are equal to a common group $G$, then we write $\W_f =
(G^\I)_f$.  

The direct sum $\W_f$ is equipped with the natural sum topology
\cite{R90}.	If all $G_k$ are discrete, then the sum topology  is simply
the discrete topology (the topology induced by the Hamming metric).  Such
a setting is purely algebraic, with no additional topological structure.  
If all symbol groups are complete, then $\W_f$ is topologically complete
under the sum topology.

	If $\I$ is finite, then there is no distinction between a  direct product
$\W^c$ and the corresponding direct sum $\W_f$, either algebraically or
topologically.  However, if $\I$ is infinite, then $\W_f$ is a proper
subset of $\W^c$, and the  sum topology of
$\W_f$ is in general different from the topology of $\W_f$ as a
subspace of $\W^c$.  In particular, $\W_f$ is not closed in
$\W^c$, and its closure is $(\W_f)^c = \W^c$, where
the first superscript ``$c$" denotes closure or completion in $\W^c$.

\subsection{Direct product/direct sum duality}

	Although an infinite direct product of LCA groups is not necessarily
LCA, the following duality theorem nevertheless holds \cite{K48}:
\begin{theorem}[Direct product/direct sum duality]  The character group
of a direct
 product $\W^c = \prod_{k\in\I} G_k$ of LCA groups is the direct
sum
$$(\W^c)\h = \bigoplus_{k\in\I} G_k\h,$$  with pairing $\inner{h}{g}$
defined by the componentwise sum
$$
		\inner{h}{g} = \sum_{k\in\I} \inner{h_k}{g_k}$$ for $h \in (\W^c)\h,
g \in \W^c.$
\end{theorem}

Note that the sum $\sum_{k\in\I} \inner{h_k}{g_k}$ is
well defined, since only finitely many $h_k$ are nonzero.

In other words, the dual of a complete sequence space is the finite
sequence space with the dual symbol groups, and \emph{vice versa}.

\subsection{Laurent sequence spaces}

	In convolutional coding theory and classical linear system theory, all 
sequences are usually semi-infinite Laurent sequences--- \ie sequences
that have only finitely many nonzero symbol values before some arbitrary
time, say $k = 0$, or equivalently that have a definite ``starting
time."  

A
natural definition of a \textbf{Laurent sequence space} is the direct
product of a finite sequence space defined on the ``past," $\I^{-} = \{k
\in \I\mid k < 0\}$ and a complete sequence space defined on the
``future," $\I^+ = \{k \in \I\mid k \ge 0\}$:
$$
		\W_{L} = \left(\bigoplus_{k\in\I^{-}} G_k\right) \times
\left(\prod_{k\in\I^{+}} G_k\right),
$$
We call $\W_L$ the \emph{Laurent product} of the symbol
groups $\{G_k, k
\in \I\}$.

Similarly, we define an \emph{anti-Laurent sequence space} by
the \emph{anti-Laurent product}
$$
		\W_{\tilde{L}} = \left(\prod_{k\in\I^{-}} G_k\right) \times
\left(\bigoplus_{k\in\I^{+}} G_k\right).
$$

	By direct product/direct sum duality, it is immediate that the dual of a
Laurent  sequence space is an anti-Laurent sequence space:
\begin{theorem}[Laurent/anti-Laurent duality].  The
anti-Laurent sequence space $\X_{\tilde{L}} = \left(\prod_{k\in\I^{-}}
G_k\h\right) \times
\left(\bigoplus_{k\in\I^{+}} G_k\h\right)$ acts as the character
group of the Laurent sequence space $\W_L = \left(\bigoplus_{k\in\I^{-}}
G_k\right) \times
\left(\prod_{k\in\I^{+}} G_k\right)$, and \emph{vice versa}:
$(\W_L)\h = \X_{\tilde{L}}$.	
\end{theorem}

	Note that in this case, for $\xb \in \X_{\tilde{L}}, \wb \in \W_L$,
the pairing $\inner{\xb}{\wb} = \sum_{k\in\I} \inner{x_k}{w_k}$ is well
defined, because only finitely many pairings $\inner{x_k}{w_k}$ are
nonzero. 

	It is customary to reverse the direction of time in the dual sequence 
space $\X_{\tilde{L}}$, so that it also becomes a Laurent
sequence space.  This yields a nice symmetry between the primal and dual
spaces, which is lacking for the complete/finite pair.

\subsection{Memorylessness}

Memorylessness is a set-theoretic property of a subset $\V$ of a
Cartesian product sequence space $\W^c = \prod_{k \in \I} G_k$.
The subset $\V$ will be called \textbf{memoryless} if for any
partition of the index set $\I$ into two disjoint subsets $\J$ and
$\I-\J$, if  $\V_{|\J}$ and $\V_{|\I-\J}$ are the corresponding
restrictions of $\V$ (see Section 4.3), then
$\V$ is the Cartesian product
$$
		 \V = \V_{|\J} \times \V_{|\I-\J}.
$$ 

In general, $\V$ will be called a
\emph{sequence space} if and only if $\V$ is memoryless.  It is easily
verified that complete, finite and Laurent sequence spaces are
memoryless. 

	Another example of a memoryless sequence space is the set
$l_{2}$ of all square-summable sequences in a real or complex complete
sequence space $\W^c$.  The character group of $l_{2}$ is the dual
square-summable sequence space $l_{2}$.  More
generally, for
$1 \le p  \le \infty$, the set $l_{p}$ of all $p$-power-summable
sequences is memoryless, and its character group is $(l_{p})\h =
l_{q}$, where $\frac{1}{p} + \frac{1}{q} = 1$ \cite{R90}.

	Given a set of symbol groups $\{G_k, k \in \I\}$, the direct product
$\W^c = \prod_{k\in\I} G_k$ is clearly the largest possible sequence
space with these symbol groups, since it consists of all possible 
 sequences $\wb$
such that $w_k \in G_k$ for all $k \in \I$.  Conversely, the direct sum
$\W_f =
\bigoplus_{k\in\I} G_k$ is the smallest memoryless sequence space $\V$
such that $\V_{|k} = G_k$ for all $k \in \I$, since by memorylessness
the finite sequence $(\prod_{j\in\J} \V_{|j})
\times {0}_{|\I-\J}$ must be in $\V$ for any finite $\J \subseteq \I$. 
It follows that if $\I$ is finite, then $\W^c = \W_f$ is the only possible
memoryless sequence space with symbol groups $\{G_k\}$.

\section{Dual group codes}

	A group code, system or behavior is a subgroup $\C$ of a sequence space
$\W$. 
In the topological group setting, it is natural to define a
\emph{topological group code} or system to be a \emph{closed} subgroup of
a topological sequence space.  Therefore, unless stated otherwise, the term
\textbf{group code} will hereafter mean a closed subgroup $\C$ of a
complete, finite or Laurent sequence space $\W$.

In this section we establish the basic duality between a closed group
code $\C$ and its dual code $\C^\perp$.  This shows that the dual code
of a complete code is a finite code, and \emph{vice versa}.  We show that
if $\C$ has certain symmetries such as linearity or time-invariance, then
so does $\C^\perp$.  We prove a basic projection/subcode duality
theorem.  A more general
principle is conditioned subcode duality, which can be regarded as a
fundamental behavioral control theorem.  We discuss the meaning of
completeness in both a topological and behavioral sense, and agree to
define completeness here as closure in a complete sequence space
(\ie closed in the product topology).  Completeness is then dual to
finiteness.  We briefly discuss Laurent completion and ``Laurentization." 
Finally, we define dual notions of controllability and observability,
based on the notions of completion and finitization.  Several example
codes are given to illustrate these concepts.

\subsection{Group code duality}

	We define the \textbf{dual code} $\C^\perp$ to a group code $\C \subseteq
\W$ as the orthogonal subgroup to $\C$ in the dual sequence
space $\W\h$.  
By orthogonal subgroup duality, we have immediately:
\begin{theorem}[Group code duality]  
\label{gcd}
If $\C \subseteq \W$ is a (closed)
group code, then its dual $\C^\perp$ is a (closed)
group code in $\W\h$, and $\C^{\perp\perp} = \C$.  	
\end{theorem}

	Thus, given $\W$, a group code $\C$ is completely
characterized by its dual code $\C^\perp$, and \emph{vice versa}.
Moreover, the dual code of a complete code is a finite code, and
\emph{vice versa}.

If all symbol groups $G_k$ are discrete, then the finite sequence space
$\W_f = \bigoplus_\I G_k$ is discrete, so every subgroup $\C$ of $\W_f$ is
closed.  In other words, this discrete setting is purely algebraic and
topology may be ignored, even when $\I$ is infinite.

	The dual sequence space of $\W_f = \bigoplus_\I
G_k$ is the complete sequence space $(\W_f)\h = \prod_\I G_k\h$.  If each $G_k$ is discrete, then each $G_k\h$ is compact and
$(\W_f)\h$ is compact.  By the orthogonal subgroup duality theorem,
the closed subgroups of $(\W_f)\h$ are precisely those subgroups
that are duals of group codes in $\W_f$.  

	Thus whereas all subgroups of $\W_f$ are closed, only certain
subgroups of $(\W_f)\h$ are closed.  This asymmetry should not be
surprising, since even if $G_k\h \cong G_k$, the complete sequence
space $(\W_f)\h$ is much larger than the finite sequence space
$\W_f$, and by Theorem \ref{gcd} there is a one-to-one correspondence
between codes in $(\W_f)\h$ and codes in $\W_f$. 

Behavioral system theory has traditionally restricted itself to
complete behaviors.\footnote{Indeed, Willems \cite[p.\ 567]{W86} has
asserted, no doubt whimsically, that ``the study of non-complete systems
does not fall within the competence of system theorists and could be
better left to cosmologists or theologians\ldots."}  But we observe that
the dual of a complete group behavior $\C \subseteq \W^c$ is a finite
behavior $\C^\perp \in (\W^c)\h$.  Thus any theory that encompasses both
complete behaviors and their duals must encompass non-complete behaviors,
particularly finite behaviors.

\subsection{Linearity and time-invariance}

In this subsection we briefly discuss the important properties of
linearity and time-invariance.  As in \cite{FT93}, linearity and
time-invariance play no essential role in our development,
although we often use linear and/or time-invariant codes as examples.
Within our group-theoretic framework, linearity and time-invariance are
simply additional symmetries of a group code, which are reflected by dual
symmetries in the dual group code.

A group code $\C \subseteq (\R^n)^\I$ over the real field $\R$ is
\emph{linear} if it is invariant under all isomorphisms $\alpha\colon 
(\R^n)^\I \to (\R^n)^\I$ defined by scalar multiplication by a nonzero
scalar $\alpha \neq 0 \in \R$.  Since  $\inner{\xb}{\alpha\wb} =
\inner{\alpha\xb}{\wb}$, the dual $\C^\perp$ of a linear code $\C$ is
linear.

Similarly, a group code $\C \subseteq \W$ is \emph{time-invariant} (or
shift-invariant) if the time axis is $\I = \Z$, if all symbol groups are
the same, and if
$\C$ is invariant under the delay isomorphism $D\colon \W \to \W$ defined by
$D(\wb)_{|k} = w_{k-1}$; \ie if $D\C = \C$.  Since  $\inner{\xb}{D(\wb)} =
\inner{D^{-1}(\xb)}{\wb}$, the dual $\C^\perp$ of a time-invariant
group code $\C$ satisfies $D^{-1}\C^\perp = \C^\perp$ and is thus
time-invariant.  

If $\C \subseteq (\R^n)^\Z$ is both linear and time-invariant, then
$\inner{\xb}{\wb} = (\tilde{\xb}\ast\wb)_0$, where
$\tilde{\xb}$ is the time-reverse of $\xb$ and ``$\ast$" denotes
convolution.  More generally,
$\inner{\xb}{D^k(\wb)} = (\tilde{\xb}\ast\wb)_k$.  It follows that $\xb$
is in $\C^\perp$ if and only if the convolution $\tilde{\xb}\ast\wb$ is
the zero sequence $\zerob$ for all $\wb \in \C$.  This shows 
that pairings of linear time-invariant code sequences may be
evaluated by sequence convolutions, and further motivates inverting the
direction of time in the dual sequence space $\W\h$.

\subsection{Restrictions, projections and subcodes}  \label{rps}

In \cite{FT93}, we asserted that projections and subcodes of a group
code $\C$ play dual roles.  This will turn out to be our key dynamical
principle.

  Let $\W$ be a sequence space defined on an index set $\I$, let $\J
\subseteq \I$ be a subset of $\I$, and let $\I - \J$ be the
complementary subset.  

  The \textbf{restriction} $R_\J\colon \W \to \W_{|\J}$ defined by 
$R_\J(\wb) = \wb_{|\J} = \{w_k, k \in \J\}$ is a continuous homomorphism. 
Since $\W$ is memoryless, $\W = \W_{|\J} \times \W_{|\I - \J}$, the image
of the homomorphism is $\W_{|\J}$ and its kernel is $\{\zerob\}_{|\J}
\times \W_{|\I-\J}$.  
The topology of $\W_{|\J}$ is induced from that of $\W$.

  The \textbf{projection} $P_\J\colon \W \to \W$ is an essentially identical
map defined by $P_\J(\wb) = (\wb_{|\J}, \zerob_{|\I - \J})$, a continuous
homomorphism with the same kernel whose image is $P_\J(\W) = \W_{|\J}
\times \{\zerob\}_{|\I -\J}$.

  Let $\C$ be a closed subgroup of $\W$.  Then the kernel of either the
restriction $R_\J\colon \C \to \W_{|\J}$ or the projection  $P_\J\colon 
\C \to \W$ is the \textbf{subcode} $\C_{:\I-\J} = \C \cap (\{\zerob\}_{|\J}
\times \W_{|\I-\J})$, namely the set of all code sequences $\wb
\in \C$ such that $w_k = 0$ when $k \in \J$.  As the  kernel of a 
continuous homomorphism of $\C$, a subcode $\C_{:\I-\J}$ is a closed
subgroup of $\C$.

  Similarly, the restriction $\C_{|:\I-\J} = (\C_{:\I-\J})_{|\I-\J}$ of the
subcode
$\C_{:\I-\J}$ to $\I - \J$, which is isomorphic to $\C_{:\I-\J}$, is a
closed subgroup of the \emph{restricted code} $\C_{|\I - \J} = R_{\I -
\J}(\C)$.

  By the fundamental homomorphism theorem, the image
$\C_{|\J}$ of %the restriction 
$R_\J\colon \C \to \W_{|\J}$ (or the image
$\C_{|\J} \times \{\zerob\}_{|\I - \J}$ of % the projection 
$P_\J\colon \C \to \W$) is algebraically isomorphic to the quotient group
$\C/\C_{:\I-\J}$.

However, we caution that in certain atypical cases the
topology of the restriction $\C_{|\J}$ as a subspace of
$\W_{|\J}$ is not necessarily consistent with the topology of the quotient
group $\C/\C_{:\I-\J}$.  In particular, even though $\C/\C_{:\I-\J}$ is
necessarily closed, $\C_{|\J}$ may not be closed in $\W_{|\J}$.

\medskip

\noindent
\textbf{Example 1}.  Let $\W = \R^2$, and let $\C$ be an irrational
lattice in $\R^2$;  \eg the lattice 
$$
		\C = \{(am + bn, -bm + an) \mid (m, n) \in \Z^2\}, 
$$
where the ratio $a/b$ is irrational.  $\C$ is discrete, and thus a closed
subgroup of $\R^2$. 
The restriction $\C_{|\J}$ of $\C$ to either coordinate is
$
		\C_{|\J} = \{am + bn \mid (m, n) \in \Z^2\}. 
$
The kernel of the restriction is $\C_{:\I-\J} = \{\zerob\}$, since $am + bn
= 0$ implies $m = n = 0$ when $a/b$ is irrational.  Thus $\C/\C_{:\I-\J}$
is discrete and  homeomorphic to $\Z^2$.  

On the other hand, as a subspace of $\W_{|\J} = \R$, the restriction
$\C_{|\J}$ is not closed, but rather is a dense subgroup of $\R$ whose
closure  is
$(\C_{|\J})\cl = \R$.   Thus these two topologies are inconsistent.

Notice that, by orthogonal subgroup duality, $(\C_{|\J})^\perp = \{0\}$
and $(\C_{|\J})^{\perp\perp} = \R$.  Therefore
projection/subcode duality (see next subsection) holds in the form
$\C_{|:\J} = (\C_{|\J})^\perp$, even though $(\C_{|:\J})^\perp \neq
\C_{|\J}$ (rather,  $(\C_{|:\J})^\perp = (\C_{|\J})\cl$).  \qed

It can be shown that a restriction $\C_{|\J}$ is closed in $\W_{|\J}$
if the sequence space $\W$ is discrete (because all subgroups are closed
in the discrete topology), or if $\W$ is compact (the dual to the discrete
case;  see Section 5.3), or if $\W = (\R^n)^\I$ and $\C$ is a subspace
(since subspaces of $\R^n$ are closed in the Euclidean
topology).  As these are the cases of most interest in coding and
system theory, the potential  pathology illustrated by Example 1 may
usually be ignored; 
\ie restrictions and projections are usually closed
subgroups of their respective sequence spaces.  We discuss this point
again in Section 5.3.

\subsection{Projection/subcode duality}

The results of this subsection follow from the simple observation that 
for $\wb \in \W, \xb \in \W\h$, the pairing $\inner{\xb}{\wb}$ may
be decomposed as follows:
$$
		\inner{\xb}{\wb} = \inner{\xb_{|\J}}{\wb_{|\J}} +
\inner{\xb_{|\I-\J}}{\wb_{|\I-\J}}.
$$
\begin{lemma}[Restricted sequence spaces]  Let $\W$ be a sequence space
defined on an index set
$\I$, let $\W\h$ be its dual sequence space, and let $\J$ be any subset of
$\I$;  then
\begin{itemize}
\item[(i)]  $(\W_{|\J})\h = (\W\h)_{|\J}$;  \ie the
character group of a restriction $(\W_{|\J})\h$ is the
corresponding restriction of $\W\h$.
\item[(ii)]  $\W = \W_{|\J} \times \W_{|\I-\J}$ implies $\W\h =
(\W\h)_{|\J} \times (\W\h)_{|\I-\J}$;  \ie if $\W$ is
memoryless, then $\W\h$ is memoryless.
\item[(iii)]  $P_\J(\W)^\perp = P_{\I-\J}(\W\h)$;  \ie the
orthogonal subgroup to the projection $P_\J(\W)$ is the complementary
projection of $\W\h$.
\end{itemize}
\end{lemma}

Our central result is then the following projection/subcode duality theorem:
\begin{theorem}[Projection/subcode duality]
Let $\C$ and $\C^\perp$ be orthogonal closed group codes in sequence
spaces $\W$ and $\W\h$, respectively. Then the orthogonal subgroup to the
restriction $\C_{|\J}$ is the restricted subcode $(\C^\perp)_{|:\J}$.
\end{theorem}

\emph{Proof}.  Since $\inner{(\xb_{|\J}, \zerob_{|\I-\J})}{\wb} =
\inner{\xb_{|\J}}{\wb_{|\J}}$, we have the following logical chain:
\begin{eqnarray*}
\xb_{|\J} \perp \C_{|\J} & \Leftrightarrow & (\xb_{|\J}, \zerob_{|\I-\J}) 
\perp \C \\ & \Leftrightarrow &  (\xb_{|\J}, \zerob_{|\I-\J}) \in \C^\perp \\
& \Leftrightarrow & \xb_{|\J} \in (\C^\perp)_{|:\J}. \qquad  \qed
\end{eqnarray*}

Note that if $\C_{|\J}$ is not closed, then the orthogonal subgroup to 
$(\C^\perp)_{|:\J}$ is the closure of  $\C_{|\J}$.

In the language of coding theory, this theorem is stated as follows:  the
dual of a punctured code is the corresponding shortened code of the dual
code.

This result immediately implies various corollaries:
\begin{corollary}[Projection/subcode duality corollaries]
\label{psdc} 
Under the same conditions:
\begin{itemize}
\item[(a)]  The orthogonal subgroup to the projection $P_{\J}(\C) =
\C_{|\J} \times \{\zerob\}_{|\I-\J}$ is $(\C^\perp)_{|:\J} \times
(\W\h)_{|\I-\J}$.
\item[(b)]  The orthogonal subgroup to the restricted subcode
$\C_{|:\J}$ is the closure of $(\C^\perp)_{|\J}$ in $(\W\h)_{|\J}$.
\item[(c)]  The orthogonal subgroup to the subcode
$\C_{:\J}$ is the closure of $(\C^\perp)_{|\J} \times
(\W\h)_{|\I-\J}$ in $\W\h$.
\item[(d)]  If $\C_{|\J}$ is closed in $\W_{|\J}$, then
$\C_{|\J}$ and $(\C^\perp)_{|:\J}$ are dual group codes.
\item[(e)]  The orthogonal subgroup to the direct product $\C_{|\J} \times
\C_{|\I-\J}$ is $(\C^\perp)_{|:\J} \times (\C^\perp)_{|:\I-\J}$.
\end{itemize}
\end{corollary}

\subsection{Conditioned code duality}

The following generalization of projection/subcode duality is the key lemma
for the graph duality results of \cite{F01}.  It is also a
fundamental result for behavioral control theory.\footnote{
We are grateful to H. Narayanan for pointing out that our conditioned code
duality theorem is closely related to his ``implicit duality theorem,"
which he has proved and used extensively in various
settings \cite{N97, N00, N000}.}
\pagebreak

Let $\C$ be a group code in a sequence space
$\W$ defined on an index set
$\I$, and let $\D$ be a group code defined on $\W_{|\I-\J}$, where $\J
\subseteq \I$.  The \emph{conditioned code} $(\C\mid\D)$ is then
defined as the set of all $\cb \in \C$ such
that $\cb_{|\I-\J} \in \D$:
$$
(\C\mid\D) = \{\cb \in \C \mid \cb_{|\I-\J} \in \D\} = \C \cap
(\W_{|\J} \times \D).
$$
Note that since $\C, \W_{|\J}$ and $\D$ are closed, $(\C\mid\D)$ is
closed.

The conditioned code may be interpreted in the behavioral control
context of Figure 4.1.  The symbols in
$\W_{|\J}$ represent to-be-controlled variables, those in $\W_{|\I-\J}$
represent control variables, and $\C$ represents a plant whose behavior
constrains both.  The symbols in $\W_{|\I-\J}$ are further constrained
by a controller $\D$.  The restricted conditioned code $(\C\mid\D)_{|\J}$
represents the controlled behavior of the  variables in $\W_{|\J}$.

\begin{figure}[t]
\setlength{\unitlength}{5pt}
\begin{center}
\begin{picture}(26,6)
\put(0,2){\line(0,1){2}}
\put(0,3){\line(1,0){6}}
\put(1,4){$\W_{|\J}$}
\put(6,0){\framebox(6,6){$\C$}}
\put(12,3){\line(1,0){8}}
\put(13,4){$\W_{|\I-\J}$}
\put(20,0){\framebox(6,6){$\D$}}
\end{picture}

Figure 4.1.  Conditioned code $(\C\mid\D)$.
\end{center}
\end{figure} 

The generalized theorem is then as follows  (see Figure
4.2):
\begin{theorem}[Conditioned code duality]
\label{FL} 
If $\CC$ and $\CC^\perp$ are dual group codes defined on $\I$, and
$\D$ and $\D^\perp$ are dual group codes defined on a subset $\I-\J
\subseteq \I$, then the restricted conditioned codes $(\CC \mid \D)_{|\J}$
and $(\CC^\perp \mid \D^\perp)_{|\J}$ are dual group codes defined on
$\J$, assuming both are closed.
\end{theorem}
\emph{Proof}.
First observe that $(\C\mid\D)_{|\J}$ may alternatively be characterized as
the restricted subcode
$$
(\C\mid\D)_{|\J} = (\C + (\{\zerob\}_{|\J} \times \D))_{|:\J},
$$
since $\cb \in (\C\mid\D)$ if and only if there is a $\db_{|\I - \J}
\in \D$ such that $(\cb + (\zerob_{|\J}, \db_{|\I - \J}))_{|\I - \J} =
\zerob_{|\I - \J}$.
Assuming that both $(\CC \mid \D)_{|\J}$
and $(\CC^\perp \mid \D^\perp)_{|\J}$ are closed, we then have
\begin{eqnarray*}
\left((\CC \mid \D)_{|\J}\right)^\perp & = & \left(\left(\CC +
(\{\zerob\}_{|\J}
\times \D)\right)_{|:\J}\right)^\perp \\
& = & \left(\left(\CC + (\{\zerob\}_{|\J} \times \D)\right)^\perp\right)_{|\J}\\
& = & \left(\CC^\perp \cap (\{\zerob\}_{|\J}
\times \D)^\perp\right)_{|\J} \\ & = &\left(\CC^\perp \cap ((\W\h)_{|\J}
\times \D^\perp)\right)_{|\J} \\ & = & (\CC^\perp \mid \D^\perp)_{|\J},
\end{eqnarray*}
where we have used projection/subcode, sum/intersection, and direct
product duality.  \qed

Notice that $(\C\mid\W_{|\I-\J}) = \C$, whereas
$(\C\mid\{\zerob\}_{|\I-\J}) =
\C_{:\J}$.  Therefore projection/subcode duality, namely
$(\C_{|\J})^\perp = (\C^\perp)_{|:\J}$, is a special
case of conditioned code duality.  

Moreover, as $\D$ ranges from $\{\zerob\}_{|\I-\J}$ to
$\W_{|\I-\J}$, the restricted conditioned code $(\C\mid\D)_{|\J}$ ranges
from the restricted subcode $\C_{|:\J}$ to the restriction
$\C_{|\J}$.  This is the essence of the ``most beautiful behavioral
control theorem" \cite{T99}.

\begin{figure}[t]
\setlength{\unitlength}{5pt}
\begin{center}
\begin{picture}(30,6)
\put(0,2){\line(0,1){2}}
\put(0,3){\line(1,0){8}}
\put(1,4){$(\W\h)_{|\J}$}
\put(8,0){\framebox(6,6){$\C^\perp$}}
\put(14,3){\line(1,0){10}}
\put(15,4){$(\W\h)_{|\I-\J}$}
\put(24,0){\framebox(6,6){$\D^\perp$}}
\end{picture}

Figure 4.2.  
Dual conditioned code $(\C^\perp\mid\D^\perp)$.
\end{center}  
\end{figure} 

\subsection{Completeness revisited}

	In behavioral system theory, the completion of a system $\C$ in a
complete sequence space $\W^c$ is defined as  \cite{W89}
$$
		\C^\mathrm{compl} = \{\wb \in \W^c \mid \wb_{|\J} \in \C_{|\J}
\mbox{~for all finite~} \J \subseteq \I\},
$$
and $\C$ is called complete if $\C^\mathrm{compl} = \C$.  In other words,
$\C$ is complete if any sequence $\wb \in \W^c$ that looks like a
sequence in $\C$ through all finite windows is actually in $\C$.

The following result characterizes the closure $\C\cl$ of a subgroup $\C
\in \W^c$, which we also call its \emph{completion} $\C^c$, in almost the
same way:
\begin{theorem}[Completion] \label{tc}
If $\C$ is a subgroup of a complete sequence space $\W^c$ defined
on an index set $\I$, then the closure (completion) of $\C$ is
\end{theorem}
$$
		\C^c =  \{\wb \in \W^c \mid \wb_{|\J} \in (\C_{|\J})\cl \mathrm{~for
~all ~finite~} \J \subseteq \I\}.
$$
\emph{Proof}.  By orthogonal subgroup duality, $\C^c$ is the dual of the
dual code $\C^\perp$ in the dual finite sequence space $(\W^c)\h$. 
Since $\C^\perp$ is finite, it is certainly generated by its subcodes
$(\C^\perp)_{:\J}$ for all finite $\J$:  
$$
		\C^\perp = \sum_{\J \mathrm{~finite}} (\C^\perp)_{:\J}.  
$$
By sum/intersection duality, $\C^c = \C^{\perp\perp}$ is the intersection
of the dual codes $((\C^\perp)_{:\J})^\perp$:  
$$
		\C^c = \bigcap_{\J \mathrm{~finite}} ((\C^\perp)_{:\J})^\perp.  
$$
The theorem follows since by Corollary \ref{psdc}(c),
\begin{eqnarray*}
		((\C^\perp)_{:\J})^\perp	& = & (\C_{|\J})\cl \times (\W^c)_{|\I-\J} \\
			& = &  \{\wb \in \W^c \mid \wb_{|\J} \in (\C_{|\J})\cl\}.	       \qquad \qed
\end{eqnarray*}

 It follows that if the restriction $\C_{|\J}$ is closed for all finite
$\J \subseteq \I$, then completeness in the behavioral system theory
sense is equivalent to closure in the product topology, which is what we
call ``completeness" in this paper.  In particular, the two concepts
coincide if all symbol groups $G_k$ are discrete.

A reviewer has pointed out that Theorem \ref{tc} may be extended to
the case in which $\C$ is merely a subset of $\W^c$.

\subsection{Completion/finitization duality}

	The \textbf{finite subset} (or ``finitization") of a subgroup
$\CC$ of a complete sequence space $\W^c$ will be denoted by
$\CC_f = \CC \cap \W_f$.   We say that $\CC$ is \emph{finite} if $\CC =
\CC_f$.  $\CC$ is evidently a subgroup of $\W_f$.  We will assume that
$\CC_f$ is closed when endowed with the topology of $\W_f$.  For example,
the finite subset of $\W^c$  or of $\W_L$ is $\W_f$. 

	The following result shows that completion and finitization are
duals:

\begin{theorem}[Completion/finitization duality]  \label{cfd}
Let $\CC$ be a closed subgroup of a
complete, finite or Laurent sequence space $\W$ with symbol groups
$\{G_k, k \in \I\}$, and let $\CC^\perp$ be the dual subgroup in the dual
sequence space $\W\h$, with symbol groups $\{G_k\h\}$.  Let
$\CC_f$ be the finite subset of $\CC$, and assume that $\CC_f$ is closed
when endowed with the topology of
$\W_f$.  Then the dual subgroup to $\CC_f$ in $(\W\h)^c = \prod_{k\in\I}
G_k\h$ is the completion of $\CC^\perp$ in $(\W\h)^c$: 
$(\CC_f)^\perp = (\CC^\perp)^c.$
\end{theorem} 
%  In particular, if $\C$ and $\C^\perp$ are dual
%group codes in dual sequence spaces $\W$ and $\W\h$, then $\C_f$ is
%a group code in $\W_f = \bigoplus_{k\in\I} G_k$, and its dual
%$(\C_f)^\perp$ is $(C^\perp)^c \in (\W\h)^c$.  Equally,
%$(C^c)^\perp = (C^\perp)_f$.
\emph{Proof}.  Following the proof of Theorem \ref{tc}, $\CC_f$ is
generated by the finite subcodes $\CC_{:\J}$ of $\CC$ for all finite $\J$: 
$$
		\C_f = \sum_{\J \mathrm{~finite}} \CC_{:\J}.  
$$
By sum/intersection duality, $(\C_f)^{\perp}$ is the intersection
of the dual codes $(\CC_{:\J})^\perp$:  
$$
(\C_f)^{\perp} = \bigcap_{\J \mathrm{~finite}} (\CC_{:\J})^\perp
$$
By projection/subcode duality, 
$$(\CC_{:\J})^\perp = \{\wb \in (\W\h)^c \mid \wb_{|\J} \in
((\CC^\perp)_{|\J})\cl\},$$
 so
$$
(\C_f)^{\perp} =  \{\wb \in (\W\h)^c \mid \wb_{|\J} \in
((\CC^\perp)_{|\J})\cl \mbox{~for all finite~} \J\},  
$$
which by Theorem \ref{tc} is $(\CC^\perp)^c$.  \qed

\subsection{Laurent codes}

	Similarly, a \emph{Laurent group code} is a closed subgroup $\C$ of a
Laurent sequence space $\W_L$.    The dual of a Laurent group code $\C$ is
an (anti-)Laurent group code
$\C^\perp$ in the dual (anti-) Laurent sequence space $(\W_L)\h$.  

	As in Theorem \ref{gcd}, if $\C$ and $\C^\perp$ are dual Laurent group
codes, then either determines the other.  Here the primal and dual codes
are symmetric.

	The \emph{Laurent completion} of a subgroup $\C$ of a Laurent
sequence space $\W_L$ is the closure of the group generated by $\C$ in
$\W_L$, denoted by $\C^L$.  $\C$ is a Laurent group code if and only if $\C
= \C^L$.  For example, the Laurent completion of $\W_f$ is $\W_L$.

	The \emph{Laurent subset} (``Laurentization") of a subgroup $\C$
of a sequence space $\W$ will be denoted by $\C^L$;  \ie 
$$\C^L = \C \cap
\W_L.$$
 $\C^L$ is endowed with the topology of $\W_L$.  $\C$ is
\emph{Laurent} if $\C = \C^L$.  For example, the Laurent subset of $\W^c$
is $\W_L$.

\subsection{Wide-sense controllability and observability}

	Fagnani \cite{F97} has proposed an elegant definition of (wide-sense)
controllability, which we restate as follows.
A complete group code $\C
\subseteq \W^c$ is \textbf{controllable} if $(\C_f)^c = \C$.  In other
words, a complete group code is controllable if it is generated by its
finite sequences.  Fagnani has shown that a complete compact
time-invariant group code that is controllable in this sense is
controllable in the sense of Willems \cite{W89}.

More generally, we say that a group code $\CC$ in a sequence space $\W$ is
controllable if $(\CC_f)^c = \CC^c$;  \ie if the completion of $\CC$ in
$\W^c$ is the completion of the finite sequences of $\CC$.  The complete
code $(\CC_f)^c$ will be called the
\emph{controllable subcode} of the complete code $\CC^c$.   Note that any
finite code $\CC$ is necessarily controllable.

%	Using Theorem \ref{cfd}, we verify that the controllable subcode
%$(\CC_f)^c$ is indeed controllable:
%\begin{eqnarray*}
%		((\CC_f)^c)^c & = & (\CC_f)^c; \\
%		((\CC_f)^c)_f)^c & = & (((\CC_f)^{\perp\perp})_f)^c =
%((\CC_f)^{\perp\perp\perp\perp})^c = (\CC_f)^c.
%\end{eqnarray*}
%In other words, after one cycle of finitization and completion, further
%cycles yield nothing new.

	We then propose the following dual definition:  a group
code $\C$ in a sequence space $\W$ is \textbf{observable} if
$(\C^c)_f = \C_f$.  In other words, completing $\C$ does not introduce any
new finite sequences beyond those already in $\C$.  The finite code
$(\C^c)_f$ will be called the \emph{observable supercode} of the finite
code $\C_f$.  Note that any complete code is necessarily observable.

%	We verify that the observable supercode $(S^c)_f$ is indeed observable:
%\begin{eqnarray*}
%		((\C^c)_f)_f & = & (\C^c)_f; \\
%		(((\C^c)_f)^c)_f & = & (((\C^c)_f)^{\perp\perp})_f =
%(((\C^c)^{\perp\perp\perp\perp})_f = (\C^c)_f.
%\end{eqnarray*}
%Again, after a cycle of completion and finitization, further cycles yield
%nothing new.

	The following shows that these two definitions are duals:

\begin{theorem}[Controllability/observability duality]  \label{cod}
  If $\C$ and $\C^\perp$ are dual group codes, then:
\begin{itemize}
\item[(a)]  $\C^c$ and $(\C^\perp)_f$ are dual group codes;
\item[(b)] The controllable subcode $(\C_f)^c$ of $\C^c$ and the
observable supercode $((\C^\perp)^c)_f$ of $(\C^\perp)_f$ are dual group
codes;
\item[(c)]  The quotient group $((\C^\perp)^c)_f/(\C^\perp)_f$ acts as the
character group of $\C^c/(\C_f)^c$; 
\item[(d)] $\C$ is controllable if and only if $\C^\perp$ is observable.
\end{itemize}
\end{theorem}
\emph{Proof}.  Part (a) is Theorem \ref{cfd}. This also implies part (b),
since 
$$((\C_f)^c)^\perp = ((\C_f)^\perp)_f = ((\C^\perp)^c)_f.$$  Part (c)
follows by quotient group duality.  Part (d) is a corollary of part (c),
since 
\begin{eqnarray*}
\C^c = (\C_f)^c & \Leftrightarrow & \C^c/(\C_f)^c = \{0\} \\
& \Leftrightarrow & ((\C^\perp)^c)_f/(\C^\perp)_f = \{0\}\h = \{0\} \\
& \Leftrightarrow & ((\C^\perp)^c)_f = (\C^\perp)_f.  	\quad \qed
\end{eqnarray*}

	Note that these notions of controllability and 
observability do not depend on $\I$ being ordered.  Therefore they
apply to systems with unordered time axes;  \eg two-D
systems \cite{RW91, VF94, FV94, FV96}.

	The core meaning of ``controllable" is that any code sequence can be
reached from any other code sequence in a finite interval.  We will
consider a strong notion of controllability below, and will prove that
strong controllability implies controllability in the sense of this
section when all symbol groups are compact.  Similarly, the core meaning
of ``observable" is that observation of a code sequence during a finite
interval gives a sufficient statistic for the future or the past.  We
will show below that strong observability in this sense implies
observability in the sense of this section when all symbol groups are
discrete. 

We say that a code is \emph{local} if it is both controllable and
observable.  By Theorem \ref{cod}, the dual of a local group code is
local.  Local codes can be completed or finitized without loss of
structure, so it does not matter much whether we consider the complete,
finite or Laurent versions of such codes.  Practical convolutional
codes are always chosen to be local, so as to avoid the pathologies
associated with uncontrollability (autonomous behavior) and
unobservability (``catastrophicity").  
%However, unobservability
%may sometimes be introduced deliberately, for example to achieve rotational
%invariance \cite{TBGM96}.

To illustrate, we now give a standard example of an uncontrollable
(autonomous) group code $\C$ that is inherently complete and cannot be
``finitized" or ``Laurentized" without losing its dynamical structure. 
Its dual $\C^\perp$ is an unobservable (catastrophic) group code that is
inherently finite and cannot be completed without losing its structure.

\medskip
\noindent
\textbf{Example 2}.  Let $G$ be an LCA group, let $\W^c$ be the complete
sequence space $G^\Z$, and let $\C \subseteq G^\Z$ be the
\emph{bi-infinite repetition code} over $G$;  \ie
$$
		\C = \{\gb = (\ldots, g, g, g, \ldots) \mid g \in G\}. 
$$
$\C$ is a complete time-invariant group code which is isomorphic to $G$.

The dual sequence space $\W\h$ to $\W^c$ is the finite sequence space
$((\Gh)^\Z)_f$, where $\Gh$ is the character group of $G$.  The dual group
code $\C^\perp$ is the \emph{bi-infinite zero-sum code} over $\Gh$, namely
the finite code defined by 
$$
		\C^\perp = \{\hb \in ((\Gh)^\Z)_f \mid \sum_{k\in\Z} h_k = 0\}.
$$
This follows since for $\gb = (\ldots, g, g, g, \ldots) \in \C$ and $\hb
\in \W\h$, the pairing $\inner{\hb}{\gb}$ is
$$
		\inner{\hb}{\gb} = \sum_{k\in\Z} \inner{h_k}{g} = \inner{\sum_{k\in\Z}
h_k}{g},
$$
which is equal to $0$ for all $g \in G$ if and only if $\sum_k h_k =
0$, the sum being well-defined because there are only finitely many
nonzero components in $\hb \in \W\h$.   $\C^\perp$ is a closed
subgroup of the finite sequence space $\W\h$, since it is the orthogonal
subgroup to the complete code $\C$.  Like $\C$, it is time-invariant.

	The repetition code $\C$ is uncontrollable, since its finite subcode
consists of only the all-zero sequence, $\C_f = \{\zerob\}$, and this
trivial subcode is complete.  The zero-sum code $\C^\perp$ is
unobservable, since its completion is the complete sequence space
$(\Gh)^\Z$.\footnote{\emph{Proof}:  let $\cb^{(m,n)}$ be the sequence in
$\C^\perp$ with
$c_m = g, c_n = -g$, and $c_k = 0$ for $k \neq m, n$;  then for fixed $m$
the ``limit" of $\cb^{(m,n)}$ as $n \to \infty$ in the product topology is
the sequence with $c_m = g$ and $c_k = 0$ for $k \neq m$, a finite
sequence that is not in $\C^\perp$.  Since such unit sequences generate
$(\Gh)^Z$, we have $(\C^\perp)^c = (\Gh)^Z$.}  Thus finitization of $\C$
or completion of $\C^\perp$ destroys dynamical structure.

Clearly $\C/\C_f \cong G$, which by Theorem \ref{cod}(c) implies that
$\W\h/\C^\perp \cong \Gh$.  The cosets of $\C^\perp$ in $\W\h$ are in fact
the subsets of $\W\h$ such that $\sum_k h_k = h$, for each $h \in \Gh$. 
$\C^\perp$ is unobservable because no finite observation can distinguish
between these cosets.  \qed

\subsection{Further examples}

We now give two more examples of dual group codes.
The first involves a standard controllable and observable (local)
time-invariant convolutional code over a finite symbol group and its
dual.
The second exhibits a curious complete time-invariant group code
that can be finitized on the past (``Laurentized") without loss of
dynamical structure, but not on the future.  Its dual has the dual
property.  These two codes were proposed in \cite{L93} and
\cite{LFMT94}, respectively, but were not recognized there as duals.

\medskip
\noindent
\textbf{Example 3}.  Let $\C$ be the complete rate-1/3 linear
time-invariant convolutional code over $\Z_4$ comprising all linear
combinations of time shifts of the generator
$$
		\gb = (\ldots, 000, 100, 010, 002, 000, \ldots)
$$
$\C$ is closed in the complete sequence space $((\Z_4)^3)^\Z$.  

The finite subcode $\C_f$ of $\C$ is generated by all finite linear
combinations of time shifts of $\gb$, and is closed in the finite sequence
space $(((\Z_4)^3)^\Z)_f$.  The completion of $\C_f$ is $\C$, so $\C$ is
controllable.  Thus $\C$ is local, since as a complete code it is
automatically observable.  

Similarly, the Laurent subcode $\C_L$ is generated by all
Laurent linear combinations of time shifts of $\gb$, and is closed in the
Laurent sequence space $(((\Z_4)^3)^\Z)_L$.

	The dual code $\C^\perp$ is the finite rate-2/3 code linear
time-invariant convolutional code over $\Z_4$ consisting of all finite
linear combinations of time shifts of the two generators
\begin{eqnarray*}
		\hb_1 & = & (\ldots, 000, 100, 030, 000, \ldots); \\
 	\hb_2 & = & (\ldots, 000, 020, 001, 000, \ldots),
\end{eqnarray*}
which are orthogonal to all time shifts of $\gb$ under the usual inner
product over $\Z_4$.  (Equivalently, the convolutions $\tilde{\hb}_1
\ast \gb$ and $\tilde{\hb}_2 \ast \gb$ of the time-reverses $\tilde{\hb}_1$ and
$\tilde{\hb}_2$ are equal to
$\zerob$.)  
\linebreak $\C^\perp$ is closed in the finite sequence space
$(((\Z_4)^3)^\Z)_f$.  

	The dual complete code $(\C_f)^\perp$ is the set of all linear
combinations of time shifts of $\hb_1$ and $\hb_2$.  $\C^\perp$ is the
finite subcode of $(\C_f)^\perp$, and $(\C_f)^\perp$ is the completion of
$\C^\perp$.  Thus $\C^\perp$ is local. 

Here there is no essential difference
between the finite, Laurent, or complete versions of $\C$ or $\C^\perp$. 
In general, the dynamical structure of a group code $\C$
is not affected by completion or finitization if and only if $\C$ is
local.  \qed

\medskip
\noindent
\textbf{Example 4}.  The following is a much more exotic example (a
``solenoid" \cite{LM95}), and is a rich source of counterexamples.

	Loeliger \cite{L93, AL95} proposed the following curious PSK-type
code.  Let $\C$ be the complete compact linear time-invariant code over
the additive circle group $\R/\Z$ that consists of all integer linear
combinations of time shifts of the Laurent generator 
$$
		\gb = (\ldots, 0, \mbox{$\frac{1}{2}, \frac{1}{4}, \frac{1}{8},$}
\ldots).
$$
Since $2\gb$ (mod $\Z$) is a shift of $\gb$, the ``input" at each time $k$
is essentially a binary variable $u_k \in \{0, \half\}$,
which may be regarded as representing the subgroup $(\half\Z)/\Z$ of
$\R/\Z$.  The ``output" symbol at time $k$ is
$$
		c_k = \frac{u_k}{2} + \frac{u_{k-1}}{4} + \frac{u_{k-2}}{8} + \cdots =
\frac{u_k}{2} + \frac{c_{k-1}}{2} \in \R/\Z.
$$
Thus $c_k$ determines the entire past input sequence.

If the output symbol is mapped onto the complex unit circle via $c_k
\mapsto e^{2\pi i c_k}$, then $\C$ is a well-defined PSK-type code that
transmits one bit per symbol and has a well-defined minimum squared
distance (6.79\ldots).  However, the symbol alphabet of $\C$ is the entire
infinite circle group $\R/\Z$, rather than a finite subgroup as with
ordinary PSK codes.  Also, the code ``state" $c_{k-1}$ lies in the
infinite state space $\R/\Z$.  Each state $c_{k-1}$ has two successors
$c_k$, but each $c_k$ has only one predecessor $c_{k-1}$.

	The dual code $\C^\perp$ to $\C$ is the finite discrete linear
time-invariant code over the integers $\Z$ (the character group of
$\R/\Z$) comprising all finite integer linear combinations of time shifts
of the finite generator
$$
		\hb = (\ldots, 0, 1, -2, 0, \ldots).
$$
It is easily verified that $\hb$ is orthogonal (mod $\Z$) to all time
shifts of $\gb$, that a sequence in $(\R/\Z)^\Z$ is in $\C$ if and only if
it is orthogonal to all shifts of $\hb$, and that a sequence in
$(\Z^\Z)_f$ is in $\C^\perp$ if and only if it is orthogonal to all shifts
of $\gb$. 

	Loeliger's code $\C$ is uncontrollable, since its finite subcode consists
only of the all-zero sequence, $\C_f = \{\zerob\}$.  Indeed, its
time-reverse $\tilde{\C}$ is a standard example of a chaotic dynamical
system whose evolution depends entirely on initial conditions
\cite{D87}.   Nevertheless, $\C$ may be generated by a causal encoder with
one input bit per unit time.  

$\C$ has a natural Laurent subcode $\C_L$ that is
generated by the input sequences that are Laurent.  Thus while
finitization destroys its structure, Laurentization does not. 
(However, Laurentization does reduce the symbol alphabet from the
uncountably infinite set $\R/\Z$ to the countably infinite set of dyadic
numbers in $\R/\Z$.)   On the other hand, the anti-Laurent subcode of $\C$
is $\{\zerob\}$, since if the output is $0$ at any time, then it must have
been $0$ at all previous times. Thus even though $\C$ is time-invariant,
its time axis has a distinct directionality.

	The finite dual code $\C^\perp$ is unobservable, since its completion is
the complete sequence space $\Z^\Z$.  Its Laurent completion is
$(\Z^\Z)_L$ (the dual of $\{\zerob\} \subseteq
((\R/\Z)^\Z)_{\tilde{L}}$).  However, its anti-Laurent completion is
simply the set of all anti-Laurent integer combinations of shifts of
$\hb$, which again indicates the directionality of the time axis.  

	Interestingly, $\C^\perp$ is a version of an example given in
\cite{LFMT94, LM96} to show that the set of all sequences generated by a
group trellis whose state space (in this case $\Z$) does not satisfy the
descending chain condition may not be a complete code.

Pontryagin suggested as a general rule that a compact group might be
best studied via its discrete character group \cite{P46}.  In this spirit,
we suggest that it might be useful in general to study compact solenoids
via their discrete duals.  In this case, for instance, the dual code
$\C^\perp$ is finite and has short integer-valued generators.  \qed

\section{Dynamical duality}

	This section develops basic dynamical dual properties of dual
group codes $\C$ and $\C^\perp$, such as:
\begin{itemize}
\item The state spaces of $\C^\perp$ act as the character groups of the
state spaces of $\C$.
\item The observability properties of $\C^\perp$ are the controllability
properties of $\C$.
\end{itemize}

\subsection{Topological state space theorems}

	The fundamental result of \cite{FT93} is the state space theorem, which
shows that for a group code $\C$ every two-way partition of the time axis
induces a certain group-theoretic minimal state space $\Sigma_\J$. 
Moreover, there exists a minimal state realization for $\C$ in which every
state space is isomorphic to the corresponding minimal state space
$\Sigma_\J$.  We now discuss this theorem for the topological group codes
of this paper.

	Given a subset $\J \subseteq \I$, the subcodes $\C_{:\J}$ and
$\C_{:\I-\J}$ and their internal direct product $\C_{:\J} \times
\C_{:\I-\J}$ are closed normal subgroups of $\C$.  The (two-sided)
\textbf{state space} of $\C$ induced by the two-way partition of $\I$ into
$\{\J, \I-\J\}$ is then well defined as the quotient group
$$
		\Sigma_{\J}(\C) = \frac{\C}{\C_{:\J} \times \C_{:\I-\J}}.
$$  
%Equivalently, the internal direct product
%$\C_{:\J} \times \C_{:\I-\J}$ may be written as the external direct product
%$\C_{|:\J} \times \C_{|:\I-\J}$.

 The proof of the following version of
the state space theorem goes through as in \cite{FT93}:
\begin{theorem}[State space theorem]  \label{sst}
Given a group code $\C$ in a sequence space defined on an index set
$\I$ and a two-way partition of $\I$ into  ``past" $\J$ and ``future"
$\I-\J$, the minimal state space of any state realization of $\C$ at the
time corresponding to this ``cut" is $\Sigma_{\J}(\C)$.
\end{theorem}

In \cite{FT93}, \emph{one-sided state spaces}
$P_{\J}(\C)/\C_{:\J}$ and $P_{\I - \J}(\C)/\C_{:\J}$ are also introduced,
and shown to be algebraically isomorphic to the state space
$\Sigma_{\J}(\C)$.  This follows from the correspondence theorem, since the
kernels of the projections of $\C$ and of $\C_{:\J} \times \C_{:\I-\J}$
onto $\J$ are the same, namely $\C_{:\I-\J}$.   
One-sided state spaces may
also be defined using restrictions since,
\eg
$$
\frac{P_{\J}(\C)}{\C_{:\J}} \cong \frac{\C_{|\J}}{\C_{|:\J}}.
$$

	As discussed in Subsection \ref{rps}, a restriction $\C_{|\J}$ is
homeomorphic to the quotient group $\C/\C_{:\J}$, provided that
$\C_{|\J}$ is closed.  With this caveat, we obtain a topological version of
the one-sided state space theorem:
\begin{theorem}[One-sided state spaces]  \label{ossst}
Under the same conditions, let $\C_{|\J}$ and $\C_{|\I-\J}$ be the
restrictions of $\C$ to $\J$ and $\I-\J$, respectively, and assume
both are closed.  Then
$$
\frac{\C_{|\J}}{\C_{|:\J}} \cong \frac{\C_{|\I-\J}}{\C_{|:\I-\J}}
\cong	\Sigma_{\J}(\C).
$$
\end{theorem}

\medskip
\noindent
\textbf{Example 1} (cont.)  Again, let $\C$ be a lattice
$\{(am + bn, -bm + an) \mid (m, n) \in \Z^2\}$, where $a/b$ is irrational.
$\C$ is isomorphic and homeomorphic to $\Z^2$.  Letting $\J$ and $\I-\J$
denote the two single-coordinate subsets,
we have $\C_{:\J} = \C_{:\I-\J} = \{\zerob\}$.  Therefore $\Sigma_{\J}(\C)
\cong \C \cong \Z^2$, as expected, since either coordinate
determines the lattice point and thus the other coordinate.

In this case, if $\C_{|\J}$ and $\C_{|\I-\J}$ are endowed with the
discrete topology, then they are homeomorphic to $\Z^2$, so
Theorem \ref{ossst} holds.  However, as subspaces of $\R$, $\C_{|\J}$ and
$\C_{|\I-\J}$ are not closed, and not homeomorphic to $\Sigma_{\J}(\C)$. 
\qed

We will continue this discussion in Section 5.3.

\subsection{The dual state space theorem}

We can now  relate the state spaces of a dual code $\C^\perp$ to those of
$\C$, using the one-sided state space theorem.  We must therefore continue
to require restrictions to be closed.
\begin{theorem}[Dual state space theorem] \label{dsst}
If $\C$ and $\C^\perp$ are dual group codes defined on $\I$, then for any subset
$\J \subseteq \I$, the corresponding one-sided state space of
$\C^\perp$ acts as the character group of the corresponding one-sided state
space of $\C$:
$$
		\left(\frac{\C_{|\J}}{\C_{|:\J}}\right)^{\h} =
\frac{(\C^\perp)_{|\J}}{(\C^\perp)_{|:\J}}.
$$
Consequently the state space of $\C^\perp$ is isomorphic to the character
group of the state space of $\C$:
\end{theorem}
$$(\Sigma_{\J}(\C))\h \cong \Sigma_{\J}(\C^\perp).$$
\emph{Proof}.  By quotient group and projection/subcode duality,
$$
\left(\frac{\C_{|\J}}{\C_{|:\J}}\right)^{\h} =
\frac{(\C_{|:\J})^\perp}{(\C_{|\J})^\perp}	= 
\frac{(\C^\perp)_{|\J}}{(\C^\perp)_{|:\J}}	\qquad \qed
$$

	In the usual cases, this simple but powerful theorem generalizes a known
result for linear codes over fields:  the state spaces of dual codes have
the same dimensions.  In particular:
\begin{itemize}
\item If $\Sigma_{\J}(\C)$ is finite, then $\Sigma_{\J}(\C)
\cong \Sigma_{\J}(\C^\perp)$.
\item If $\Sigma_{\J}(\C)$ is a finite-dimensional real vector space, then
$\dim
\Sigma_{\J}(\C) = \dim \Sigma_{\J}(\C^\perp)$.
\end{itemize}

The following examples show that when the restrictions $\C_{|\J}$ and
$\C_{|\I-\J}$ are closed, the dual state space theorem gives a
satisfactory system-theoretic result, even when $\C$ is uncontrollable,
unobservable, or solenoidal.

\medskip
\noindent
\textbf{Example 2} (cont.)  For the bi-infinite repetition code $\C$ over
$G$, given any proper subset $\J \subseteq \Z$, we have $\C_{:\J} = \C_{:\I-\J} =
\{\zerob\}$, so the state space $\Sigma_\J(\C)$ is
isomorphic to $\C \cong G$.  For the dual bi-infinite zero-sum code
$\C^\perp$ over $\Gh$, $(\C^\perp)_{:\J}$ is the set of all finite
sequences $\hb$ with support in $\J$ whose component sum is 0,
$\sum_{k\in\J} h_k = 0$, whereas $(\C^\perp)_{|\J}$ is the set
$((\Gh)^\J)_f$ of all finite sequences with support in $\J$, so
$$
		\Sigma_\J(\C^\perp) \cong \frac{(\C^\perp)_{|\J}}{(\C^\perp)_{|:\J}}
\cong \Gh,
$$
where the cosets of $(\C^\perp)_{|:\J}$ in $(\C^\perp)_{|\J}$ correspond to
the different possible component sums $\sum_{k\in\J} h_k \in \Gh$.  Hence
$\Sigma_\J(\C^\perp) \cong (\Sigma_\J(\C))\h$. The dual state spaces are 
isomorphic if and only if $G \cong \Gh$.

Note that $\C^\perp$ has nontrivial state spaces, even
though its completion is the memoryless sequence space $\W\h$.  The
unobservability of $\C^\perp$ is reflected in the fact that the state of a
sequence $\hb \in \C^\perp$ cannot be observed from any finite segment
$\hb_{|\J}$ of $\hb$.  \qed

\noindent
\textbf{Example 3} (cont.)  
For any partition of the time axis into past $k^-$
and future $k^+$, the state spaces of both time-invariant codes $\C$
and $\C^\perp$ of Example 3 are isomorphic to $\Z_2 \times \Z_4$, which as
a finite abelian group is isomorphic to its character group.  Generators
for representatives of the cosets of $\C_{|:k^+}$ in $\C_{|k^+}$ are
$
		|010, 002, 000, \ldots) \mbox{~and~}
		|002, 000, 000, \ldots), 
$
which generate cyclic groups of orders 4 and 2, respectively.  Generators 
for representatives of the cosets of $(\C^\perp)_{|:k^+}$ in
$(\C^\perp)_{|k^+}$ are
$
		|030, 000, 000, \ldots) \mbox{~and~}
		|001, 000, 000, \ldots); 
$
the first has order 4, but the order of the second is only  2, since $|002,
000, 000, \ldots)$ is a code sequence in $(\C^\perp)_{|:k^+}$. \qed

\medskip
\noindent
\textbf{Example 4} (cont.)  The state of Loeliger's code $\C$ at time $k$
is the output $c_k \in \R/\Z$, since $\C_{:k^-} = \{\zerob\}$ (if the
future is all-zero, then $c_k = 0$, which implies that the past
$\cb_{|k^-}$ is all-zero).   Since the dual code $\C^\perp$ is the set of
all finite integer combinations of $\hb = (\ldots, 0, 1, -2, 0,
\ldots)$, the state of $\C^\perp$ at time $k$ is essentially its most
recent input $u_{k-1} \in \Z$
\linebreak  (representatives of the cosets of
$(\C^\perp)_{|:k^+}$ in $(\C^\perp)_{|k^+}$ are generated by $|-2, 0, 0,
\ldots)$).  The dual state spaces are thus $\R/\Z$ and $\Z$, which are
indeed each other's character groups, but which are not isomorphic.  \qed

\subsection{Non-closed restrictions}

However, in the exceptional cases where restrictions are not closed, the
dual state space theorem can fail.

\medskip
\noindent
\textbf{Example 1} (cont.)  As shown above, the
irrational lattice $\C = \{(am + bn, -bm + an) \mid (m, n) \in \Z^2\}$ is
isomorphic and homeomorphic to $\Z^2$, and so is its state space
$\Sigma_{\J}(\C)$ corresponding to splitting the two coordinates.  The two
restrictions $\C_{|\J}$ and $\C_{|\I-\J}$ are isomorphic and
homeomorphic to $\Z^2$ under the discrete topology, but not under the
subspace topology.  

The definition of the dual code $\C^\perp$ depends on the sequence space
in which $\C$ is considered to lie.  If $\C$ is regarded as a subspace of
$\R^2$, then the dual sequence space is $\R^2$, with pairing equal to the
usual inner product mod $\Z$.  Let us write $\C = A \Z^2$, where 
$
A = \small{ \left[\begin{array}{rr}
a & b \\
-b & a 
\end{array}\right] }.
$
The dual code is then the irrational lattice $\C^\perp =
A^{-1}\Z^2$ in $\R^2$, whose state space is again isomorphic to $\Z^2$.  Thus,
under the usual subspace topologies, the dual state space theorem fails.

However, suppose we regard $\C$ as a subspace of the
trimmed sequence space $\C_{|\J} \times  \C_{|\I-\J}$ under the discrete
topology;  then this sequence space is isomorphic and homeomorphic to $\Z^2
\times \Z^2$, and the dual sequence space is isomorphic to $(\R/\Z)^2
\times (\R/\Z)^2$.  As $\C$ is isomorphic to a repetition code over
$\Z^2$, the dual code $\C^\perp$ in this dual sequence space is isomorphic
to a zero-sum code over
$(\R/\Z)^2$, whose state space is isomorphic to $(\R/\Z)^2$ (see Example
2, above).  Thus, under these topologies, the dual state space
theorem holds. 
\qed

We conjecture that the dual state space theorem, and all later duality results,
hold when the symbol groups $G_k$ are taken as the restrictions $\C_{|\{k\}}$,
with the appropriate topologies.

However, as we see from this example, although use of nonstandard
topologies may lead to results which are formally correct, they may not be
consistent with the usual conventions, which are often based on subspace
topologies.  For instance, the usual definition of a dual lattice is with
respect to
$\R^n$; then the dual lattice of any full-rank lattice, even an irrational
lattice, is itself a lattice (a discrete subgroup of
$\R^n$), not some weird continuous compact group like $(\R/\Z)^2$.

One drawback of a Laurent sequence space is that in general it is neither
discrete nor compact, so we may expect Laurent codes to provide further
counterexamples, such as the following one.

\medskip
\noindent
\textbf{Example 5}.
Let $\CC \subseteq (\Z_2)^\Z$ be the binary \emph{mirror-image code}
consisting of all binary sequences $\xb \in (\Z_2)^\Z$ that exhibit mirror
symmetry;  \ie $x_k = x_{-k}$ for all $k \in \Z$.  $\CC$ is complete (a
closed  subgroup of the complete sequence space
$\W^c = (\Z_2)^\Z$) and controllable ($\CC$ is generated by
its finite sequences, $\CC = (\CC_f)^c$).  Its dual code $\CC^\perp$ in
$\W_f = ((\Z_2)^\Z)_f$ is the set of all finite binary sequences in
$\CC$ with $x_0 = 0$; \ie  $$\CC^\perp = (\CC_f)_{:\Z - \{0\}}.$$

$\C^\perp$ may also be regarded as a Laurent code $\C_L$ in the Laurent
sequence space $\W_L = ((\Z_2)^\Z)_L$, where it remains closed.  Its dual
$(\C_L)^\perp$ in this setting is an anti-Laurent code in $\W_{\tilde{L}}$,
which as a set is equal to the finite subcode $\CC_f$.

\begin{figure*}[t]
\setlength{\unitlength}{5pt}
\begin{center}
\begin{picture}(66,26)(0,-3)
\put(0,10){$\{\zerob\}$}
\put(4,12){\vector(2,1){8}}
\put(13,16){$\C_{|\J}$}
\put(12,15){\vector(-2,-1){8}}
\put(4,9){\vector(2,-1){8}}
\put(13,4){$\C_{|:\J}$}
\put(12,6){\vector(-2,1){8}}
\put(15,6){\line(0,1){8}}
\put(16,10){$\Sigma_J(\C)$}
\put(17,18){\vector(2,1){8}}
\put(25,21){\vector(-2,-1){8}}
\put(26.5,22){$\C_{|\J} \times \C_{|\I-\J}$}
\put(20,15){$R_\J$}
\put(31,10){$\C$}
\put(29,12){\vector(-3,1){12}}
\put(32,13){\line(0,1){8}}
\put(33,16){$\Sigma_J(\C)$}
\put(17,6){\vector(3,1){12}}
\put(26.5,-2){$\C_{|:\J} \times \C_{|:\I-\J}$}
\put(17,3){\vector(2,-1){8}}
\put(25,0){\vector(-2,1){8}}
\put(32,1){\line(0,1){8}}
\put(33,4){$\Sigma_J(\C)$}
\put(47,18){\vector(-2,1){8}}
\put(39,21){\vector(2,-1){8}}
\put(39,15){$R_{\I-\J}$}
\put(35,12){\vector(3,1){12}}
\put(47,6){\vector(-3,1){12}}
\put(47,3){\vector(-2,-1){8}}
\put(39,0){\vector(2,1){8}}
\put(61,10){$\{\zerob\}$}
\put(60,12){\vector(-2,1){8}}
\put(47,16){$\C_{|\I-\J}$}
\put(52,15){\vector(2,-1){8}}
\put(60,9){\vector(-2,-1){8}}
\put(47,4){$\C_{|:\I-\J}$}
\put(52,6){\vector(2,1){8}}
\put(49,6){\line(0,1){8}}
\put(50,10){$\Sigma_J(\C)$}
\end{picture}

Figure 5.1.  Tableau illustrating state space and reciprocal
state space theorems.
\end{center}
\end{figure*}

Whereas $\W_f$ is discrete and $\W^c$ is compact, the sequence spaces
$\W_L$ and $\W_{\tilde{L}}$ are neither discrete nor compact.  Thus whereas
the restrictions of $\CC \subseteq \W^c$ and $\CC^\perp 
\subseteq \W_f$ to the past interval $\PP = (-\infty, 0)$ are necessarily
closed, the restrictions $(\C_L)_{|\PP}$ and  $((\C_L)^\perp)_{|\PP}$
are not necessarily closed.  In fact, $(\C_L)_{|\PP}$ is closed in
$\W_L$, but $((\C_L)^\perp)_{|\PP}$ is not closed in  $\W_{\tilde{L}}$,
even though they are identical as sets (both are equal to $((\Z_2)^\P)_f$).

This shows again that the validity of our topological results
depends very much on the topologies of the sequence spaces in which codes
are regarded as being defined, and in particular on whether restrictions
are necessarily closed.  \qed

In order not to have to continually deal with such pathological cases, we
therefore impose from now on the following \textbf{closed-projections
assumption}:  
\begin{quote}
The topology induced by every restriction or projection onto
a subset $\J \subseteq \I$ is consistent with the topology of $\W_{|\J}$. 
In particular, projections of closed subgroups are closed in the subspace
topology. 
\end{quote}

A reviewer has pointed out that the closed-projections assumption is
satisfied for a complete sequence space $\W$ if all symbol groups $G_k$ are
compact metric spaces, and in particular if all $G_k$ are finite.  Then
$\W$ is a compact metrizable space, so every closed and
thus compact subset of $\W$ has a compact and thus closed image under the
continuous restriction map $R_\J:  \W \to \W_{|\J}$.

 Under the closed-projections assumption, we can apply our duality results
freely without continual consideration of topological issues.  The reader
must therefore use our results with caution whenever topological
subtleties are suspected.

\subsection{The reciprocal state space theorem}

	What is the character group of the two-sided state space
$\Sigma_{\J}(C)$?  The following theorem shows that it is the (two-sided)
\textbf{reciprocal state space} of $\C^\perp$, defined as
$$
\Sigma^{\J}(\C^\perp) = 
\frac{(\C^\perp)_{|\J} \times (\C^\perp)_{|\I-\J}}{\C^\perp}.
$$  
(The reciprocal state space was introduced in a different context in
\cite{FW92}.)
\begin{theorem}[Reciprocal state space theorem] \label{rsst}
  If $\C$ and $\C^\perp$ are dual group codes, then the reciprocal state
space $\Sigma^{\J}(\C^\perp)$
acts as the character group of the two-sided state space $\Sigma_{\J}(C)$.
\end{theorem}
\emph{Proof}.  Using quotient group, direct product, and
projection/subcode duality, we have 
\begin{eqnarray*}
\left(\frac{\C}{\C_{|:\J} \times \C_{|:\I-\J}}\right)^{\h} & = &
\frac{(\C_{|:\J} \times \C_{|:\I-\J})^\perp}{\C^\perp} \\ & = &
\frac{(\C_{|:\J})^\perp \times (\C_{|:\I-\J})^\perp}{\C^\perp} \\ & = &
\frac{(\C^\perp)_{|\J} \times (\C^\perp)_{|\I-\J}}{\C^\perp}.
 \qed
\end{eqnarray*}
%Notice that if C is not
%an abelian group code, then C need not be a normal subgroup of C|J \times
%C|I-J, so the reciprocal state space may not be well-defined.

The reciprocal state space theorem has an immediate corollary, which
yields a fourth state space for the group codes that we are
considering:

\begin{corollary}
The reciprocal state space $\Sigma^{\J}(\C)$ is isomorphic to the
state space $\Sigma_{\J}(\C)$.
\end{corollary}
\emph{Proof}.  By the reciprocal state space and dual state space theorems,
$$
		\Sigma^{\J}(\C) = (\Sigma_{\J}(\C^\perp))^{\h} \cong \Sigma_{\J}(\C).
\qquad \qed
$$

We caution the reader that this result depends on the closed-projections
assumption.  Moreover, as we will
discuss further below, it applies only when $\C$ is abelian. 
Nonetheless, it rounds out the state space theorem nicely when it applies.

	When the reciprocal state space theorem holds, there is a chain 
$$
		\C_{|:\J} \times \C_{|:\I-\J} \subseteq \C \subseteq \C_{|\J} \times
\C_{|\I-\J}, 
$$
in which both quotients are isomorphic to $\Sigma_{\J}(\C)$.  The dual
chain is 
$$
		(\C^\perp)_{|:\J} \times (\C^\perp)_{|:\I-\J} \subseteq \C^\perp
\subseteq (\C^\perp)_{|\J} \times (\C^\perp)_{|\I-\J}, 
$$
which has quotients isomorphic to $\Sigma_{\J}(\C^\perp) \cong
\Sigma_{\J}(\C)\h$, as illustrated by the dual diagrams below.

$$
\begin{array}{clcl}
\C_{|\J} \times \C_{|\I-\J} && (\C^\perp)_{|\J} \times
(\C^\perp)_{|\I-\J} & \\ | & \Sigma^{\J}(\C) & | & \Sigma^{\J}(\C^\perp)
%= \Sigma_{\J}(\C)\h 
\\
\C && \C^\perp & \\
| & \Sigma_{\J}(\C) & | & \Sigma_{\J}(\C^\perp)
% = \Sigma^{\J}(\C)\h 
\\
\C_{|:\J} \times \C_{|:\I-\J} && (\C^\perp)_{|:\J} \times
(\C^\perp)_{|:\I-\J} & 
\end{array}
$$
%\begin{center}
%Figure 5.1.  Dual diagrams, illustrating the reciprocal state space
%theorem.
%\end{center}
%\end{figure}
\medskip

	Figure 5.1 exhibits a related tableau of homomorphisms, in which all
quotient groups are isomorphic to the state space $\Sigma_{\J}(\C)$.  Note
that every left-to-right or right-to-left chain of four maps in this
tableau is a short exact sequence (a sequence in which the image of each
map is the kernel of the next).  Moreover, this tableau is self-dual,
in the sense that the dual diagram is the corresponding tableau for
$\C^\perp$.  

\subsection{The abelian dynamics theorem}

In this subsection we give a purely algebraic proof that the reciprocal
state space $\Sigma^{\J}(\C)$ is isomorphic to the state space
$\Sigma_\J(\C)$ when $\Sigma_\J(\C)$ is abelian.  When $\Sigma_\J(\C)$ is
not abelian, $\Sigma^\J(\C)$ is not well defined, but on the
other hand the situation is not essentially different.  Finally, we show
that these results are a special case of the abelian dynamics theorem.

  The one-sided state space theorem shows that we can compute the state
$\sigma_\J(\cb) \in \Sigma_\J(\C)$ of a code sequence $\cb \in \C$ from
either its ``past" $\cb_{|\J}$ or its ``future" $\cb_{|\I-\J}$;  \ie there
exist homomorphic \emph{state maps} $\sigma_{|\J}:  \C_{|\J} \to
\Sigma_\J(\C)$ and $\sigma_{|\I-\J}:  \C_{|\I-\J} \to \Sigma_\J(\C)$, whose
images are the state space $\Sigma_\J(\C)$ and whose kernels are the
restricted subcodes $\C_{|:\J}$ and $\C_{|:\I-\J}$, respectively.  For
$\cb \in \C$, the images of these maps must agree:  $\sigma_\J(\cb_{|\J}) =
\sigma_{\I-\J}(\cb_{|\I-\J})$.

A general pair $(\wb_{|\J}, \wb_{|\I-\J}) \in \C_{|\J} \times 
\C_{|\I-\J}$ is in $\C$ if and only if $\sigma_\J(\wb_{|\J}) =
\sigma_{\I-\J}(\wb_{|\I-\J})$ \cite{FT93}.  Therefore we can test
whether $(\wb_{|\J}, \wb_{|\I-\J})$ is in
$\C$ by forming the state difference (``syndrome")
$$
d(\wb_{|\J}, \wb_{|\I-\J}) = \sigma_\J(\wb_{|\J}) -
\sigma_{\I-\J}(\wb_{|\I-\J}).
$$
Then $(\wb_{|\J}, \wb_{|\I-\J}) \in \C$ if and only if $d(\wb_{|\J},
\wb_{|\I-\J}) = 0$.  In other words, $\C$ is the kernel of the state
difference map $d: \C_{|\J} \times \C_{|\I-\J} \to \Sigma_\J(\C)$.

When $\Sigma_\J(\C)$ is abelian, the state difference map is a
homomorphism.  Since $\C$ is its kernel, it follows that $\C$ is a closed
normal subgroup of $\C_{|\J} \times \C_{|\I-\J}$, and therefore that the
quotient group
$(\C_{|\J} \times \C_{|\I-\J})/\C$ (\ie the reciprocal state space) is
well defined.

When $\Sigma_\J(\C)$ is not abelian, $\C$ is still the kernel of the state
difference map. The following theorem shows that in this case $\C$ cannot
be a normal subgroup of $\C_{|\J} \times \C_{|\I-\J}$, and therefore the
state difference map cannot be a homomorphism.
\begin{theorem}[algebraic reciprocal state space theorem]  \label{arsst}
If $\C$ is an 
 algebraic group code in the sense of \cite{FT93}, then
the state space $\Sigma_\J(\C)$ is abelian if and only if $\C$ is a normal
subgroup of $\C_{|\J} \times \C_{|\I-\J}$.
\end{theorem}
\emph{Proof}.  On the one hand, if $\Sigma_\J(\C)$ is abelian, then the
state difference map $d: \C_{|\J} \times \C_{|\I-\J} \to \Sigma_\J(\C)$ is
a homomorphism with kernel $\C$, so $\C$ is a normal subgroup of $\C_{|\J}
\times
\C_{|\I-\J}$. 

On the other hand, if $\C$ is a normal subgroup of $\C_{|\J} \times
\C_{|\I-\J}$, then the reciprocal state space $\Sigma^\J(\C) = (\C_{|\J}
\times \C_{|\I-\J})/\C$ is abelian, which implies that $\Sigma_\J(\C) \cong
\Sigma^\J(\C)$ is abelian.  
Let $\wb \in P_\J(\C)$; then $\wb \in \C_{|\J} \times \{\zerob\}_{|\I-\J}
\subseteq \C_{|\J} \times \C_{|\I-\J}$, so by normality $\wb \cb \wb^{-1}
\in \C$ and thus $\wb \cb \wb^{-1} \cb^{-1} \in \C$ for any $\cb \in \C$. 
Now $\wb$ has support $\J$, so $P_{|\I - \J}(\wb \cb \wb^{-1} \cb^{-1}) =
\zerob$, which implies $\wb \cb \wb^{-1} \cb^{-1} \in \C_{:\J}$ and $\wb
\cb \wb^{-1} \cb^{-1} = \wb P_\J(\cb) \wb^{-1} (P_\J(\cb))^{-1}$.  As $\wb$
and $P_\J(\cb)$ run through $P_\J(C)$, the commutators $\wb P_\J(\cb)
\wb^{-1} (P_\J(\cb))^{-1} \in \C_{:\J}$ therefore run
through the generators of the commutator subgroup $[P_\J(C), P_\J(C)]$. 
Therefore $[P_\J(C), P_\J(C)] \subseteq \C_{:\J}$.  By a general
property of commutator subgroups \cite[Ex.\ 2.52]{R88}, $(\C_{|\J} \times
\C_{|\I-\J})/\C$ is thus abelian. \qed
\medskip

  Theorem \ref{arsst} shows that there is a distinct algebraic difference
between the abelian and nonabelian cases.   However, the two
cases are otherwise not fundamentally different.  Even when $\C$ is not a
normal subgroup of $\C_{|\J} \times \C_{|\I-\J}$, we can still partition
$\C_{|\J}
\times \C_{|\I-\J}$ into ``cosets" corresponding to the distinct possible
state differences in $\Sigma_\J(\C)$ under the state difference map, thus
establishing a one-to-one map between the ``cosets" of $\C$ in $\C_{|\J}
\times \C_{|\I-\J}$ and the state space $\Sigma_\J(\C)$.
Thus the basic idea of a correspondence
between syndrome equivalence classes of $\C_{|\J} \times \C_{|\I-\J}$ and
$\Sigma_\J(\C)$ still holds.

	There is a nice generalization of the above theorem, as follows.  Given an
algebraic group code $\C$ in the sense of \cite{FT93}, the \emph{label
groups} of $\C$ are defined as the quotient groups
$\{\C_{|\{k\}}/\C_{|:\{k\}} \cong \Sigma_{\{k\}}(\C), k \in \I\}$.  The
group code $\C$ then has the same dynamical structure as its label code
$\qb(\C)$, obtained by the natural map $q_k:  \C_{|\{k\}} \to
\C_{|\{k\}}/\C_{|:\{k\}}$ of each \emph{output group}
$\C_{|\{k\}}$ of $\C$ onto its label group.   $\C$ is said to have
\emph{abelian dynamics} if all label groups are abelian, for then and only
then all state spaces $\Sigma_\J(\C)$ are abelian \cite{FT93}.

We define the \emph{output sequence space} of $\C$ as the direct product
(or whatever product/sum is appropriate) of the output groups, 
$\W(\C) =  \prod_{k\in \I} \C_{|\{k\}}$, and the \emph{nondynamical
sequence space} as the product $\V(\C) =  \prod_{k\in \I} \C_{|:\{k\}}$.
\begin{theorem}[abelian dynamics theorem]  If $\C$ is an algebraic group 
code in the sense of \cite{FT93}, then $\C$ has abelian dynamics if
and only if $\C$ is normal in its output sequence space $\W(\C)$.
\end{theorem}

\emph{Proof}.  If $\C$ has abelian dynamics, then $\W(\C)/\V(\C) =
\prod_{k\in\I} \C_{|\{k\}}/\C_{|:\{k\}}$ is abelian.  Thus $\C/\V(\C)$ is
an abelian and normal subgroup.  By the correspondence theorem,
$\C$ is normal in $\W(\C)$.

Conversely, if $\C$ is a normal subgroup of $\W(\C) = \C_{|\{k\}} \times 
\prod_{k'\in\I-\{k\}} \C_{|\{k'\}}$, then \emph{a fortiori} $\C$ is normal
in $\C_{|\{k\}} \times \C_{|\I-\{k\}}$, since $\C_{|\I-\{k\}} \subseteq
\prod_{k'\in\I-\{k\}} \C_{|\{k'\}}$.  Therefore, by the previous
theorem, the label group $\C_{|\{k\}}/\C_{|:\{k\}}$ (which is the state
space $\Sigma_{\{k\}}(\C)$) is abelian, for any $k \in \I$.  Since all
label groups are abelian, $\C$ has abelian dynamics.
\qed

A \emph{syndrome-former} for $\C$ is a dynamical map defined on the output
sequence space $\W(\C)$ (or a larger sequence space) whose kernel is $\C$. 
It follows from this theorem that a syndrome-former can be homomorphic if
and only if $\C$ has abelian dynamics.  However, as we see from the example
of a state difference map, a syndrome-former can be non-homomorphic while
still being straightforward and essentially group-theoretic.  Thus our
assumption of abelian dynamics in this paper is not fundamental, as the
syndrome-former constructions of Fagnani and Zampieri \cite{FZ99} show.

\section{Notions of finite memory}

In this section we discuss several notions of finite memory, and study
their duality properties in a group-theoretic context.  Most of these
notions have been introduced previously in behavioral system theory
\cite{W89} in a set-theoretic context.

We first introduce $L$-controllability and $L$-observability, which turn
out to be duals.  We give two characterizations of each, which are also
duals.  We then introduce $L$-finiteness and $L$-completeness, also duals,
and show that they are equivalent to $L$-controllability and
$L$-observability, respectively, in appropriate settings.

To discuss memory, we must assume that the time index set $\I$ is ordered; 
\ie without loss of generality, $\I \subseteq \Z$.
  We will use the notation of
\cite{FT93} for subintervals of $\I$;  \eg
\begin{eqnarray*}
		[m, n) & = & \{k \in \I \mid  m \le k < n\}; \\
		m^- & = & \{k \in \I \mid k < m\};  \\
		n^+ & = & \{k \in \I \mid k \ge n\}.
\end{eqnarray*}
Thus $\I$ is the disjoint union of the three subintervals $\{m^-, [m, n),
n^+\}$.

\subsection{Strong controllability and observability}  \label{sco}

We now study the duality between notions of strong controllability and
observability.  Our definition of strong controllability is the same as
that  of Willems \cite{W89}.  Our definition of strong observability
(introduced in \cite{LFMT94}) corresponds to Willems' definition of
``finite memory."\footnote{In \cite[p.\ 336]{W97},
Willems calls this notion ``insightful" for discrete-time behaviors.}  We
show that these two notions are duals.  We also show that strong
controllability or observability implies controllability or observability,
respectively, as defined earlier.

	  Given a finite interval $[m, n)$, a code $\C$ is
\textbf{[\emph{m, n})-controllable} if for any $\cb,\cb' \in \C$ there
exists a $\cb'' \in \C$ such that $\cb''_{|m^-} = \cb_{|m^-}$ and
$\cb''_{|n^+} =
\cb'_{|n^+}$.  A code is \textbf{\emph{L}-controllable} if it is $[m, m +
L)$-controllable for every length-$L$ interval $[m, m + L)$, and
\textbf{strongly controllable} if it is $L$-controllable for some $L$. 
The least such $L$ is the
\textbf{controller memory} of $\C$.

	The following controllability test follows directly from the definition.

\begin{theorem}[first [m, n)-controllability test]  A code $\C$ is
$[m, n)$-controllable if and only if $\C_{|\I-[m,n)} = \C_{|m^-} \times
\C_{|n^+}$.
\end{theorem}

\emph{Proof}.  This merely restates the definition;  it says that $\C$ is
$[m, n)$-controllable if and only if any past in $\C_{|m^-}$ can be linked
to any future in $\C_{|n^+}$.  	\qed

	If $\C$ is a group code, then we have an alternative controllability test:
\begin{theorem}[second [m, n)-controllability test]
A group code $\C$
is  $[m, n)$-controllable if and only if $\C = \C_{:n^-} + \C_{:m^+}$.
\end{theorem}

\emph{Proof}.  If $\C$ is generated by $\C_{:n^-}$ and $\C_{:m^+}$, then
any past $\cb_{|m^-}$ can be linked to any future $\cb_{|n^+}$ as
follows:  find any $\cb^- \in \C_{:n^-}$  and $\cb^+ \in \C_{:m^+}$ such
that $(\cb^-)_{|m^-} = \cb_{|m^-}$ and $(\cb^+)_{|n^+} = \cb_{|n^+}$; 
then $\cb^- + \cb^+$ is the desired linking sequence. 
Conversely, if $\C$ is $[m, n)$-controllable, then any $\cb^- \in
\C_{:m^-}$ can be linked to $\zerob \in \C_{:n^+}$, and any $\cb^+
\in \C_{:n^+}$ can be linked to $\zerob \in \C_{:m^-}$, which implies that
$\C = \C_{:n^-} + \C_{:m^+}$.  \qed  

\begin{figure}[t]
\setlength{\unitlength}{5pt}
\begin{center}
\begin{picture}(32,10)(0,2)
\put(0,4){\line(1,0){32}}
\put(3,2){$\cb$}
\put(3,5){$\cb''$}
\put(0,10){\line(1,0){32}}
\put(3,11){$\cb'$}
\put(8,2){\dashbox(0,10){}}
\put(7,13){$m$}
\qbezier(10,4)(16,4)(16,7)
\qbezier(22,10)(16,10)(16,7)
\put(17,6){$\cb''$}
\put(24,2){\dashbox(0,10){}}
\put(23,13){$n$}
\put(27,2){$\cb$}
\put(27,8){$\cb''$}
\put(27,11){$\cb'$}
\end{picture}

Figure 6.1.  Illustration of $[m, n)$-controllability \\ with code sequences
$\cb, \cb'$ and $\cb''$.
\end{center}
\end{figure}

	The definition of $[m, n)$-controllability, illustrated in Figure 6.1,
involves a notion of finite reachability:  from any state (set of past
trajectories) at time $m$ we can reach any state (set of future
trajectories) at time $n$.  The first $[m, n)$-controllability test
translates this into a notion of memorylessness:  the state at time $n$ is
not constrained by the trajectory before time $m$.  The second $[m,
n)$-controllability test relies on the group property, by which it suffices
to show that every state at time $m$ can reach the zero state at time $n$
and every state at time $n$ can be reached from the zero state at time
$m$;  it then translates this observation into the statement that every
code sequence can be decomposed into a code sequence in $C_{:n^-}$ and a
code sequence in $C_{:m^+}$, which is a generatability criterion.

We define a code $\C$ to be \textbf{[\emph{m, n})-observable} if whenever
$\cb_{|[m,n)} = \cb'_{|[m,n)}$ for $\cb,\cb' \in \C$,
then the concatenation of $\cb_{|m^-}, \cb_{|[m,n)} = \cb'_{|[m,n)}$, and
$\cb'_{|n^+}$ is in $\C$.  A code is \textbf{\emph{L}-observable} if it is
$[m, m + L)$-observable for every length-$L$ interval $[m, m + L)$, and
\textbf{strongly observable} if it is $L$-observable for some $L$.  The
least such $L$ is the \textbf{observer memory} of
$\C$.

	The following observability test follows directly from this definition:
\begin{theorem}[first [m, n)-observability test]  A code $\C$ in a
sequence space $\W$ is  $[m, n)$-observable if and only if 
\end{theorem}
$$\C = \{\wb \in
\W \mid \wb_{|n^-} \in \C_{|n^-}, \wb_{|m^+} \in \C_{|m^+}\}.$$

\emph{Proof}.  If $\C = \{\wb \in \W \mid \wb_{|n^-} \in \C_{|n^-},
\wb_{|m^+} \in \C_{|m^+}\}$ and $\cb,\cb' \in \C$ have a common central
segment $\cb_{|[m,n)}$, then $\wb = (\cb_{|m^-}, \cb_{|[m,n)},
\cb'_{|n^+})$ satisfies the constraints  $\wb_{|n^-} \in \C_{|n^-},
\wb_{|m^+} \in \C_{|m^+}$ and is therefore in $\C$, so $\C$ is $[m,
n)$-observable.  Conversely, if $\C$ is $[m, n)$-observable, then the
fact that if $\cb,\cb' \in \C$ have a common central segment
$\cb_{|[m,n)}$ then $\wb = (\cb_{|m^-}, \cb_{|[m,n)},
\cb'_{|n^+})$ is a code sequence implies that any sequence $\wb \in
\W$ whose restrictions $\wb_{|n^-}$ and $\wb_{|m^+}$ equal restricted code
sequences $\cb_{|n^-} \in \C_{|n^-}$ and $\cb_{|m^+} \in \C_{|m^+}$,
respectively, is a valid code sequence.	\qed

	If $\C$ is a group code, then we have an alternative observability test:
\begin{theorem}[second [m, n)-observability test]  A group code $\C$
is $[m, n)$-observable if and only if $\C_{:I-[m,n)} = \C_{:m^-} \times
\C_{:n^+}$.
\end{theorem}

\emph{Proof}.  In general, $\C_{:m^-} \times \C_{:n^+} \subseteq
\C_{:I-[m,n)}$.  If $\cb \in \C_{:I-[m,n)}$, then $\cb_{|[m,n)} =
\zerob_{|[m,n)}$.  Since $\zerob \in \C$, if $\C$ is $[m, n)$-observable,
then the concatenations $(\cb_{|m^-}, \zerob_{|m^+}) = P_{m^-}(\cb)$
and $(\zerob_{|n^-}, \cb_{|n^+}) = P_{n^+}(\cb)$ are in $\C$, and thus in
$\C_{:m^-}$ and $\C_{:n^+}$, respectively.  So $\C_{:I-[m,n)} \subseteq
\C_{:m^-} \times \C_{:n^+}$, which implies that $\C_{:I-[m,n)} =
\C_{:m^-} \times \C_{:n^+}$.

	Conversely, if $\cb,\cb' \in \C$ are such that $\cb_{|[m,n)} =
\cb'_{|[m,n)}$, then $\cb - \cb' \in \C_{:I-[m,n)}$.  If $\C_{:I-[m,n)} =
\C_{:m^-} \times \C_{:n^+}$, then $\cb - \cb'$ may be written as $\cb -
\cb' = \cb^- + \cb^+$, where $\cb^- \in C_{:m^-}$ and $\cb^+ \in
C_{:n^+}$.  It follows that 
$$
		P_{m^-}(\cb) + P_{m^+}(\cb') = \cb' + P_{m^-}(\cb - \cb') = \cb' + \cb^-,
$$
which by the group property of $\C$ is in $\C$.  So $\C$ is $[m,
n)$-observable.   \qed 

\begin{figure}[t]
\setlength{\unitlength}{5pt}
\begin{center}
\begin{picture}(32,10)(0,2)
\put(1,2){$\cb$}
\qbezier(0,12)(0,7)(8,7)
\qbezier(0,2)(0,7)(8,7)
\put(10,8){$\cb = \cb' = \cb''$}
\put(8,7){\line(1,0){16}}
\put(1,11){$\cb'$}
\put(8,2){\dashbox(0,10){}}
\put(7,13){$m$}
\put(0,5){$\cb''$}
\put(24,2){\dashbox(0,10){}}
\put(23,13){$n$}
\qbezier(32,12)(32,7)(24,7)
\qbezier(32,2)(32,7)(24,7)
\put(30,2){$\cb$}
\put(31,8){$\cb''$}
\put(30,11){$\cb'$}
\end{picture}

Figure 6.2.  Illustration of $[m, n)$-observability \\ with code sequences 
$\cb, \cb'$ and $\cb''$.
\end{center}
\end{figure}

Our definition of $[m, n)$-observability, illustrated in Figure 6.2, is
implicitly a notion of state observability:  given a segment of a code
sequence $\cb_{|[m,n)}$, the states at time $m$ and $n$ (and indeed during
the entire interval $[m, n)$) are determined.  The first $[m,
n)$-observability test translates this into a checkability criterion:  if
a sequence looks like a code sequence during the overlapping intervals
$n^-$ and $m^+$, then it is a code sequence.  The second $[m,
n)$-observability test relies on the group property, by which it suffices
to show that $\cb_{|[m,n)} =
\zerob_{|[m,n)}$ implies that $\cb \in \C$ passes through the zero state at
times $m$ and $n$ (and therefore during the entire interval $[m, n)$);  it
then translates this observation into the statement that every code
sequence with $\cb_{|[m,n)} = \zerob_{|[m,n)}$ can be decomposed into a
code sequence in $\C_{:m^-}$ and a sequence in $\C_{:n^+}$, which is
another notion of memorylessness.

	Our desired duality theorem then follows directly from projection/subcode 
duality, applied to either of two dual pairs of tests.  
	The first proof shows that the first $[m, n)$-controllability test and the
second $[m, n)$-observability test are duals, whereas the second proof
shows that the second $[m, n)$-controllability test and the first $[m,
n)$-observability test are duals.
\begin{theorem}[strong controllability/observability duality]  Given dual
group codes $\C, \C^\perp$ and a finite interval $[m, n) \subseteq
\I$, $\C$ is $[m, n)$-controllable if and only if $\C^\perp$ is 
$[m, n)$-observable.
\end{theorem}

\emph{First proof}.  By the first $[m, n)$-controllability test, $\C$ is
$[m, n)$-controllable if and only if $\C_{|I-[m,n)} = \C_{|m^-} \times
\C_{|n^+}$.  By projection/subcode duality, the duals of the left and right
sides of this equation are $(\C^\perp)_{|:I-[m,n)}$ and $(C^\perp)_{|:m^-}
\times (\C^\perp)_{|:n^+}$, respectively.  Therefore $\C_{|I-[m,n)} =
\C_{|m^-} \times \C_{|n^+}$ if and only if $(\C^\perp)_{|:I-[m,n)}
= (C^\perp)_{|:m^-} \times (\C^\perp)_{|:n^+}$, which is effectively the
second $[m, n)$-observability test for $\C^\perp$.  	 	\qed

\emph{Second proof}.  By the second $[m, n)$-controllability test, $\C$ is
$[m, n)$-controllable if and only if $\C = \C_{:n^-} + \C_{:m^+}$.  By
projection/subcode and sum/intersection duality, the duals of these two codes
are $\C^\perp$ and $(\C_{:n^-})^\perp \cap (\C_{:m^+})^\perp$, respectively. 
Furthermore, by projection/subcode duality, 
$$(\C_{:n^-})^\perp =
\{\xb \in \W\h \mid \xb_{|n^-}
\in (C^\perp)_{|n^-}\};$$  $$(\C_{:m^+})^\perp = \{\xb \in \W\h \mid
\xb_{|m^+} \in (C^\perp)_{|m^+}\};$$
so $(\C_{:n^-})^\perp \cap
(\C_{:m^+})^\perp$ is equal to
 $$\{\xb \in \W\h \mid \xb_{|n^-}
\in (C^\perp)_{|n^-},
\xb_{|m^+} \in (C^\perp)_{|m^+}\}.$$  But this is $\C^\perp$ if and only if
$\C^\perp$ is $[m,n)$-observable, by the first
$[m, n)$-observability test for $\C^\perp$.  	\qed

As immediate corollaries, we have:

\begin{corollary}  Given dual group codes $\C$ and $\C^\perp$, 
\begin{itemize}
\item[(a)]  $\C$ is $L$-controllable $\Leftrightarrow$ $\C^\perp$ is
$L$-observable;
\item[(b)]  $\C$ is strongly controllable  $\Leftrightarrow$ $\C^\perp$ is
strongly observable;
\item[(c)]  controller memory of $\C$ $=$ observer memory of
$\C^\perp$.
\end{itemize}
\end{corollary}

This fundamental duality result provides strong support for our use of
the term ``observability" rather than ``finite memory" in \cite{LFMT94} and
here.  Also, it is desirable to distinguish between controller and observer
memory.

All notions of zero memory coincide:  a code is 0-controllable or
0-observable or memoryless if for any time $m$ and any $\cb, \cb' \in \C$,
the concatenation
$(\cb_{|m^-},
\cb'_{|m^+})$ is in $\C$.  

However, if $\C$ is not memoryless, then there is no necessary relationship
between its controller memory and its observer memory;  these are two
distinct (and dual) notions of the memory of $\C$.  The controller memory
measures the maximum time needed to link any past to any future.  The
observer memory measures the maximum observation time needed to obtain a
``sufficient statistic" for predicting the future (resp.\ the past) from
the past (resp.\ the future).

Finally, we now verify that strong controllability (resp.\ observability)
implies wide-sense controllability (resp.\ observability) as defined
earlier.  For observability, we will consider only the case in which all
symbol groups are discrete, in which case the topology of the complete
sequence space $\W^c$ is the topology of pointwise convergence.  Under our
standing assumptions, the corresponding controllability result then holds
when all symbol groups are compact.
\begin{theorem}
Let $\C$ and $\C^\perp$ be dual group codes in sequence spaces $\W$ and
$\W\h$, respectively.  Let all symbol groups $G_k$ of $\W^c$ be
discrete, and all symbol groups $G_k\h$ of $(\W^c)\h$ be compact.  Then
$\C$ is observable if $\C$ is strongly observable, and $\C^\perp$ is
controllable if $\C^\perp$ is strongly controllable.
\end{theorem}
\emph{Proof}.  Suppose $\C$ is strongly observable but not observable;  \ie
$(\C^c)_f \neq \C_f$.  Then there exists some finite sequence $\wb \in
(\C^c)_f$ that is not in
$\C_f$.  Since the topology of $\C^c$ is the topology of pointwise
convergence, this means that there is some series $\{\cb^n\}$ of code
sequences $\cb^n \in \C$ that converges pointwise to $\wb$ as $n \to
\infty$.  Now $\C$ is $L$-observable for some integer $L$, and the support
of $\wb$ is some finite interval $[k, k')$.  Pointwise convergence then
implies that $\cb^n_{|[k-L,k'+L)} = \wb_{|[k-L,k'+L)}$ for all sufficiently
large $n$.  But $L$-observability then implies that $\wb$ is a finite code
sequence in $\C$, since $\cb^n_{|[k-L,k'+L)}$ is a code sequence that
agrees with the all-zero sequence $\zerob$ during the length-$L$ intervals
$[k - L, k)$ and $[k', k + L)$; contradiction.  Thus $\C$ must
be observable.

	Finally, $\C^\perp$ is controllable if and only if $\C$ is observable by
Theorem \ref{cod}, and $\C^\perp$ is strongly controllable if and only if
$\C$ is strongly observable by the corollary above.	\qed

%Fagnani \cite{F97} proved that wide-sense controllability implies
%strong controllability when $\C^\perp$ is complete and
%time-invariant. This theorem extends Fagnani's controllability result to
%non-time-invariant codes $\C^\perp$ in a finite sequence space $(\W^c)\h$,
%provided that all symbol groups $G_k\h$ are compact.  Note that in this
%case $(\W^c)\h$ is compact.

	On the other hand, the following
example shows that a controllable code need not be strongly controllable, 
and an observable code need not be strongly observable.

\medskip
\noindent
\textbf{Example 6}.  Let $\I = \{1, 2, \ldots\}$, and let $\C$ be the group
code over a group $G$ in which the symbols $c_k$ are chosen freely from
$G$ at times $k = 2^n$ for all $n \in \{0, 1, \ldots\}$, but at all other
times $c_k = c_{k-1}$.  Then $\C$ is generated by finite sequences of the
form $(\ldots, 0, g, g, \ldots, g, 0, \ldots)$ with support $[2^n,
2^{n+1})$ and is thus controllable, but $\C$ is not $L$-controllable for
any $L \in
\Z$.  The dual subcode $\C^\perp$ is thus observable but not strongly
observable.
\qed

\subsection{$L$-finiteness and $L$-completeness}

In this subsection we introduce $L$-finiteness and
$L$-completeness, which turn out to be duals.  Our definition of
$L$-completeness is the same as that of Willems
\cite{W89}, except for the modification that we made earlier when defining
completeness; it is a notion of finite checkability in complete sequence
spaces.  We define
$L$-finiteness in a dual way as a notion of finite generatability
that applies to group codes in finite sequence spaces. We show that in
these restricted contexts $L$-finiteness is equivalent to
$L$-controllability, and $L$-completeness is equivalent to
$L$-observability.

	We define a group code $\C$ in a finite sequence space $\W_f$ to be
\textbf{\emph{L}-finite} if it is generated by its finite code sequences of
length $L + 1$:
$$
		\C = \sum_{k\in\Z} \C_{:[k,k+L]}.
$$
In other words, $\C$ is $L$-finite if and only if any $\cb \in \C$ may be
decomposed into a sum of code sequences $\cb_{[k,k+L]} \in \C_{:[k,k+L]}$
whose supports are intervals of length $L + 1$:
$$
		\cb = \sum_{k\in\Z} \cb_{[k,k+L]}.
$$
Notice that this definition makes sense only in the setting of group
codes;  no analogue exists for set-theoretic codes.

	The following theorem shows that for such group codes, $L$-finiteness is
equivalent to $L$-controllability:
\begin{theorem}[L-finite = L-controllable + finite] 
\label{LfLcf} If $\C$
is a group code in a finite sequence space $\W_f$, then $\C$ is $L$-finite
if and only if $\C$ is $L$-controllable.
\end{theorem}

\emph{Proof}.  If $\C$ is $L$-finite, then we may write any $\cb \in \C$ as
$\cb = \sum_{j\in\Z} \cb_{[j,j+L]}$, so for any $k$, $\cb$ may be written
as a sum $\cb = \cb_{(k+L)^-} + \cb_{k^+}$ with $\cb_{(k+L)^-} \in
\C_{:(k+L)^-}$ and $\cb_{k^+} \in \C_{:k^+}$, as follows:
$$
		\cb = \sum_{j<k} \cb_{[j,j+L]} + \sum_{j \ge k} \cb_{[j,j+L]} =
\cb_{(k+L)^-} + \cb_{k^+}.
$$
Thus $\C = \C_{:(k+L)^-} + \C_{:k^+}$, so by the second
$[m,n)$-controllability test $\C$ is $[k,k+L)$-controllable for all $k$,
and thus $L$-controllable.

	Conversely, let $\C$ be $L$-controllable.  Since all code sequences are
finite, the support of any $\cb \in \C$ is a finite interval, say $[k, k' +
L]$.  By $L$-controllability, for any  $j \in \Z$ there exists a $\cb_{j^+}
\in \C_{:j^+}$ such that $(\cb_{j^+})_{|j^-} = \zerob_{|j^-}$ and
$(\cb_{j^+})_{|(j+L)^+} = \cb_{|(j+L)^+}$.  Then $\cb_{[j,j+L]} = \cb_{j^+}
- (\cb_{(j+1)^+})_{|(j+L)^+}$ has support $[j, j + L]$.  Thus for any
$\cb \in \C$ we have $\cb = \sum_{j\in [k,k']} \cb_{[j,j+L]}$; so $\C$ is
$L$-finite.		\qed

Dually, a group code $\C$ in a complete sequence space $\W^c$ will be
defined as
\textbf{\emph{L}-complete} if 
$$
		\C = \{\wb \in \W^c \mid \wb_{|[k,k+L]} \in (\C_{|[k,k+L]})\cl
\mbox{~for all~} k \in \Z\}.
$$
As in our definition of completeness, this definition uses closed
restrictions $(\C_{|[k,k+L]})\cl$.  If the closed-projections assumption
holds, then this reduces to Willems' definition \cite{W89}. In other
words, $\C$ is
$L$-complete  if whenever
$\wb \in \W^c$ looks like a code sequence through all windows of length
$L + 1$, then $\wb$ is in fact a code sequence.

The duality of $L$-completeness
and $L$-finiteness then follows directly from projection/subcode duality:
\begin{theorem}[L-finiteness/L-completeness duality]
If $\C$ and $\C^\perp$ are dual group codes in dual finite and complete
sequence spaces $\W_f$ and $(\W_f)\h$, then $\C$ is $L$-finite if and only
if $\C^\perp$ is $L$-complete.
\end{theorem}

\emph{Proof}.  By sum/intersection duality, $\C = \sum_{k\in\Z}
\C_{:[k,k+L]}$ if and only if $\C^\perp =
\bigcap_{k\in\Z} (\C_{:[k,k+L]})^\perp.$  By projection/subcode duality,
$(\C_{:[k,k+L]})^\perp$  is the closure of 
$$\{\xb \in (\W_f)\h \mid
\xb_{|[k,k+L]}
\in (\C^\perp)_{|[k,k+L]}\}.$$  Then 
$$\C^\perp = \{\xb \in (\W_f)\h \mid
(\xb_{|[k,k+L]})\cl \in (\C^\perp)_{|[k,k+L]}  \mbox{~for all~} k\},$$
 which is the definition of $L$-completeness for $\C^\perp$.		\qed

\begin{corollary}[L-complete = L-observable + complete]
\label{LcLoc}  If
$\C$ is a (complete) group 
 code in a complete sequence space $\W^c$, then
$\C$ is $L$-complete if and only if $\C$ is $L$-observable.
\end{corollary}

\emph{Proof}.  We have now shown that the following are equivalent:  
\begin{eqnarray*}
\mbox{
$\C$ is $L$-complete} & \Leftrightarrow & \mbox{$\C^\perp$ is $L$-finite} 
 \Leftrightarrow \\ \mbox{$\C^\perp$ is $L$-controllable}  & \Leftrightarrow
& \mbox{$\C$ is
$L$-observable.} 	\qed
\end{eqnarray*}

This is a group-theoretic version of Willems' set-theoretic theorem
\cite{W89} that a complete code is $L$-complete if and only if it has
$L$-finite memory (is $L$-observable).

	While $L$-finiteness and $L$-controllability are equivalent (resp.\
$L$-completeness and
$L$-observability), the tests that they imply are different
in practice, as we show by revisiting the
controllability and observability tests of Subsection \ref{sco}, and then
applying these tests to our examples.

	The tests of Subsection \ref{sco} involve a three-way partition of the
time axis $\I$, namely $\I = \{m^-, [m, n), n^+\}$.  We may correspondingly
identify $\I$ with an equivalent finite time axis $\I' = \{1, 2, 3\}$ of
length 3, and we may regard any code $\C$ defined on $\I$ as a
code $\C'$ defined on $\I'$.  Note that the
equivalent length-3 sequence space $\W' = \W_{|m^-} \times \W_{|[m,n)}
\times \W_{|n^+}$ is both complete and finite, assuming that each of the
restrictions $\W_{|m^-}, \W_{|[m,n)}$ and $\W_{|n^+}$ is complete (closed).

	Now in terms of the equivalent code $\C'$  on $\I'$, we have:
\begin{itemize}
\item $\C$ is $[m, n)$-controllable $\Leftrightarrow$ $\C'$ is 1-controllable;
\item $\C'$ is 1-controllable $\Leftrightarrow$ $\C'_{|\{1,3\}} = \C'_{|\{1\}}
\times \C'_{|\{3\}}$;
\item $\C'$ is 1-finite $\Leftrightarrow$ $\C' = \C'_{:\{1,2\}} +
\C'_{:\{2,3\}}$.
\end{itemize}
The latter two tests correspond to our first and second $[m,
n)$-controllability tests, respectively, and their equivalence follows from
Theorem \ref{LfLcf}.  Similarly,
\begin{itemize}
\item $\C$ is $[m, n)$-observable $\Leftrightarrow$ $\C'$ is 1-observable;
\item $\C'$ is 1-observable $\Leftrightarrow$ $\C'_{:\{1,3\}} = \C'_{:\{1\}}
\times \C'_{:\{3\}}$;
\item $\C'$ is 1-complete $\Leftrightarrow$ \\ $\C' = \{\wb \in \W' \mid
\wb_{|\{1,2\}} \in\C'_{|\{1,2\}}, \wb_{|\{2,3\}} \in \C'_{|\{2,3\}}\}.$
\end{itemize}
These two tests correspond to our second and first $[m, n)$-observability
tests, respectively, and their equivalence follows from Corollary
\ref{LcLoc}, or by duality from our $[m, n)$-controllability
tests.

	Now let us see how these various tests apply to some of our example
codes.

\medskip
\noindent
\textbf{Example 2} (cont.)  A bi-infinite repetition code $\C$ over $G$ is
1-observable, because two code sequences that agree anywhere
agree everywhere.  It is 1-complete, because a sequence $\wb$ is in $\C$ if
and only if the two components of every length-2 restriction
$\wb_{|[k,k+1]}$ are equal.  The zero-sum code $\C^\perp$ over $\Gh$
is 1-controllable, because for any two finite sequences $\xb, \xb'$ and
any $k \in \Z$, there is an $h \in \Gh$ such that
$(\xb_{|k^-}, h, (\xb')_{|(k+1)^+})$ is in $\C^\perp$.  It is
1-finite, because it is generated by its length-2 sequences $(\ldots, 0, g,
-g, 0, \ldots)$. 	\qed

\medskip
\noindent
\textbf{Example 3} (cont.)  The finite subcode $\C_f$ of the rate-1/3
linear time-invariant convolutional code $\C$ over $\Z_4$ comprising all
linear combinations of time shifts of $\gb = (\dots, 000, 100, 010,
002, 000, \ldots)$ is by definition 2-finite and evidently 2-controllable,
since it has a feedbackfree encoder with memory 2.  The finite subset
$(\C^\perp)_f$ of its dual rate-2/3 code $\C^\perp$ comprising all linear
combinations of time shifts of the generators $\hb_1 = (\ldots, 000,
100, 030, 000, \ldots)$, $\hb_2 = (\ldots, 000, 020, 001, 000, \ldots)$
is by definition 1-finite and evidently 1-controllable.  

$\C$ is
1-complete, because it is the set of all sequences orthogonal to all
shifts of the length-2 sequences $\hb_1$ and $\hb_2$.  $\C$ is
1-observable, because a zero symbol 000 can be observed only if $\C$ is in
the zero state.  Similarly, $\C^\perp$ is 2-complete, because it is the set
of all sequences orthogonal to all shifts of the length-3 sequence $\gb$,
and it is 2-observable since two successive zero symbols $(000, 000)$ can
be observed only if $\C^\perp$ is in the zero state, as the reader may verify.	
\qed

\medskip
\noindent
\textbf{Example 4} (cont.)  Loeliger's code $\C$ is 1-observable, since two
code sequences with the same output $c_k$ have a uniquely determined past
and the same set of possible futures.  It is 1-complete, because a sequence
$\wb$ is in $\C$ if and only if the first component of every length-2
restriction $\wb_{|[k,k+1]}$ is twice the second component (mod $\Z$). Its
dual $\C^\perp$ is generated by the time shifts of the length-2 generator
$\hb = (\ldots, 0, 1, -2, 0, \ldots)$, and thus is by definition 1-finite; 
it is 1-controllable since it evidently has a feedbackfree
 encoder with memory 1.	\qed

\section{Dual granule decompositions}

	The development of \cite{FT93} is based on a decomposition of an
$L$-controllable group code $\C$ according to a chain of $j$-controllable
subcodes $\C_j$,
$$
		\C_0 \subseteq \C_1 \subseteq \cdots \subseteq \C_L = \C,
$$
and then a further decomposition of the quotients $\C_j/\C_{j-1}$ into
direct products of $j$th-level granules, defined (in additive notation) as
$$
		\Gamma_{[k,k+j]}(\C) = \frac{\C_{:[k,k+j]}}{\C_{:[k,k+j)} +
\C_{:(k,k+j]}}.
$$

	We now give a dual decomposition of an $L$-observable group code $\C$
according to the $j$-observable supercode chain,
$$
		\C = \C^L \subseteq \C^{L-1} \subseteq \cdots \subseteq \C^0,
$$
and then a further decomposition of the quotients $\C^{j-1}/\C^j$ into
products of $j$th-level observer granules $\Phi_{[k,k+j]}(\C)$.  Here the
``granules" $\Gamma_{[k,k+j]}(\C)$ of \cite{FT93} will be called
``controller granules."

	We will show that the $j$-observable supercode $\C^j$ of $\C$ is the dual
of the $j$-controllable subcode $(\C^\perp)_j$ of its dual $\C^\perp$, and
that the observer granules of $\C$ act as the character groups of the
corresponding controller granules of $\C^\perp$.

	In the following section, we will give examples of how this observability
structure can be used to construct minimal observer-form encoders, state
observers and syndrome-formers.  A general construction of syndrome-formers
for non-topological group codes over finite, possibly nonabelian groups
that uses this observability structure is given in \cite{FZ99}.

\subsection{Controller decomposition}

	We review the results of \cite{FT93} in our topological group setting, to
prepare for dualizing them.

From here on, for simplicity, when we denote a sequence subspace in a
sequence space $\W$ by a Cartesian product, \eg $\prod_{k\in\I} A_k$, we
imply that the product is of the same type as that of $\W$--- \eg a direct
product, Laurent product, or direct sum.

	As in \cite{FT93}, we define the \textbf{\emph{j}-controllable subcode}
$\C_j$ of a group code $\C$ in a sequence space $\W$ as the code generated
by the length-$(j + 1)$ subcodes $\C_{:[k,k+j]}$ of $\C$:
$$
		\C_j = \sum_{k\in\Z} \C_{:[k,k+j]}.
$$
If $\W$ is finite, then $\C_j$ by definition is $j$-finite.  By a
proof like that of Theorem \ref{LfLcf}, $\C_j$ is $j$-controllable, and
$\C$ is $L$-controllable if and only if $\C = \C_L$.

	If $\C$ is $L$-controllable, then we have a chain of $j$-controllable
subcodes
$$
		\{\zerob\} \subseteq \C_0 \subseteq \C_1 \subseteq \cdots \subseteq \C_L
= \C.
$$
For consistency in indexing, we may denote the trivial subcode $\{\zerob\}$
by $\C_{-1}$.

The 0-controllable subcode $C_0$ (called the \emph{parallel transition
subcode} of $\C$) is a memoryless sequence space of the same type as $\W$,
whose symbol groups are the length-1 subcodes $\C_{:\{k\}}$.  Since it is
memoryless, it has trivial dynamics (\ie trivial state spaces).

	The controller granules $\Gamma_{[k,k+j]}(\C)$ are defined by
$$
		\Gamma_{[k,k+j]}(\C) = \frac{(\C_j)_{:[k,k+j]}}{(\C_{j-1})_{:[k,k+j]}}, 
\quad k \in \Z, 0 \le j \le L.
$$
Since $(\C_j)_{:[k,k+j]} = \C_{:[k,k+j]}$ and $(\C_{j-1})_{:[k,k+j]} =
\C_{:[k,k+j)} + \C_{:(k,k+j]}$, this is equivalent to the definition of
\cite{FT93}:
\begin{eqnarray*}
		\Gamma_{[k,k+j]}(\C) & = & \frac{\C_{:[k,k+j]}}{\C_{:[k,k+j)} +
\C_{:(k,k+j]}}, \quad 1 \le j \le L; \\
		\Gamma_{[k,k]}(\C) & = & \C_{:[k,k]} = \C_{:\{k\}}.
\end{eqnarray*}
The cosets of $(\C_{j-1})_{:[k,k+j]}$ in $(\C_{j})_{:[k,k+j]}$ are
represented by sequences in $\C_{:[k,k+j]}$ that are not in the $(j -
1)$-controllable subcode $\C_{j-1}$.  The zeroth-level controller granules
$\Gamma_{[k,k]}(\C)$ are called ``nondynamical granules" and are equal to
the \emph{parallel transition subgroups} $\C_{:\{k\}}$.

	As in the code granule theorem of \cite{FT93}, we can then show that 
$$
		\frac{\C_j}{\C_{j-1}} \cong \prod_{k\in\Z} \Gamma_{[k,k+j]}(\C),  \quad 0
\le j \le L,
$$
where the product is a direct product, Laurent product, or direct sum 
according to the character of the sequence space $\W$ in which $\C$ lies. 
The proof essentially follows from the facts that $\C_j/\C_{j-1}$ is
generated by the sequences in $\C_{:[k,k+j]}$ that are not in $\C_{j-1}$
for all $k \in \Z$, and that $(\C_j)_{:[k,k+j+1]}$ is the direct product of
$(\C_j)_{:[k,k+j]}$ and $(\C_j)_{:[k+1,k+j+1]}$ modulo $\C_{j-1}$, since
the intersection of $(\C_j)_{:[k,k+j]}$ and $(\C_j)_{:[k+1,k+j+1]}$ is
$(\C_j)_{:(k,k+j]} \subseteq \C_{j-1}$.

	The restrictions of the future subcodes $(\C_j)_{:k^+}$ to time $k$ are
defined as the \textbf{\emph{j}th-level first-output groups}
$$
		F_{j,k}(\C) = ((\C_j)_{:k^+})_{|\{k\}}, \quad 0 \le j \le L,
$$
which form a chain
$$
		\{\zerob\} \subseteq F_{0,k}(\C) \subseteq F_{1,k}(\C)
\subseteq \cdots \subseteq F_{L,k}(\C) = F_k(\C),
$$
where $F_k(\C) = (\C_{:k^+})_{|\{k\}}$ is the \textbf{first-output group}
of $\C$ at time $k$ (also called the \emph{input group} \cite{FT93}). 
Since
$F_{j,k}(\C) = (\C_{:[k,k+j]})_{|\{k\}}$ and the kernels of the
restrictions to $\{k\}$ of $(\C_j)_{:[k,k+j]}$ and
$(\C_{j-1})_{:[k,k+j]}$ are both equal to $\C_{:(k,k+j]}$, it follows from
the correspondence theorem that the quotients of this chain are isomorphic
to the corresponding controller granules:
$$
		\frac{F_{j,k}(\C)}{F_{j-1,k}(\C)} \cong \Gamma_{[k,k+j]}(\C).
$$

Similarly, we define the \textbf{\emph{j}th-level last-output groups} as
$$ 
		L_{j,k}(\C) = ((\C_j)_{:(k+1)^-})_{|\{k\}}, \quad 0 \le j \le L,
$$
which form a chain up to $L_k(\C) = (\C_{:(k+1)^-})_{|\{k\}}$, the
\textbf{last-output group} of $\C$:
$$
\{\zerob\} \subseteq L_{0,k}(\C) \subseteq L_{1,k}(\C)
\subseteq \cdots \subseteq L_{L,k}(\C) = L_k(\C),
$$
with quotients also isomorphic to controller granules:
$$
		\frac{L_{j,k}(\C)}{L_{j-1,k}(\C)} \cong \Gamma_{[k-j,k]}(\C).
$$

	The \textbf{state code} of $\C$ is the group code $\sigmab(\C)$, where the
state map $\sigmab$ is the Cartesian product of the state maps $\sigma_k$;
\ie
$$
		\sigmab(\cb) = \{\sigma_k(\cb), k \in \Z\},
$$
where $\sigma_k(\cb) \in \Sigma_k(\C)$ is the state of the code sequence
$\cb \in \C$ at time $k$.  The kernel of the state code map is the parallel
transition subcode $\C_0$.  

As in \cite{FT93}, the state spaces and state
code of $\C$ may be decomposed according to the chains
\begin{eqnarray*}
		\{0\} = \sigma_k(\C_0) & \subseteq & \sigma_k(\C_1) \subseteq \cdots
\subseteq \sigma_k(\C_L) = \Sigma_k(\C); \\
		\{\zerob\} = \sigmab(\C_0) & \subseteq & \sigmab(\C_1) \subseteq \cdots
\subseteq \sigmab(\C_L) = \sigmab(\C),
\end{eqnarray*}
where $\sigma_k(\C_j)$ is the state space of the $j$-controllable subcode
$\C_j$ at time $k$ and $\sigmab(\C_j)$ is the state code of $\C_j$.  The
zeroth-level state code $\sigmab(\C_0)$ is trivial since $\C_0$ is
memoryless.

The quotients of the latter chain are isomorphic to
$\C_j/\C_{j-1}$:
$$
		\frac{\sigmab(\C_j)}{\sigmab(\C_{j-1})} \cong \frac{\C_j}{\C_{j-1}} \cong
\prod_{k\in\Z} \Gamma_{[k,k+j]}(\C), \quad  1 \le j \le L.
$$
The quotients of the former chain are isomorphic to direct products of the 
$j$th-level controller granules that are ``active" at time $k$:
$$
		\frac{\sigma_k(\C_j)}{\sigma_k(\C_{j-1})} \cong \prod_{i\in[k-j,k)}
\Gamma_{[i,i+j]}(\C), \quad  1 \le j \le L.
$$

Each single controller granule $\Gamma_{[k,k+j]}(\C)$ may be implemented
by a little state machine with an input group and a state space isomorphic
to $\Gamma_{[k,k+j]}(\C)$ which is active during the interval $(k,k+j]$,
as follows. An input in $F_{j,k}(\C)/F_{j-1,k}(\C) \cong
\Gamma_{[k,k+j]}(\C)$ arrives at time $k$ and determines a corresponding
first output, which is the time-$k$ output symbol of a representative of
the corresponding coset of
$(\C_{j-1})_{:[k,k+j]}$ in $(\C_j)_{:[k,k+j]}$, as well as a corresponding
state in a state space isomorphic to $\Gamma_{[k,k+j]}(\C)$ at time $k +
1$.  During the interval $(k, k + j]$, the state is constant, and
determines the remaining output symbols of the representative sequence. 
At time  $k + j$, the last output is emitted (a representative of
$L_{j,k+j}(\C)/L_{j-1,k+j}(\C) \cong \Gamma_{[k,k+j]}(\C)$), and the
granule becomes ``inactive;"  \ie no further memory is required.

	A $j$th-level encoder for $\C_j/\C_{j-1}$ in controller form may then be
implemented by combining the outputs of encoders for $\Gamma_{[k,k+j]}(\C)$
for all $k \in \Z$ (with finite or Laurent constraints, if appropriate; \eg
that there be only finitely many nonzero inputs for $k < 0$).  The
resulting encoder implements an isomorphism from the ``input sequence
space"  $\prod_{k} F_{j,k}(\C)/F_{j-1,k}(\C)  \cong \prod_k
\Gamma_{[k,k+j]}(\C)$ to the output space
$\C_j/\C_{j-1} \cong \prod_k \Gamma_{[k,k+j]}(\C)$.  The state space of the
$j$th-level encoder at any time is isomorphic to
$\sigma_k(\C_j)/\sigma_k(\C_{j-1})$ and is thus minimal.

Since the parallel transition subcode $\C_0 = \prod_k \C_{:\{k\}}$ is
a memoryless sequence space, the zeroth-level encoder for $\C_0$ requires
no memory;  the output is simply a complete, Laurent or finite sequence of
elements from the parallel transition subgroups
$F_{0,k}(\C) = \C_{:\{k\}}, k \in \Z$.

	Finally, a minimal encoder for $\C$ in controller form may be implemented
by adding the outputs of all $j$th-level encoders, $0 \le j \le L$.  Such
an encoder implements a one-to-one correspondence from $\prod_k F_k(\C)$ to
$\C$, but not necessarily an isomorphism \cite{FT93} (see also
\cite{LFMT94, LM96}).

\subsection{Controllable and uncontrollable codes}

	We now extend the results above to codes $\C \subseteq \W$ that are not
necessarily strongly controllable.  For simplicity, we take $\W$ to be a
complete sequence space.

	We then have a chain of subcodes
$$
		\{\zerob\} \subseteq \C_0 \subseteq \C_1 \subseteq \cdots \subseteq
(\C_f)^c
\subseteq \C,
$$
where $\C_j$ is the $j$-controllable subcode of $\C$, $j \ge 0$, and
$(\C_f)^c$ is the controllable subcode of $\C$.  We recall that $\C$ is
controllable if and only if $(\C_f)^c = \C$.  

	Since $(\C_f)^c$ is the (closure of the) code generated by all finite
subcodes of $\C$, 
$$
		(\C_f)^c = \sum_{\mathrm{finite~} \J} \C_\J,
$$
it is clear in the topological setting that $(\C_f)^c$ may be regarded as
the ``limit" of the $j$-controllable subcodes $\C_j$ as $j \to \infty$. 
For instance, if the symbol groups are discrete, then $(\C_f)^c$ is the
code consisting of the limits of all finite sequences in $\C$ in the
topology of pointwise convergence.

$\C$ is therefore controllable if and only if every sequence in $\C$ can be
expressed as such a limit of finite code sequences.  If $\C$ is
uncontrollable, then the code sequences not in $(\C_f)^c$ not only are
not finite, but also are not the limit of any series of finite sequences in
$\C$.  For example, in Examples 2 and 4, the only finite sequence in $\C$
is $\zerob$.

	Again, there are state space and state code chains as follows:
\begin{eqnarray*}
		\{0\} & \subseteq & \sigma_k(\C_1) \subseteq \cdots
\subseteq \sigma_k((\C_f)^c) \subseteq \sigma_k(\C) = \Sigma_k(\C); \\
		\{\zerob\} & \subseteq & \sigmab(\C_1) \subseteq \cdots
\subseteq \sigmab((\C_f)^c) \subseteq \sigmab(\C).
\end{eqnarray*}
Since $\C/(\C_f)^c \cong \sigmab(\C)/\sigmab((\C_f)^c)$ (because the
kernel $\C_0$ of the state map is a subcode of both $(\C_f)^c$ and $\C$),
it follows that
$(\C_f)^c = \C$ if and only if $\sigmab((\C_f)^c) = \sigmab(\C)$: 
\begin{theorem}[dynamical controllability test] \label{dct}
A complete group code $\C$
is controllable if and only if $\sigmab((\C_f)^c) = \sigmab(\C)$.
\end{theorem}

	In other words, finitization preserves the dynamics of $\C$ if and only if
$\C$ is controllable.

Let $\C$ be time-invariant, so that all state spaces $\Sigma_k(\C)$ are
congruent to $\Sigma_0(\C)$, and suppose that $\Sigma_0(\C)$ satisfies the
descending chain condition (DCC).  Then $\C$ is complete \cite[Prop.\
3.6]{LM96};  moreover, by the DCC, there can be only a finite number
of steps in the state space chain, which implies that the controllable
subcode $(\C_f)^c$ is strongly controllable (see
\cite{LFMT94, LM96}).  Thus we have:
\begin{theorem} \label{tidcc}
If $\C$ is a
time-invariant group code whose state space $\Sigma_0(\C)$ satisfies
the descending chain condition, then $(\C_f)^c$ is strongly controllable. 
Thus $\C$ is controllable if and only if $\C$ is strongly
controllable.	
\end{theorem}  

Example 6, in which $\C$ is controllable but not
strongly controllable even though the state space is never larger than 
$G$, shows that time-invariance
is essential.

Theorem \ref{tidcc} has been extended to time-invariant group codes over
finitely generated abelian symbol groups in \cite{F97}, and to
time-invariant ring codes over finitely generated modules over a principal
ideal domain in
\cite{FZ97}.  All versions of Theorem \ref{tidcc} depend on some sort of
finiteness condition.

	The basic structure theorem of Miles and Thomas \cite{MT78} (see
\cite{F97}) says that if $G$ is a compact abelian Lie group and $\C
\subseteq G^\Z$ is a closed time-invariant group code over $G$, then there
exists a finite chain
$$
		\{\zerob\} \subseteq \C_0 \subseteq \C_1 \subseteq \cdots \subseteq \C_L
= (\C_f)^c \subseteq \C^s \subseteq \C
$$
of closed normal time-invariant subcodes of $\C$, where $\C_j/\C_{j-1}
\cong (G_j)^\Z$ for some compact Lie group $G_j$ (in our setting, $G_j$
may be identified with the $j$th-level controller granule
$\Gamma_{[k,k+j]}$); $\C^s/(\C_f)^c$ is a solenoid (a compact connected
abelian group of finite topological dimension), as in our Example 4; and
$\C/\C^s$ is autonomous (a semi-simple Lie group), as in our Example 2. 
Using this theorem, Kitchens and Schmidt \cite{KS89} show that if $\C$ is a
compact controllable time-invariant group code whose state space
$\Sigma_0(\C)$ satisfies the DCC, then $\C$ is
strongly controllable.

\subsection{Observable and unobservable codes}

	An observer decomposition of $\C$ may be obtained by simply ``dualizing"
the controller decomposition just described.

	The \textbf{\emph{j}-observable supercode} $\C^j$ of a group code $\C$ in
a sequence space $\W$ is defined as
$$
		\C^j = \{\wb \in \W \mid \wb_{|[k,k+j]} \in \C_{|[k,k+j]} \mbox{~for
all~} k \in \Z\}.
$$
If $\W$ is complete, then  $\C^j$ by definition is $j$-complete.

	By projection/subcode duality, we have:
\begin{theorem}[subcode/supercode duality]  \label{ssd} 
  If $\C$ and $\C^\perp$ are dual group codes, then the 
 $j$-observable supercode of $\C^\perp$ is the
dual of the $j$-controllable subcode of $\C$:
$$
		(\C^\perp)^j = (\C_j)^\perp.		
$$
\end{theorem}

	It follows that $C^j$ is $j$-observable, and that $\C$ is $L$-observable
if and only if $\C = \C^L$.

	Since the dual of the controllable subcode $(\C_f)^c$ of a complete code
$\C$ is the observable supercode $((\C^\perp)^c)_f$ of the finite code
$\C^\perp$, we have for a finite code $\C$ a supercode chain dual to the
subcode chain of $\C^\perp$:
$$
		\C \subseteq (\C^c)_f \subseteq \cdots \subseteq \C^1 \subseteq \C^0 =
\W(\C) \subseteq \W_f.
$$
Again, by duality $(\C^c)_f$ is the ``limit" of the $j$-observable
supercodes $\C^j$ as $j \to \infty$.  

Each of these supercodes has well-defined state spaces $\Sigma_k$, which
are trivial in the case of the memoryless supercodes $\C^0 =
\W(\C)$ and $\W_f$, and well-defined state maps $\sigma_k$ and $\sigmab
= \{\sigma_k, k \in \I\}$.  By dualizing Theorem
\ref{dct}, we obtain:

\begin{theorem}[dynamic observability test]  \label{dot}
A finite group code $\C$ is observable  if and only if
$\sigmab((\C^c)_f) =
\sigmab(\C)$.
\end{theorem}

	In other words, completion preserves the dynamics of $\C$ if and only if
$\C$ is observable.

%	By the dual state space theorem and Theorem \ref{ssd}, the $j$-observable
%state space $\sigma_k(\C^j)$ acts as the character group of the
%$j$-controllable state space $\sigma_k((\C^\perp)_j)$, and similarly
%$\sigmab(\C^j) = \sigmab((\C^\perp)_j)\h$.  From the discussion in Section
%2.6
%$$
%		\{\zerob\} = \sigmab(\C^0) \subseteq \sigmab(\C^1) \subseteq \cdots
%\subseteq \sigmab((\C^c)_f) \subseteq \sigmab(\C).
%$$

	Also, by dualizing Theorem \ref{tidcc}, we obtain:
\begin{theorem} \label{tidcco}
If $\C$ is a finite time-invariant group code whose state space
$\Sigma_0(\C)$ satisfies the descending chain condition, then $(\C^c)_f$ is
strongly observable.  Consequently, $\C$ is observable if and only if $\C$
is strongly observable.	
\end{theorem}

\subsection{Observer granule decomposition}

	Now let $\C$ be a general finite, Laurent or complete $L$-observable group
code.  Then we have an ascending $j$-observable supercode chain:
$$
		\C = \C^L \subseteq \C^{L-1} \subseteq \cdots \subseteq \C^0 = \W(\C)
\subseteq \W.
$$
For indexing consistency, we denote $\W$ by $\C^{-1}$.

	The 0-observable supercode $\C^0 = \prod_k \C_{|\{k\}}$ is the output
sequence space $\W(\C)$, a memoryless sequence space of the same type as
$\W$.

	By Theorem \ref{ssd}, this chain is dual to the subcode chain of the dual
code $\C^\perp$:
$$
		(\C^\perp)_{-1} = \{\zerob\} \subseteq (\C^\perp)_0 \subseteq
(\C^\perp)_1 \subseteq \cdots \subseteq (\C^\perp)_L = \C^\perp.
$$
By quotient group duality, the quotients of the latter chain act as the 
character groups of the quotients of the former:
$$
		\left(\frac{\C^{j-1}}{\C^j}\right)^{\h} =
\frac{(\C^\perp)_j}{(\C^\perp)_{j-1}}, \quad 0 \le j \le L.
$$
Note that the output sequence space $\W(\C) = \C^0$ acts as the character
group of the parallel transition subcode $(\C^\perp)_0$, and that $\W(\C) =
\W$ if and only if $(\C^\perp)_0 = \{\zerob\}$.  Dynamically, $\C$ should
be regarded as lying between the memoryless sequence spaces $\C_0$ and
$\W(\C)$, rather than between $\{\zerob\}$ and $\W$.  Trimming the sequence
space $\W$ to $\W(\C)$ is dual to factoring out the parallel transition
subcode to yield the dynamically equivalent ``label code" $\qb(\C) \cong
\C/\C_0$ \cite{FT93}.

	Since 
$$
		\frac{(\C^\perp)_j}{(\C^\perp)_{j-1}} \cong \prod_k
\Gamma_{[k,k+j]}(\C^\perp),
$$
it follows from direct product/direct sum duality that 
$$
		\frac{\C^{j-1}}{\C^j} \cong  \prod_k
\Gamma_{[k,k+j]}(\C^\perp)\h
$$
where as usual the indicated product denotes a direct product, Laurent
product, or  direct sum according to the character of $\W$.

	Since the controller granule $\Gamma_{[k,k+j]}(\C^\perp)$ is defined as a
quotient group, it is natural to define the \textbf{observer granule}
$\Phi_{[k,k+j]}(\C)$ as the dual quotient group:
$$
		\Phi_{[k,k+j]}(\C) = \frac{(\C^{j-1})_{|[k,k+j]}}{(\C^j)_{|[k,k+j]}},
\quad  k \in \Z, 0 \le j \le L.
$$
For $0 \le j \le L$, $(\C^j)_{|[k,k+j]} = \C_{|[k,k+j]}$, so the cosets of
$(\C^j)_{|[k,k+j]}$ in $(\C^{j-1})_{|[k,k+j]}$ are represented by sequences
in $(\C^{j-1})_{|[k,k+j]}$ that are not in $\C_{|[k,k+j]}$.  

For $1 \le j \le L$, we may write $(\C^{j-1})_{|[k,k+j]}$ as
\begin{eqnarray*}
&		\{\wb \in \W \mid \wb_{|[k,k+j)} \in
 \C_{|[k,k+j)}, \wb_{|(k,k+j]} \in \C_{|(k,k+j]}\} &  \\
& = (\C_{|[k,k+j)} \times \W_{|\{k+j\}}) \cap (\W_{|\{k\}} \times
\C_{|(k,k+j]}). &
\end{eqnarray*}
In other words, $\Phi_{[k,k+j]}(\C)$ is the quotient of the subset of
sequences in $\W_{|[k,k+j]}$ that satisfy the checks of $\C$ on the
intervals $[k, k + j)$ and $(k, k + j]$ with the subset that checks on the
entire interval $[k, k + j]$.

	For $j = 1$, we have $(\C^0)_{|[k,k+1]} = \C_{|\{k\}} \times
\C_{|\{k+1\}}$, so
$$
		\Phi_{[k,k+1]}(\C) = \frac{\C_{|\{k\}} \times
\C_{|\{k+1\}}}{\C_{|[k,k+1]}}.
$$
In other words, $\Phi_{[k,k+1]}(\C)$ is the reciprocal state space of
$\C_{|[k,k+1]}$ as a length-2 code.

	For $j = 0$, we have $(\C^{-1})_{|[k,k]} = G_k$ and $(\C^0)_{|[k,k]} =
\C_{|\{k\}}$, so the zeroth-level (nondynamical) time-$k$ observer granule is
$$
		\Phi_{[k,k]}(\C) = \frac{G_k}{\C_{|\{k\}}}.
$$

Now in summary, having defined observer granules to be dual to controller
granules, we obtain our main duality and decomposition theorems:
\begin{theorem}[granule duality] \label{gd}
  If $\C$ and $\C^\perp$ are dual
group codes, then the observer granule $\Phi_{[k,k+j]}(\C)$ acts as the
character group of the controller granule $\Gamma_{[k,k+j]}(\C^\perp)$:
\end{theorem}
$$
		\Gamma_{[k,k+j]}(\C^\perp)\h = \Phi_{[k,k+j]}(\C).
$$

\emph{Proof}.  Follows from quotient group duality, projection/subcode
duality, and subcode/supercode duality.		\qed

\begin{corollary}[observer granule decomposition theorem] \label{ogdt}
 If $\C$ is a complete (resp.\ 
 Laurent, finite) group code, then  $\C^{j-1}/\C^j$ is isomorphic
to the direct product (resp.\ Laurent product, direct sum) of the observer
granules $\Phi_{[k,k+j]}(\C)$:
\end{corollary}
$$
		\frac{\C^{j-1}}{\C^j} \cong \prod_k \Phi_{[k,k+j]}(\C), \quad j \ge 0.
$$

	Thus we may decompose a sequence in $\W$ according to the $j$-observable
supercode chain into a sequence in $\C$ and representatives of
$\C^{j-1}/\C^j, 0 \le j \le L$, and then decompose each of these into a
product of observer granule representatives.

\medskip
\noindent
\textbf{Example 2} (cont.)  The bi-infinite zero-sum code $\C^\perp$ over
$\Gh$ is 1-controllable.  Its 0-controllable subcode is $\{\zerob\}$, and
its 1-controllable subcode is itself.  Its only nontrivial controller
granules are therefore the first-level granules $\Gamma_{[k,k+1]}(\C^\perp)
= (\C^\perp)_{:[k,k+1]}$, each of which is a group of length-2 sequences
of the form $(\ldots, 0, h, -h, 0, \ldots)$ with  $h \in \Gh$, and is
isomorphic to $\Gh$.  $\C^\perp$ is the code generated by all finite sums
of such sequences, and thus is isomorphic to the finite sequence space
$((\Gh)^\Z)_f$.  

	The dual bi-infinite repetition code $\C$ over $G$ is 1-observable.  Its
only nontrivial observer granules are the first-level granules
$\Phi_{[k,k+1]}(\C) = (\C_{|\{k\}} \times \C_{|\{k+1\}})/\C_{|[k,k+1]}$. 
Now $\C_{|\{k\}} \times \C_{|\{k+1\}}$ is the set of all pairs $\{(g, h),
g, h \in G\}$, while $\C_{|[k,k+1]}$ is the set of all repeated pairs
$\{(g, g), g \in G\}$, so $\Phi_{[k,k+1]}(\C) \cong G$.  Sets of coset
representatives for $\Phi_{[k,k+1]}(\C)$ are $\{(0, g), g \in G\}$ or
$\{(g, 0), g \in G\}$.  The quotient $\W/\C$ is isomorphic to the direct
product of these granules--- \ie to the complete sequence space $G^\Z$.
\qed

\medskip
\noindent
\textbf{Example 3} (cont.)  Here the rate-1/3 convolutional code $\C$ is
2-controllable and 1-observable, while its dual rate-2/3 code $\C^\perp$ is
1-controllable and 2-observable.  The zeroth-level controller granules of
$\C^\perp$ are generated by $(\ldots, 000, 002, 000, \ldots)$ and are
isomorphic to $\Z_2$;  the first-level controller granules are generated
by  generators $\hb_1 = (\ldots, 000, 100, 030, 000, \ldots)$, which has
order 4, and $\hb_2 = (\ldots, 000, 020, 001, 000, \ldots)$, which has
order 2 modulo the zeroth-level granules, so they are isomorphic to $\Z_4
\times \Z_2$.  This confirms that the state space $\Sigma_0(\C^\perp)$ is
isomorphic to $\Z_4 \times \Z_2$.

	It follows that $\C$ has nontrivial observer granules at levels 0 and 1
isomorphic to $\Z_2$ and to $\Z_4 \times \Z_2$, respectively.  Indeed,
$\C_{|\{k\}}$ is the 32-element subgroup $\Z_4 \times \Z_4 \times 2\Z_4$ of
the 64-element group $G_k = (\Z_4)^3$, so the nondynamical length-1
granules $G_k/\C_{|\{k\}}$ are isomorphic to $\Z_2$; a nonzero coset
representative is 001.  We verify that the length-2 observer granules
$\Phi_{[k,k+1]}(\C) = (\C_{|\{k\}} \times \C_{|\{k+1\}})/\C_{|[k,k+1]}$
have order 8, since $|\C_{|\{k\}} \times \C_{|\{k+1\}}| = 32 \times 32$,
whereas $|\C_{|[k,k+1]}| = 8 \times 4 \times 4$ (the number of states times
the number of input pairs).  A set of coset representatives
for $\C_{|[k,k+1]}$ in $\C_{|\{k\}} \times \C_{|\{k+1\}}$ is generated by
$(000, 010)$ and $(000, 002)$, so the length-2 observer granules are indeed
isomorphic to $\Z_4 \times \Z_2$.  

Similarly, the first-level controller granules of $\C$ are generated by
sequences such as $(\ldots, 000$, $200, 020, 000, \ldots)$ and are
isomorphic to $\Z_2$, while the second-level controller granules are
generated by $\gb = (\ldots, 000, 100, 010, 002, 000, \ldots)$, modulo the
first-level granules, and thus are also isomorphic to $\Z_2$.  It follows
that $\C^\perp$ has nontrivial observer granules for $j = 1$
and $j = 2$, all isomorphic to $\Z_2$, as the reader may verify.  Since
first-level granules are active for 1 time unit and second-level granules
are active for 2 time units, this implies a state space of size 8.	\qed

\medskip
\noindent
\textbf{Example 4} (cont.)  The dual code $\C^\perp$ over $\Z$ is again
1-controllable.  As in Example 2, the only nontrivial controller granules
are the first-level granules $\Gamma_{[k,k+1]}(\C^\perp) =
(\C^\perp)_{:[k,k+1]}$, which are generated by time shifts of $\hb =
(\ldots, 0, 1, -2, 0, \ldots)$, and are isomorphic to $\Z$.  $\C^\perp$ is
generated by all finite sums of such sequences, and thus is
isomorphic to the finite sequence space $(\Z^\Z)_f$.  

	The primal code $\C$ over $\R/\Z$ is 1-observable.  Its only nontrivial
observer granules are the first-level granules $\Phi_{[k,k+1]}(C) =
(\C_{|\{k\}} \times \C_{|\{k+1\}})/\C_{|[k,k+1]}$.  Now $\C_{|\{k\}} \times
\C_{|\{k+1\}}$ is the set of all pairs $\{(g, h), g, h \in \R/\Z\}$,
whereas $\C_{|[k,k+1]}$ is the set of all pairs $\{(g, h) \mid g \equiv 2h 
\mod \Z\}$.  Since $g$ is determined by $h$, $\C_{|[k,k+1]}
\cong \R/\Z$ and $\Phi_{[k,k+1]}(\C) \cong \R/\Z$.  Sets of coset
representatives for $\C_{|[k,k+1]}$ are $\{(g, 0), g \in \R/\Z\}$ or $\{(0,
h/2), h \in \R/\Z\}$.  $\W/\C$ is isomorphic to the direct product of these
granules--- \ie to $(\R/\Z)^\Z$.	\qed

\subsection{Observer granule decomposition of state spaces}

	We now obtain an observer decomposition of the state spaces and state code
of an $L$-observable code $\C$ by dualizing the controller decomposition of an
$L$-controllable code $\C^\perp$, again using the chain of
$j$-observable supercodes $\C^j$.

Combining the reciprocal state space theorem with
subcode/supercode duality, we obtain the following basic result:
\begin{theorem}[state space duality] 
  If $\C$ and $\C^\perp$ are group codes, then the
reciprocal state space
$\Sigma^k(\C^j)$ of the $j$-observable supercode $\C^j$ at time
$k$ acts as the character group of the state space
$\Sigma_k((\C^\perp)_j)$ of the $j$-controllable subcode $(\C^\perp)_j$
at time $k$:
$
		\Sigma^k(\C^j) = \Sigma_k((\C^\perp)_j)\h.
$
\end{theorem}

Thus the $j$-observable state space $\sigma_k(\C^j)$ is isomorphic to the
character group of the $j$-controllable state space
$\sigma_k((\C^\perp)_j)$, and $\sigmab(\C^j) \cong
\sigmab((\C^\perp)_j)\h$.

For the state spaces of the ascending chain of the $j$-controllable
subcodes $(\C^\perp)_j$ of the $L$-controllable dual code $\C^\perp$, we
have chains of inclusion maps as follows:
\begin{eqnarray*}
		\{0\} & \to & \sigma_k((\C^\perp)_1) \to
\cdots
\to \sigma_k((\C^\perp)_L) = \sigma_k(\C^\perp); \\
		\{\zerob\} & \to & \sigmab((\C^\perp)_1)
\to \cdots
\to \sigmab((\C^\perp)_L) = \sigmab(\C^\perp).
\end{eqnarray*}
This shows that $\sigma_k(\C^\perp)$ and $\sigmab(\C^\perp)$ may be
regarded as being composed of the quotient groups
$\sigma_k((\C^\perp)_j)/\sigma_k((\C^\perp)_{j-1})$ and 
$\sigmab((\C^\perp)_j)/\sigmab((\C^\perp)_{j-1})$, respectively.

 As discussed in Section II-F, although $\sigma_k((\C^\perp)_{j-1})$ is a
subgroup of $\sigma_k((\C^\perp)_j)$, the dual state space (character
group) $\sigma_k(\C^{j-1})$ is not in general a subgroup of
$\sigma_k(\C^j)$.  Nevertheless, there still exists a decomposition into
dual quotient groups.  The adjoint chains of the above inclusion map
chains are chains of natural maps, as follows:
\begin{eqnarray*}
\sigma_k(\C^L) = \Sigma_k(\C) \to & \cdots & \to
\sigma_k(\C^1)  \to 		\{0\}; \\
\sigmab(\C^L) = \sigmab(\C)  \to & \cdots & \to 
\sigmab(\C^1)
\to 		\{\zerob\}.
\end{eqnarray*}
Moreover, $\Sigma_k(\C)$ and $\sigmab(\C)$ may be
regarded as being composed of the respective kernels of these maps, 
$(\sigma_k((\C^\perp)_j)/\sigma_k((\C^\perp)_{j-1}))\h$ and 
$(\sigmab((\C^\perp)_j)/\sigmab((\C^\perp)_{j-1}))\h$.

Dualizing our granule decompositions of these quotient groups and
using direct product/direct sum duality, we have
\begin{eqnarray*}
		\left(\frac{\sigma_k((\C^\perp)_j)}{\sigma_k((\C^\perp)_{j-1})}\right)\h
& \cong & \prod_{i\in[k-j,k)} \Phi_{[i,i+j]}(\C), \quad  1 \le j \le L; \\
		\left(\frac{\sigmab((\C^\perp)_j)}{\sigmab((\C^\perp)_{j-1})}\right)\h &
\cong & \prod_{k\in\Z} \Phi_{[k,k+j]}(\C), \quad  1 \le j \le L.
\end{eqnarray*}
As we have already seen, the latter is isomorphic to
$C^{j-1}/C^j$.

	In summary:
\begin{theorem}[dual state granule theorem]  \label{dsgt}  
Let $\C$ be an $L$-observable group code, let $\Sigma_k(\C)$ be its state
space at time $k$, and let $\sigma_k(\C^j)$ be the state space at time
$k$ of its $j$-observable supercode $\C^j$.  Then there exists a
chain of natural maps
$$
		\sigma_k(\C^L) = \Sigma_k(\C) \to \cdots \to
\sigma_k(\C^1)  \to 		\{0\} = \sigma_k(\C^0),
$$
whose kernels are isomorphic to direct products of the $j$ observer
granules $\Phi_{[i,i+j]}(\C), k - j \le i < k$, for  $1 \le j \le L$. 
Consequently there is a one-to-one correspondence between the state space
$\Sigma_k(\C)$ and the Cartesian product of the observer granules
$\Phi_{[i,i+j]}(\C), k - j \le i < k, 1 \le j \le L$.
\end{theorem}

\subsection{Dual first-output and last-output groups}

What is the dual to the $j$th first-output group $F_{j,k}(\C) =
(\C_{:[k,k+j]})_{|\{k\}}$ (which can also be thought of as the input
group at level $j$ at time $k$)?  By projection/subcode duality, it is
the parallel transition subgroup at time $k$ of $(\C^\perp)_{|[k,k+j]}$. 
In other words, $F_{j,k}(\C)^\perp$ is the set
$((\C^\perp)_{:\I-(k,k+j]})_{|\{k\}}$ of time-$k$ symbols in all sequences
in $\C^\perp$ whose components are all zero during $(k, k + j]$.  

	We therefore define the \textbf{\emph{j}th-level dual last-output group}
of $\C$ at time $k$ as
$$
		L^{j,k}(\C) = F_{j,k}(\C)^\perp = (\C_{:\I-(k,k+j]})_{|\{k\}}.
$$
In other words, $L^{j,k}(\C)$ is the set of time-$k$ symbols that can be
followed by a sequence of $j$ consecutive zeroes, or equivalently that can
precede the zero state in $\sigma_k(\C^j)$.

	Note that $L^{0,k}(\C) = \C_{|\{k\}}$.  Moreover, if $\C$ is
$L$-observable, then $L^{L,k}(\C) = L_k(\C)$, because by the second $[m,
n)$-observability test $\C_{:\I-(k,k+L]} = \C_{:(k+1)^-} \times
\C_{:(k+L+1)^+}$.   

	Thus the \emph{time-k dual last-output chain} of $\C$,
$$
		L_k(\C) = L^{L,k}(\C) \subseteq L^{L-1,k}(\C) \subseteq \cdots \subseteq
L^{0,k}(\C) = \C_{|\{k\}},
$$
is dual to the time-$k$ first-output chain of $\C^\perp$.  By quotient
group duality, the quotients of this chain are the character groups of the
quotients of the dual chain.  These quotients are isomorphic to the
controller granules of $\C^\perp$, whose character groups act as the
observer granules of $\C$:
$$
		\frac{L^{j-1,k}(\C)}{L^{j,k}(\C)} \cong \Phi_{[k,k+j]}(\C), \quad 0 \le 
j \le L.
$$
	Similarly, we define the \textbf{\emph{j}th-level dual first-output group}
of $\C$ at time $k$ as
$$
		F^{j,k}(\C) = L_{j,k}(\C)^\perp = (\C_{:I-[k-j,k)})_{|\{k\}}.
$$
In other words, $F^{j,k}(\C)$ is the set of time-$k$ symbols that can
follow a sequence of $j$ consecutive zeroes, or equivalently that can
follow the zero state in $\sigma_k(\C^j)$.

	Again we have $F^{0,k}(\C) = \C_{|\{k\}}$, and if $\C$ is $L$-observable,
then $F^{L,k}(\C) = F_k(\C)$, since $\C_{:\I-[k-L,k)} = \C_{:(k-L)^-}
\times \C_{:k^+}$.   The \emph{time-k dual first-output chain} of $\C$,
$$
		F_k(\C) = F^{L,k}(\C) \subseteq F^{L-1,k}(\C) \subseteq \cdots \subseteq
F^{0,k}(\C) = \C_{|\{k\}},
$$
is dual to the time-$k$ last-output chain of $\C^\perp$, and the quotients
are isomorphic to observer granules:
$$
	\frac{F^{j-1,k}(\C)}{F^{j,k}(\C)} \cong \Phi_{[k-j,k]}(\C), \quad 0 \le j
\le L.
$$
	The quotient $G_k/F_k(\C)$ will be called the \textbf{syndrome group}
$S_k(\C)$ of $\C$ at time $k$, and the quotient $F^{j-1,k}(\C)/F^{j,k}(\C)$
will be called the \textbf{\emph{j}th-level syndrome group} $S_{j,k}(\C)$
at time $k, 0 \le j \le L$.  The syndrome group at time $k$ may be
decomposed according to the dual first-output chain at time $k$ into an
element of the first-output group $F_k(\C)$ and representatives of the
quotients $F^{k,j-1}(\C)/F^{k,j}(\C)$, which are isomorphic to the observer
granules that ``end" at time $k$.  The syndrome group $S_k(\C)$ acts as the
character group of $L_k(\C^\perp)$.

In summary:
\begin{theorem}[first-output/last-output duality] \label{folod}
  If $\C$ and $\C^\perp$ are dual group codes and $\C$ is $L$-observable,
then the dual first-output (resp.\ last-output) chain of $\C$ at time $k$
is dual to the last-output (resp.\ first-output) chain of $\C^\perp$ at
time $k$.  In particular, 
\begin{eqnarray*}
		F^{0,k}(\C) & = & L^{0,k}(\C) = \C_{|\{k\}} =
((\C^\perp)_{|:\{k\}})^\perp; \\
		F^{L,k}(\C) & = & F_k(\C) = (L_k(\C^\perp))^\perp;   \\
		L^{L,k}(\C) & = & L_k(\C) = (F_k(\C^\perp))^\perp.
\end{eqnarray*}
The quotients of the dual chains of $\C$  act as the character groups of 
the corresponding quotients of the  primal chains of $\C^\perp$, and are
isomorphic to observer granules as follows: 
\begin{eqnarray*}
\frac{L^{j-1,k}(\C)}{L^{j,k}(\C)} & \cong & \Phi_{[k,k+j]}(\C), \quad 0 
\le j \le L; \\
\frac{F^{j-1,k}(\C)}{F^{j,k}(\C)} & \cong & \Phi_{[k-j,k]}(\C), \quad 0
\le j \le L.
\end{eqnarray*}
\end{theorem}

\section{Minimal observer-form encoders and syndrome-formers}

	One original objective of this paper was to develop a minimal 
syndrome-former construction based on observer granules for a strongly
observable code $\C$ that would be dual to the minimal controller-form
encoder construction of \cite{FT93} for strongly controllable codes, with
memory equal to the observer memory $L$.  Such a syndrome-former is easily
found in many cases:  for Examples 2-4 of this paper, for codes and
systems over fields \cite{F73, K80}, and we dare say for most codes that
the reader is likely to imagine.  However, finding a general minimal
syndrome-former construction that has all of the properties that one might
desire turns out to be quite difficult.  

	This problem has now been solved satisfactorily by Fagnani and
Zampieri \cite{FZ99}. Interestingly, their construction works equally well
for nonabelian codes and, although it is based on the observer granule
decomposition of the previous section, it does not make any use of
duality.  

In this section we construct minimal syndrome-formers and observer-form
encoders for Examples 2-4 of this paper, and also for the main example of
\cite{FZ99}.  Our approach uses the observability granules of $\C$
directly, and seems simpler than the general methods of
\cite{FZ99} for these simple codes.

	A \textbf{minimal syndrome-former} for $\C$ is a dynamical map from $\W$
to the syndrome sequence space $\prod_k S_k(\C)$ that has at least the
following properties:

 (a) The kernel of the map is the code $\C$;

 (b) If $\C$ is $L$-observable, then the map has memory $L$;

 (c) The time-$k$ state space corresponds in some way to the
active observer granules at time $k$.

	We also desire that the inverse images of the syndrome sequences 
form a disjoint partition of $\W$ in which each inverse image is in some
sense isomorphic to $\C$ (see \cite{FZ99}). However, we ignore here the
behavior of the syndrome-former for input sequences not in $\C$. 
Nevertheless, in all our examples, our syndrome-former construction turns out to
have this property.

An encoder for $\C$ is a dynamical one-to-one map from the memoryless input
sequence space $\prod_k F_k(\C)$ to $\C$.  A \textbf{minimal observer-form
encoder} for $\C$ is an encoder for $\C$ whose state space at time $k$
corresponds in some sense to the active observer granules at time $k$.

	Our constructions will be based on the construction of a minimal state 
observer for $\C$.  If $\C$ is $L$-observable, then a \textbf{state
observer} for $\C$ with memory $L$ is a system that maps $\cb_{|[k-L,k)} 
\in \C_{|[k-L,k)}$ to the state $\sigma_k(\cb)$ of $\cb \in \C$ at time $k$
for each time $k$.  In other words, the state observer dynamically
implements the state map $\sigmab:  \C \to \sigmab(\C)$ using a ``sliding
window" of width $L$.  

In view of the dual state granule theorem, the state $\sigma_k(\cb)$ is
determined by the values of the observer granules $\Phi_{[i-j,i]}(\C), k 
\le i < k + j, 1 \le j \le L$; namely, the observer granules that are
``active" at time $k$.  A state observer is minimal if its state space at
time $k$ corresponds in some sense to the active observer granules at time
$k$.

Our approach to realizing such a minimal state observer is as
follows.  If $\cb \in \C$, then \emph{a} \emph{fortiori} $\cb \in \C^j$ for
all
$j$-observable supercodes $\C^j, 0 \le j \le L$.  Given $\cb \in \C^{j-1}$,
the $j$th-level observer granule $\Phi_{[i-j,i]}(\C)$ may be computed by
determining the character table column (``check")
$$
		\inner{\Gamma_{[i-j,i]}(\C^\perp)}{\cb} = \{\inner{\xb}{\cb} \mid \xb \in
\Gamma_{[i-j,i]}(\C^\perp)\}
$$
since $\Gamma_{[i-j,i]}(\C^\perp)$ acts as the character group of
$\Phi_{[i-j,i]}(\C)$.  This requires the calculation of the pairing
$\inner{\xb}{\cb}$ only for a set of generators of
$\Gamma_{[i-j,i]}(\C^\perp)$.  

Since the pairing $\inner{\xb}{\cb}$ is a componentwise sum over the
interval $[i - j, i]$, and since the character table column
$\inner{\Gamma_{[i-j,i]}(\C^\perp)}{\cb}$ specifies an element of
$\Phi_{[i-j,i]}(\C)$, implementation of such a pairing requires only a
memory element storing an element of $\Phi_{[i-j,i]}(\C)$ that is active
during the interval $(i - j, i]$.  At each time during this interval, the
memory element stores a ``partial sum" in $\Phi_{[i-j,i]}(\C)$.  The
values of all of the partial sums corresponding to all active observer
granules is then the observer state at time $k$.

	Given a minimal state observer for $\C$, a minimal observer-form encoder
for $\C$ may then be realized as follows.  Assume that at time $k$ the
encoder has generated the past $\cb_{|k^-}$ of a code sequence $\cb \in
\C$.  A minimal state observer that tracks this past will indicate the
current state $\sigma_k(\cb)$ by the stored values of its currently active
observer granules.  The next output is then determined by an ``input" in
the first-output group $F_k(\C)$ and the current state $\sigma_k(\cb)$.  

	Specifically, the next output $c_k \in G_k$ must be chosen so that all
observer granules $\Phi_{[k-j,k]}(\C),$ $0 \le j \le L$, that end at time
$k$ take on the value 0, since $\wb \in \C$ if and only if the values of
all quotients in the chain $$\C = \C^L \subseteq \C^{L-1} \subseteq \cdots
\subseteq C^0 \subseteq \W$$ are equal to zero.  In view of the dual
first-output chain $$F_k(\C) = F^{L,k}(\C) \subseteq F^{L-1,k}(\C) \subseteq
\cdots \subseteq F^{0,k}(\C) = \C_{|\{k\}},$$ given
representatives of each quotient group $F^{j-1,k}(\C)/F^{j,k}(\C) \cong
\Phi_{[k-j,k]}(\C)$, this can be done by subtracting representatives from
an arbitrary ``free" input in $\C_{|\{k\}}$ according to the current partial sums
of the ending granules $\Phi_{[k-j,k]}(\C)$, leaving a residual free input
in $F_k(\C)$.  This produces a next output $c_k$ such that $\cb_{|(k+1)^-}
\in \C_{|(k+1)^-}$, which determines the next state, and so forth.

Similarly, a minimal syndrome-former can simply check whether the next
output is in the appropriate set determined by  $\sigma_k(\cb)$. If so, it
continues.  If not, then it needs to make
some ``correction" to reduce it to this set, so that the state observer
can continue.

	We now give some applications of this approach.

\smallskip
\noindent
\textbf{Example 2} (cont.)  For the bi-infinite repetition code $\C$ over
$G$, the state space at any time $k$ is $G$, and consists of a single
first-level observer granule $\Phi_{[k-1,k]}(\C) \cong G$.  The
first-output (input) group of $\C$ is trivial, $F_k(\C) = \{0\}$, and the
syndrome group $S_k(\C)$ is $G$.   

	The check corresponding to $\Phi_{[k-1,k]}(\C)$ is orthogonality to the
dual first-level controller granule $\Gamma_{[k-1,k]}(\C^\perp) =
\{(\ldots, 0, h, -h, 0, \ldots) \mid h \in \Gh\}$.  For $\wb \in 
G^\Z$, we have
$$
		\inner{(\ldots, 0, h, -h, 0, \ldots)}{\wb} = \{\inner{h}{w_{k-1} - w_k}
\mid h \in \Gh\},
$$
which is equal to zero for all $h \in \Gh$ if and only if $w_k = w_{k-1}$. 
A minimal state observer for $\C$ therefore needs only to store the partial
sum $w_{k-1} \in G$ at time $k$, so it has memory 1.

	A minimal observer-form encoder for $\C$ stores the previous output
$c_{k-1} \in G$ and enforces the constraint $c_k = c_{k-1}$, as shown in
Figure 8.1(a); \ie there is no nontrivial input, and the state space is
$G$.  The initial condition of the memory element is unspecified, and its
effect persists indefinitely. 

\begin{figure}[t]
\setlength{\unitlength}{5pt}
\begin{center}
\begin{picture}(48,11)(-3,-2)
\put(0,0){\framebox(6,6){$c_{k-1}$}}
\put(6,3){\vector(1,0){9}}
\put(10,4){$c_k$}
\put(9,3){\line(0,1){5}}
\put(9,8){\line(-1,0){12}}
\put(-3,8){\line(0,-1){5}}
\put(-3,3){\vector(1,0){3}}
\put(7,-2){(a)}
\put(27,0){\framebox(6,6){$w_{k-1}$}}
\put(33,3){\vector(1,0){4.5}}
\put(36,3.5){$-$}
\put(39,3){\circle{3}}
\put(40.5,3){\vector(1,0){4.5}}
\put(42,4){$s_k$}
\put(18,8){$w_k$}
\put(39,8){\vector(0,-1){3.5}}
\put(39.5,4.5){$+$}
\put(39,8){\line(-1,0){18}}
\put(24,8){\line(0,-1){5}}
\put(24,3){\vector(1,0){3}}
\put(34,-2){(b)}
\end{picture}

Figure 8.1.  Minimal (a) observer-form encoder and \\ (b) syndrome-former
for  bi-infinite repetition code $\C$.
\end{center}
\end{figure}

	A minimal syndrome-former for $\C$ may simply be constructed by
implementing this check dynamically, as shown in Figure 8.1(b).
(Conversely, the minimal encoder of Figure 8.1(a) may be obtained by
forcing $s_k = 0$ in Figure 8.1(b).)  The value of each check is the
syndrome $s_k = w_k - w_{k-1} \in G$.  The syndrome sequence is $\zerob$ if
and only if $\wb \in \C$, and in this case the syndrome-former acts as a
state observer for $\C$.  In this example each coset $\C + \sb$ of $\C$ in
$\W$ maps to a unique syndrome sequence $\sb
\in G^\Z$.  \qed

\smallskip
\noindent
\textbf{Example 3} (cont.)  We now consider our 1-observable rate-1/3
convolutional code $\C$ over $\Z_4$.  

For an element $g \in \Z_4$, it will often be useful to consider a two-bit
representation $(g_1, g_0) \in (\Z_2)^2$ such that $g = 2g_1 + g_0$;  \ie
$g_1$ is the ``high-order bit" and $g_0$ is the ``low-order bit."

We recall that $\C$ has nontrivial observer granules at levels 0 and 1
isomorphic to $\Z_2$ and $\Z_4 \times \Z_2$, respectively.  The
zeroth-level observer granule corresponds to the constraint that $c_{k,3}
\in 2\Z_4$--- \ie the low-order bit $c_{0,k,3}$ equals 0. 
Thus $\C_{|\{0\}} = \Z_4 \times \Z_4 \times 2\Z_4$.  The first-level
granule corresponds to the constraint of orthogonality with the shifts of
the generators $\hb_1 = (\ldots, 000, 100, 030, 000, \ldots)$ and $\hb_2 =
(\ldots, 000, 020, 001, 000, \ldots)$ of $\C^\perp$.  The inner product 
with $\hb_1$ yields the constraint $c_{k-1}\cdot(100) + c_k\cdot(030) = 0$,
or $c_{k,2} = c_{k-1,1}$.  The inner product with $\hb_2$ yields 
$c_{k-1}\cdot(020) + c_k\cdot(001) = 0$, or $c_{1,k,3} = c_{0,k-1,2}$.

In short, $c_{0,k,3} = 0$, $c_{1,k,3} = c_{0,k-1,2}$, and $c_{k,2} =
c_{k-1,1}$. Thus we obtain a minimal observer-form encoder with ``free
input" $c_{k,1} \in \Z_4$, as shown in Figure 8.2(a).  Note that this
encoder is feedbackfree, and is also a minimal controller-form encoder
with controller memory 2.

\begin{figure*}[t]
\setlength{\unitlength}{5pt}
\begin{center}
\begin{picture}(68,20)(0,-1)
\put(0,19){\vector(1,0){30}}
\put(0,16){\vector(1,0){30}}
\put(25,20){$c_{1,k,1}$}
\put(25,17){$c_{0,k,1}$}
\put(1,19){\line(0,-1){5.5}}
\put(2,16){\line(0,-1){5.5}}
\put(1,13.5){\vector(1,0){3}}
\put(2,10.5){\vector(1,0){2}}
\put(4,12){\framebox(8,3){$c_{1,k-1,1}$}}
\put(4,9){\framebox(8,3){$c_{0,k-1,1}$}}
\put(12,13.5){\vector(1,0){18}}
\put(12,10.5){\vector(1,0){18}}
\put(25,14.5){$c_{1,k,2}$}
\put(25,11.5){$c_{0,k,2}$}
\put(14,10.5){\line(0,-1){6}}
\put(14,4.5){\vector(1,0){2}}
\put(16,3){\framebox(8,3){$c_{0,k-1,2}$}}
\put(24,4.5){\vector(1,0){6}}
\put(25,5.5){$c_{1,k,3}$}
\put(24,1.5){\vector(1,0){6}}
\put(25,2.5){$c_{0,k,3}$}
\put(22,1){0}
\put(14,-1){(a)}
\put(40,19){\vector(1,0){6}}
\put(40,16){\vector(1,0){6}}
\put(40,20){$w_{1,k,1}$}
\put(40,17){$w_{0,k,1}$}
\put(46,17.5){\framebox(8,3){$w_{1,k-1,1}$}}
\put(46,14.5){\framebox(8,3){$w_{0,k-1,1}$}}
\put(54,19){\line(1,0){6}}
\put(54,16){\line(1,0){3}}
\put(60,19){\vector(0,-1){4}}
\put(57,16){\vector(0,-1){4}}
\put(40,13.5){\vector(1,0){18.5}}
\put(40,10.5){\vector(1,0){15.5}}
\put(40,14.5){$w_{1,k,2}$}
\put(40,11.5){$w_{0,k,2}$}
\put(60,13.5){\circle{3}}
\put(58.5,13.5){\line(1,0){3}}
\put(60,12){\line(0,1){3}}
\put(57,10.5){\circle{3}}
\put(55.5,10.5){\line(1,0){3}}
\put(57,9){\line(0,1){3}}
\put(61.5,13.5){\vector(1,0){7}}
\put(58.5,10.5){\vector(1,0){10}}
\put(63,14.5){$s_{1,k,2}$}
\put(63,11.5){$s_{0,k,2}$}
\put(43,7.5){\line(0,1){3}}
\put(43,7.5){\vector(1,0){3}}
\put(46,6){\framebox(8,3){$w_{0,k-1,2}$}}
\put(40,4.5){\vector(1,0){15.5}}
\put(57,4.5){\circle{3}}
\put(55.5,4.5){\line(1,0){3}}
\put(57,3){\line(0,1){3}}
\put(57,7.5){\vector(0,-1){1.5}}
\put(54,7.5){\line(1,0){3}}
\put(58.5,4.5){\vector(1,0){10}}
\put(40,5.5){$w_{1,k,3}$}
\put(63,5.5){$s_{1,k,3}$}
\put(40,1.5){\vector(1,0){28.5}}
\put(40,2.5){$w_{0,k,3}$}
\put(63,2.5){$s_{0,k,3}$}
\put(53,-1){(b)}
\end{picture}

Figure 8.2  Minimal (a) observer-form encoder and (b) syndrome-former 
for  rate-1/3 convolutional code over $\Z_4$.
\end{center}
\end{figure*}

Similarly, a minimal syndrome-former for $\C$ has two levels.  The zeroth
(nondynamical) level checks whether $w_{k,3} \in 2\Z_4$, or equivalently
whether $w_{0,k,3} = 0$, and, if not, ``corrects" to meet this constraint. 
This can be done simply by regarding $w_{0,k,3}$ as the zeroth-level
syndrome, and ignoring it thereafter.  The next (first) level checks the
constraints $c_{k,2} = c_{k-1,1}$ and $c_{1,k,3} = c_{0,k-1,2}$ by forming
the syndromes $s_{k,2} = w_{k,2} - w_{k-1,1} \in \Z_4$ and $s_{1,k,3} =
w_{1,k,3} - w_{0,k-1,2} \in \Z_2$, as shown in Figure 8.2(b).  For
simplicity, we merely compare the two bits of $w_{k,2}$ and $w_{k-1,1}$; 
this makes the syndrome-former linear over $\Z_2$.

The syndrome-former is evidently feedbackfree and has memory 1.
Its output sequence $\sb$ is $\zerob$ if and only if
$\wb \in \C$, and in this case the syndrome-former acts as a state observer
for $\C$.

For the dual 2-observable rate-2/3 code $\C^\perp$, recall that
$\C^\perp$ has nontrivial observer granules at levels 1 and 2 isomorphic
to $\Z_2$ and $\Z_2$, respectively.  The first-level observer
granules correspond to the constraint of orthogonality with the shifts of 
$2\gb = (\ldots, 000, 200, 020, 000, \ldots)$, which yields the
constraint $2c_{k,2} = 2c_{k-1,1}$, or $c_{0,k,2} = c_{0,k-1,1}$.  The
second-level observer granules correspond to
orthogonality with the shifts of 
$\gb = (\ldots, 000, 100, 010, 002, 000, \ldots)$, which yields
$2c_{k,3}
=$

\noindent
$c_{k-1,2} + c_{k-2,1}$.  If
$c_{0,k-1,2} = c_{0,k-2,1}$, which is guaranteed by the first-level
constraint, then this is equivalent to
$c_{0,k,3} = c_{1,k-1,2} + c_{1,k-2,1} + c_{0,k-2,1}$, where $c_{0,k-2,1}$
is a ``carry bit."

Thus we obtain a minimal observer-form encoder with ``free" binary 
inputs $c_{1,k,1}, c_{0,k,1}, c_{1,k,2},$ $c_{1,k,3}$, shown in
Figure 8.3(a).  The encoder is feedbackfree with memory 2, and is
$\Z_2$-linear.

\begin{figure*}[t]
\setlength{\unitlength}{5pt}
\begin{center}
\begin{picture}(83,19)(-7,1)
\put(-7,19){\vector(1,0){37}}
\put(-7,16){\vector(1,0){37}}
\put(25,20){$c_{1,k,1}$}
\put(25,17){$c_{0,k,1}$}
\put(-6,19){\line(0,-1){14.5}}
\put(-5,16){\line(0,-1){5.5}}
\put(-6,4.5){\vector(1,0){2}}
\put(-5,10.5){\vector(1,0){2}}
\put(-4,3){\framebox(8,3){$c_{1,k-1,1}$}}
\put(4,4.5){\vector(1,0){2}}
\put(-3,9){\framebox(8,3){$c_{0,k-1,1}$}}
\put(-7,13.5){\vector(1,0){37}}
\put(5,10.5){\vector(1,0){25}}
\put(25,14.5){$c_{1,k,2}$}
\put(25,11.5){$c_{0,k,2}$}
\put(7.5,13.5){\vector(0,-1){7.5}}
\put(12.5,10.5){\vector(0,-1){4.5}}
\put(7.5,4.5){\circle{3}}
\put(6,4.5){\line(1,0){3}}
\put(7.5,3){\line(0,1){3}}
\put(9,4.5){\vector(1,0){2}}
\put(12.5,4.5){\circle{3}}
\put(11,4.5){\line(1,0){3}}
\put(12.5,3){\line(0,1){3}}
\put(14,4.5){\vector(1,0){2}}
\put(16,3){\framebox(8,3){$c_{0,k-1,3}$}}
\put(-7,7.5){\vector(1,0){37}}
\put(25,8.5){$c_{1,k,3}$}
\put(24,4.5){\vector(1,0){6}}
\put(25,5.5){$c_{0,k,3}$}
\put(15,1){(a)}
\put(40,18.5){\vector(1,0){6}}
\put(40,15.5){\vector(1,0){6}}
\put(40,19.5){$w_{1,k,1}$}
\put(40,16.5){$w_{0,k,1}$}
\put(46,17){\framebox(8,3){$w_{1,k-1,1}$}}
\put(46,14){\framebox(8,3){$w_{0,k-1,1}$}}
\put(54,18.5){\line(1,0){6}}
\put(54,15.5){\vector(1,0){4.5}}
\put(60,18.5){\vector(0,-1){1.5}}
\put(60,15.5){\circle{3}}
\put(58.5,15.5){\line(1,0){3}}
\put(60,14){\line(0,1){3}}
\put(60,14){\vector(0,-1){2}}
\put(57,15.5){\vector(0,-1){6.5}}
\put(40,10.5){\vector(1,0){18.5}}
\put(40,7.5){\vector(1,0){15.5}}
\put(40,11.5){$w_{1,k,2}$}
\put(40,8.5){$w_{0,k,2}$}
\put(60,10.5){\circle{3}}
\put(58.5,10.5){\line(1,0){3}}
\put(60,9){\line(0,1){3}}
\put(57,7.5){\circle{3}}
\put(55.5,7.5){\line(1,0){3}}
\put(57,6){\line(0,1){3}}
\put(58.5,7.5){\vector(1,0){20}}
\put(73,8.5){$s_{0,k,2}$}
\put(61.5,10.5){\vector(1,0){3}}
\put(64.5,9){\framebox(8,3){$t_{0,k-1,3}$}}
\put(68,9){\vector(0,-1){3}}
\put(68,4.5){\circle{3}}
\put(66.5,4.5){\line(1,0){3}}
\put(68,3){\line(0,1){3}}
\put(40,4.5){\vector(1,0){26.5}}
\put(69.5,4.5){\vector(1,0){9}}
\put(40,5.5){$w_{0,k,3}$}
\put(73,5.5){$s_{0,k,3}$}
\put(53,1){(b)}
\end{picture}

Figure 8.3  Minimal (a) observer-form encoder and (b) syndrome-former 
for  rate-2/3 convolutional code over $\Z_4$.
\end{center}
\end{figure*}

	A minimal syndrome-former for $\C^\perp$ again has two levels.  The first
level checks whether $w_{0,k,2} = w_{0,k-1,1}$ by forming the syndrome
$s_{0,k,2} = w_{0,k,2} + w_{0,k-1,1} \in \Z_2$.  The second level checks
whether
$w_{0,k,3} = w_{1,k-1,2} + w_{1,k-2,1}$ by forming the syndrome $w_{0,k,3}
= w_{0,k,3} + w_{1,k-1,2} + w_{1,k-2,1} + w_{0,k-2,1} = w_{0,k,3} +
t_{0,k-1,3} \in
\Z_2$, as shown in Figure 8.3(b).  The syndrome-former is feedbackfree with
memory 2, and is $\Z_2$-linear.   \qed

\medskip
\noindent
\textbf{Example 4} (cont.)   For Loeliger's code $\C$, the state space at
any time $k$ is $\R/\Z$, and it consists of a single first-level observer
granule $\Phi_{[k-1,k]}(\C) \cong \R/\Z$.  The first-output (input) group
of $\C$ is binary, $F_k(C) = \{0, \half\} = (\half\Z)/\Z$, and the
syndrome group $S_k(\C)$ is $(\R/\Z)/F_k(\C) \cong \R/\Z$. A set of
representatives for $(\R/\Z)/F_k(\C)$ is the interval $[0, 1/2)$. 

	The check corresponding to $\Phi_{[k-1,k]}(\C)$ is orthogonality to the
dual first-level controller granule $\Gamma_{[k-1,k]}(\C^\perp)$, which is
generated by $\hb = (\ldots, 0, 1, -2, 0, \ldots)$.  For $\wb \in \W =
(\R/\Z)^\Z$, we have
%the pairing $\inner{\hb}{\wb}$ is given by
$$
		\inner{\hb}{\wb} = w_{k-1} - 2w_k \in \R/\Z. 
$$

A minimal state observer for $\C$ therefore needs only to store the partial
sum $w_{k-1} \in \R/\Z$ of this check at time $k$, and thus has memory 1.

	A minimal observer-form encoder for $\C$ stores the previous output
$c_{k-1} \in \R/\Z$ and enforces the constraint $2c_k = c_{k-1} \mod \Z$. 
The set of $c_k \in \R/\Z$ that satisfy this constraint is the set $c_k =
\{u_k + \half c_{k-1} \mid u_k \in \{0, \half\}\}$, so the encoder has a
binary input $u_k \in F_k(\C)$ and a state space of $\R/\Z$, as shown in
Figure 8.4(a).  The initial condition of the memory element decays to zero
(but is still visible forever in $c_k$). 

\begin{figure*}[t]
\setlength{\unitlength}{5pt}
\begin{center}
\begin{picture}(80,12)(-9,-2)
\put(2,9){\framebox(6,2){$\div 2$}}
\put(2,0){\framebox(6,6){$c_{k-1}$}}
\put(8,3){\line(1,0){4}}
\put(-9,10){$u_k$}
\put(-9,7.5){$\{0, \half\}$}
\put(12,10){\line(0,-1){7}}
\put(12,10){\vector(-1,0){4}}
\put(2,10){\vector(-1,0){3.5}}
\put(-6.5,10){\vector(1,0){2}}
\put(-6,10.5){$+$}
\put(-2,10.5){$+$}
\put(-3,10){\circle{3}}
\put(-3,8.5){\vector(0,-1){8.5}}
\put(-5.5,3){$c_k$}
\put(-3,3){\vector(1,0){5}}
\put(3,-2){(a)}
\put(30,9){\framebox(10,2){$\times 2 \mod \Z$}}
\put(30,0){\framebox(6,6){$w_{k-1}$}}
\put(36,3){\vector(1,0){4.5}}
\put(39,3.5){$-$}
\put(42,3){\circle{3}}
\put(43.5,3){\vector(1,0){4.5}}
\put(45,4){$s_k$}
\put(21,10){$w_k$}
\put(42,10){\vector(0,-1){5.5}}
\put(42.5,4.5){$+$}
\put(42,10){\line(-1,0){2}}
\put(24,10){\vector(1,0){6}}
\put(27,10){\line(0,-1){7}}
\put(27,3){\vector(1,0){3}}
\put(33,-2){(b)}
\put(60,9){\framebox(10,2){$\times 2 \mod \Z$}}
\put(60,0){\framebox(6,6){$\tilde{c}_{k-1}$}}
\put(66,3){\line(1,0){6}}
\put(54.5,3){$\tilde{c}_k$}
\put(72,10){\line(0,-1){7}}
\put(72,10){\vector(-1,0){2}}
\put(57,10){\line(1,0){3}}
\put(57,10){\vector(0,-1){10}}
\put(57,3){\vector(1,0){3}}
\put(63,-2){(c)}
\end{picture}

Figure 8.4.  Minimal (a) observer-form encoder and (b) syndrome-former for
Loeliger's code $\C$, and \\ (c) minimal observer-form encoder for the chaotic
time-reversed code $\tilde{\C}$.
\end{center}
\end{figure*}

A minimal syndrome-former for $\C$ may be constructed by implementing this
check dynamically, as shown in Figure 8.4(b).
(Conversely, the minimal encoder of Figure 8.4(a) may be obtained by
forcing $s_k = 0$ in Figure 8.4(b).  Note that there are two
values of $w_k$ that satisfy $2w_k = w_{k-1} \mod \Z$, namely $w_k =
\{u_k + \half w_{k-1} \mid u_k \in \{0, \half\}\}$.)   The value of each
check is the syndrome
$s_k = 2w_k - w_{k-1} \in \R/\Z$.  The syndrome sequence is
$\zerob$ if and only if $\wb \in \C$, and in this case the syndrome-former
acts as a state observer for $\C$.  In this example also each coset $\C +
\sb$ of $\C$ in $\W$ maps to a unique syndrome sequence $\sb \in
(\R/\Z)^\Z$.

Finally, consider the chaotic time-reversed code $\tilde{\C}$.  The state
space at any time $k$ is $\R/\Z$, and consists of a single first-level
observer granule $\Phi_{[k-1,k]}(\C) \cong \R/\Z$.  The first-output
(input) group of $\tilde{\C}$ is trivial, $F_k(\tilde{\C}) = L_k(\C) =
\{0\}$, and the syndrome group $S_k(\C)$ is $\R/\Z$.  A minimal
observer-form encoder for $\tilde{\C}$ stores the previous output
$\tilde{c}_{k-1} \in \R/\Z$ and enforces the constraint $\tilde{c}_k =
2\tilde{c}_{k-1} \mod \Z$, which completely determines $\tilde{c}_k$, as
shown in Figure 8.4(c);  \ie there is no nontrivial input, and the state
space is $\R/\Z$.  Since the map $\tilde{c}_{k-1} \to 2\tilde{c}_{k-1}
\mod \Z$ is ``expansive," the behavior of $\tilde{\C}$ is not only
uncontrollable, but in fact chaotic.	 \qed

\medskip
\noindent
\textbf{Example 7}.  This code was the main example in \cite{FZ99}.  It
turns out that our construction method yields a simpler syndrome-former
than the general construction given in \cite{FZ99}. 

Let $\C$ be the set of sequences in $(\Z_4)^\Z$
that (a) are either all odd or all even, and (b) have period 1 or 2.  In
other words, a code sequence is the bi-infinite repetition of one of the 8
pairs $\{00, 22, 02, 20, 11, 33, 13, 31\}$;  therefore $\C \cong \Z_4
\times \Z_2$.  The dual is the finite linear code $\C^\perp$ over $\Z_4$
generated by shifts of $\hb_1 = (\ldots, 0, 2, 2, 0, \ldots)$ and $\hb_2 =
(\ldots, 0, 1, 0, 1, 0, \ldots)$.

	$\C$ is clearly linear, time-invariant, autonomous and 2-observable.  Its
first-level observer granules $\Phi_{[k-1,k]}(\C)$ check
orthogonality to $\hb_1$ ($2c_k = 2c_{k-1}$) and are isomorphic to $\Z_2$; 
its second-level observer granules $\Phi_{[k-2,k]}(\C)$ check orthogonality
to
$\hb_2$ ($c_k = c_{k-2}$) and are isomorphic to
$\Z_2$ (assuming $\cb \in \C^1$).  Its first-output (input) group is
$\{0\}$, and its syndrome group is $\Z_4$. 

	A minimal observer-form encoder for $\C$ may store the previous output
$c_{k-1} \in \Z_4$ in the two-bit form $(c_{0,k-1}, c_{1,k-1})$.  At the
first level, it enforces the constraint $2c_k = 2c_{k-1}$, which determines
the low-order bit $c_{0,k}$ of $c_k$.  Given this constraint, it need only
store the high-order bit $c_{1,k-1}$ to enforce the second-level
constraint $c_{1,k} = c_{1,k-2}$, which determines $c_{1,k}$.

\begin{figure*}[t]
\setlength{\unitlength}{5pt}
\begin{center}
\begin{picture}(70,20)(-3,-3)
\put(0,9){\framebox(8,6){$c_{0,k-1}$}}
\put(8,12){\vector(1,0){22}}
\put(26,13){$c_{0,k}$}
\put(12,17){\line(0,-1){5}}
\put(12,17){\line(-1,0){15}}
\put(-3,17){\line(0,-1){5}}
\put(-3,12){\vector(1,0){3}}
\put(0,-1){\framebox(8,6){$c_{1,k-1}$}}
\put(8,2){\vector(1,0){3}}
\put(12,7){\line(-1,0){15}}
\put(-3,7){\line(0,-1){5}}
\put(-3,2){\vector(1,0){3}}
\put(11,-1){\framebox(8,6){$c_{1,k-2}$}}
\put(19,2){\vector(1,0){11}}
\put(26,0){$c_{1,k}$}
\put(23,7){\line(0,-1){5}}
\put(23,7){\line(-1,0){16}}
\put(8,-3){(a)}
\put(40,9){\framebox(8,6){$w_{0,k-1}$}}
\put(48,12){\vector(1,0){2.5}}
\put(50.5,12){\line(1,0){3}}
\put(52,10.5){\line(0,1){3}}
\put(52,12){\circle{3}}
\put(53.5,12){\vector(1,0){16.5}}
\put(66,13){$s_{0,k}$}
\put(36,18){$w_{0,k}$}
\put(52,17){\vector(0,-1){3.5}}
\put(52,17){\line(-1,0){16}}
\put(37,17){\line(0,-1){5}}
\put(37,12){\vector(1,0){3}}
\put(40,-1){\framebox(8,6){$w_{1,k-1}$}}
\put(48,2){\vector(1,0){3}}
\put(36,8){$w_{1,k}$}
\put(52,7){\line(-1,0){16}}
\put(37,7){\line(0,-1){5}}
\put(37,2){\vector(1,0){3}}
\put(51,-1){\framebox(8,6){$w_{1,k-2}$}}
\put(59,2){\vector(1,0){2.5}}
\put(61.5,2){\line(1,0){3}}
\put(63,0.5){\line(0,1){3}}
\put(63,2){\circle{3}}
\put(64.5,2){\vector(1,0){5.5}}
\put(66,3){$s_{1,k}$}
\put(63,7){\vector(0,-1){3.5}}
\put(63,7){\line(-1,0){16}}
\put(48,-3){(b)}
\end{picture}

Figure 8.5  Minimal (a) observer-form encoder and (b) syndrome-former \\
for the Fagnani-Zampieri periodic code over $\Z_4$ \cite{FZ99}.
\end{center}
\end{figure*}

A minimal memory-2 syndrome-former for $\C$ may store the
low-order bit $w_{k,0}$ for one time unit and the high-order bit
$w_{k,1}$ for two time units, so as to compute the first-level
syndrome $s_{0,k} = w_{0,k} - w_{0,k-1}$ and the second-level
syndrome $s_{1,k} = w_{1,k} - w_{1,k-2}$, as shown in Figure
8.5(b).  The syndrome-former is feedbackfree with memory 2, and is
$\Z_2$-linear.  Again, the encoder may be derived from the
syndrome-former simply by forcing the syndromes to 0. 
\qed

\section{The end-around theorem}

In this section we show that every observer granule of a group code C may
be viewed purely algebraically as an ``end-around" controller granule, and
vice versa.  As consequences of this observation, we develop:
\begin{itemize}
\item A definition of observer granules for nonabelian group codes;
\item Simple, purely algebraic alternative proofs of some previous results;
\item Myriad further isomorphisms.
\end{itemize}

	An interval $\I - [m, n), n > m$, may be viewed as an ``end-around"
interval that ``starts" at time $n$, wraps around from $+\infty$ to
$-\infty$, and finally ``ends" at time $m - 1$. We denote such an interval
by $[n, m)$.   Similarly, we define $[n, m] = \I - (m, n)$ for $n > m$, an
interval which ``starts" at time $n$ and ``ends" at time $m < n$.  If $n =
m + 1$, then $(m, n)$ is the empty set and $\I - (m, n)$ is the entire time
axis $\I$.  Finally, we define $(n, m] = \I - (m, n], n > m$, the
end-around interval from time $n + 1$ to time $m$.

	We then define an \textbf{end-around controller granule} on $[n, m], n >
m$, analogously to an ordinary controller granule, as follows:
$$
		\Gamma_{[n,m]}(\C) = \frac{\C_{:[n,m]}}{\C_{:[n,m)} + \C_{:(n,m]}}, \quad
n > m.
$$
Then we obtain the following interesting isomorphism:
\begin{theorem}[end-around theorem]  \label{eot}
For $n > m$, the end-around controller
granule $\Gamma_{[n,m]}(\C)$ is isomorphic to the observer granule
$\Phi_{[m,n]}(\C)$.
\end{theorem}

\emph{Proof}.  The restrictions of $\C_{:[n,m]}$ and $\C_{:[n,m)} +
\C_{:(n,m]}$ onto time $n$ both have kernel $\C_{:(n,m]}$, with images
$(\C_{:[n,m]})_{|\{n\}} = F^{n-m-1,n}(\C)$ and $(\C_{:[n,m)})_{|\{n\}} =
F^{n-m,n}(\C)$, respectively, so by the correspondence theorem,
$$
		\Gamma_{[n,m]}(\C)	\cong \frac{F^{n-m-1,n}(\C)}{F^{n-m,n}(\C)}.
$$
By Theorem \ref{folod}, this is isomorphic to
$\Phi_{[m,n]}(\C)$.   \qed

We may similarly define an end-around observer granule $\Phi_{[n,m]}(\C)$
for $n > m$, and show that it is isomorphic to $\Gamma_{[m,n]}(\C)$.

	One consequence of the end-around theorem is that all dynamical observer 
granules may be expressed as end-around controller granules.  But
controller granules, unlike observer granules, are well-defined for
nonabelian group codes.  Therefore it is possible to define the dynamical
observer granules of a nonabelian code $\C$ by $\Phi_{[m,n]}(\C) =
\C_{:[n,m]}/(\C_{:[n,m)}\C_{:(n,m]})$ (in multiplicative notation), which
opens the door to extending our observer dynamics results to nonabelian
codes.  We regard this as an important topic for further study, especially
in view of the successful constructions of \cite{FT93} and \cite{FZ99} in
the nonabelian case.

We now sketch a few applications of the end-around theorem.  These involve
partitioning the time axis $\I$ into 2, 3, or 4 subintervals, which we then
regard as a new finite time axis $\I'$ of length 2, 3, or 4, respectively
(as in Subsection VI-B).

	The state space theorem involves a two-way partition of $\I$ into disjoint
subsets $\J$ and $\I-\J$.  We may regard a code $\C$ defined on $\I$ as a code
defined on the length-2 time axis $\I' = \{\J, \I-\J\}$, which we identify with
the length-2 interval $[1, 2]$.  

	Now the nondynamical controller granules of $\C$ are
$\Gamma_{[1,1]}(\C) = \C_{:\J}$ and $\Gamma_{[2,2]}(\C) = \C_{:\I-\J}$, and
the sole dynamical controller granule is the first-level granule
$\Gamma_{[1,2]}(\C) = \C/(\C_{:\J} + \C_{:\I-\J})$, which is the two-sided
state space $\Sigma_\J(\C)$.  The nondynamical observer granules of $\C$
are $\Phi_{[1,1]}(\C) = \W_{|\J}/\C_{|\J}$ and $\Phi_{[2,2]}(\C) =
\W_{|\I-\J}/\C_{|\I-\J}$, and the sole dynamical observer granule is the
first-level granule $\Phi_{[1,2]}(\C) = (\C_{|\J} + \C_{|\I-\J})/\C$, which
we recognize as the two-sided reciprocal state space $\Sigma^\J(\C)$.  The
end-around controller granule $\Gamma_{[2,1]}(\C)$ is $\C/(\C_{:\J} +
\C_{:\I-\J}) = \Sigma_\J(\C)$.  The end-around theorem
therefore implies $\Sigma_\J(\C) \cong \Sigma^\J(\C)$, an important
isomorphism that we derived previously as a corollary of the reciprocal
state space theorem, as well as purely algebraically.

The $[m, n)$-controllability and $[m, n)$-observability tests involve a
three-way partition of $\I$ into disjoint subsets $m^-, [m, n)$, and
$n^+$.  We may regard a code $\C$ defined on $\I$ as a code
defined on the length-3 time axis $\I' = \{m^-, [m, n), n^+\}$, which we
identify with the length-3 interval $[1, 3]$.  

	Now by the first $[m, n)$-observability test, $\C$ is $[m, n)$-observable
if and only if $\C$ is 1-observable on $\I'$;  \ie if and only if the
second-level observer granule
$
\Phi_{[1,3]}(\C) = \C^1/\C^2
$
is trivial, where $\C^2 = \C$ and
$$
		\C^1	= \{\wb \in \W \mid \wb_{|[1,2]} \in \C_{|[1,2]}, \wb_{|[2,3]}
\in \C_{|[2,3]}\}.
$$
By the end-around theorem, 
$$
		\Phi_{[1,3]}(\C) \cong \Gamma_{[3,1]}(\C) =
\frac{\C_{:[3,1]}}{\C_{:\{3\}} + \C_{:\{1\}}} =
\frac{\C_{:[n,m)}}{\C_{:n^+} \times \C_{:m^-}},
$$
so $\C$ is $[m, n)$-observable if and only if $\C_{:[n,m)} = \C_{:n^+} \times
\C_{:m^-}$.  Thus the first $[m, n)$-observability test is equivalent to
the second by the end-around theorem.

	Similarly, by our second $[m, n)$-controllability test, $\C$ is $[m,
n)$-controllable if and only if $\C$ is 1-controllable on $\I'$;  \ie if
and only if the second-level controller granule
$$
		\Gamma_{[1,3]}(\C) = \frac{\C}{\C_{:[1,2]} + \C_{:[2,3]}}
$$
is trivial.  By the dual to the end-around theorem, $\Gamma_{[1,3]}(\C)$ is
isomorphic to the end-around observer granule
$$
		\Phi_{[3,1]}(\C)	= \frac{\C_{|\{3\}} \times \C_{|\{1\}}}{\C_{|[3,1]}}
			= \frac{\C_{|n^+} \times \C_{|m^-}}{\C_{|[n,m)}},
$$
so $\C$ is $[m, n)$-controllable if and only if $\C_{|[n,m)} = \C_{|n^+}
\times \C_{|m^-}$.  Thus the first $[m, n)$-controllability test is
equivalent to the second by the dual end-around theorem.  

	Projections of these quotients onto the subintervals $m^-, [m, n)$, and
$n^+$ yield still further tests in terms of trivial quotients of primal and
dual first-output and last-output chains, which are cumbersome to write but
which have the advantage of being testable on a single interval.  Even on
a time axis of length 3, there are a great many isomorphisms that can be
derived from the general granule isomorphisms, since the system dynamical
structure is determined by only three dynamical controller granules
$\Gamma_{12}, \Gamma_{23}$ and $\Gamma_{123}$ and three dynamical observer
granules $\Phi_{12}, \Phi_{23}$ and $\Phi_{123}$ (or equivalently three
end-around controller granules $\Gamma_{231}, \Gamma_{312}$ and
$\Gamma_{31}$).

A set of such isomorphisms is illustrated in Figure 9.1.  Moreover, every
permutation of the three indices $\{1, 2, 3\}$ yields a similar set of
further isomorphisms.  Here $\C_{:1} \times \C_{:2} \times \C_{:3}$ is the
0-controllable subcode of $\C$, and $\C_{:12} + \C_{:23}$ is the
1-controllable subcode of $\C$.  The figure shows how $\C/(\C_{:1} \times
\C_{:2} \times \C_{:3})$ decomposes into the controllability granules
$\Gamma_{12}, \Gamma_{23}$ and $\Gamma_{123} \cong \Phi_{31}$.  Similarly,
$\C_{|1} \times \C_{|2} \times \C_{|3}$ is the 0-observable supercode of
$\C$, $(\C_{|12} \times \C_{|3}) \cap (\C_{|1} \times \C_{|23})$ is the
1-observable supercode of $\C$, and the figure shows how $(\C_{|1} \times
\C_{|2} \times \C_{|3})/\C$ decomposes into the observability granules
$\Phi_{12}, \Phi_{23}$ and $\Phi_{123} \cong \Gamma_{31}$.  The diagram is
self-dual.

\begin{figure}[t]
\setlength{\unitlength}{5pt}
\begin{center}
\begin{picture}(48,52)(2,-1)
\put(19,0){$\C_{:1} \times \C_{:2} \times \C_{:3}$}
\put(1,7){$\C_{:1} \times \C_{:23}$}
\put(9,6){\line(3,-1){12}}
\put(16,4){$\Gamma_{23}$}
\put(29,2){\line(3,1){12}}
\put(31,4){$\Gamma_{12}$}
\put(42,7){$\C_{:12} \times \C_{:3}$}
\put(9,8){\line(3,1){12}}
\put(16,8){$\Gamma_{12}$}
\put(29,12){\line(3,-1){12}}
\put(31,8){$\Gamma_{23}$}
\put(20,13){$\C_{:12} \times \C_{:23}$}
\put(9,9){\line(1,1){14}}
\put(16,15){$\Sigma_{1}$}
\put(41,9){\line(-1,1){14}}
\put(31,15){$\Sigma_{2}$}
\put(25,15){\line(0,1){8}}
\put(26,18){$\Phi_{31}$}
\put(24.5,24){$\C$}
\put(25,26){\line(0,1){8}}
\put(26,29){$\Gamma_{31}$}
\put(14,35){$(\C_{|12} \times \C_{|3}) \cap (\C_{|1} \times \C_{|23})$}
\put(9,40){\line(1,-1){14}}
\put(16,29){$\Sigma_{1}$}
\put(41,40){\line(-1,-1){14}}
\put(31,29){$\Sigma_{2}$}
\put(9,41){\line(3,-1){12}}
\put(16,39){$\Phi_{12}$}
\put(41,41){\line(-3,-1){12}}
\put(31,39){$\Phi_{23}$}
\put(1,42){$\C_{|1} \times \C_{|23}$}
\put(41,42){$\C_{|12} \times \C_{|3}$}
\put(9,44){\line(3,1){12}}
\put(16,45){$\Phi_{23}$}
\put(41,44){\line(-3,1){12}}
\put(31,45){$\Phi_{12}$}
\put(19,49){$\C_{|1} \times \C_{|2} \times \C_{|3}$}
\put(4,9){\line(0,1){32}}
\put(5,24){$\Sigma_{1} \times \Sigma_1$}
\put(46,9){\line(0,1){32}}
\put(38,24){$\Sigma_{2} \times \Sigma_2$}
\end{picture}

Figure 9.1.  Granule isomorphisms on a length-3 time axis.
\end{center}
\end{figure}

	Finally, our $j$th dual first-output and last-output group results for $j
\ge 1$ involve a four-way partition of I into disjoint subsets $\{k - j\},
(k- j, k), \{k\}, (k, k - j)$, which we regard as a length-4 time axis
$\I'$ and identify with the length-4 interval $[1, 4]$.

	In this point of view, the dual first-output group $F^{j,k}(\C)$ is the
set of time-2 symbols in the subcode $\C_{:[3,4]}$ that ``starts" at time
$k$ and ``ends" at time $k - j - 1$, 
$$
		F^{j,k}(\C) = (\C_{:[3,4]})_{|\{3\}},
$$
while $F^{j-1,k}(\C)$ is the set of time-2 symbols in the subcode
$\C_{:[3,1]}$ that ``starts" at time $k$ and ``ends" at time $k - j$:
$$
		F^{j-1,k}(\C) = (\C_{:[3,1]})_{|\{3\}}.
$$
Since the end-around controller granule $\Gamma_{[3,1]}(\C)$ is
$\C_{:[3,1]}/(\C_{:[3,4]} + \C_{:[4,1]})$, we have by projection onto time
2 and the end-around theorem 
$$
		\frac{F^{j-1,k}(\C)}{F^{j,k}(\C)} \cong \Gamma_{[3,1]}(\C) \cong
\Phi_{[1,3]}(\C),
$$
where $\Phi_{[1,3]}(\C)$ denotes the observer granule
$\Phi_{[k-j,k]}(\C)$.   The isomorphism $L^{j-1,k}(\C)/L^{j,k}(\C) \cong
\Phi_{[k,k+j]}(\C)$ may be derived similarly.

\section{Conclusion}

	In this paper we have extended the duality principles that have proved to 
be so useful in coding and system theory to abelian group codes.  We have
introduced a bit of topology in order to make use of Pontryagin duality,
but topology is not used in any essential way other than to clarify
duality principles when the time axis is infinite.  We have also
introduced a few technical ``well-behavedness" conditions, principally the
closed-projections assumption.  Since this assumption holds when symbol
groups are compact, or \emph{a fortiori} finite, we do not believe that it
will prove to be restrictive in practical applications.

We have generalized the dual state space theorem of linear system theory,
which shows in a precise sense that the state complexity of dual codes or
systems is dual in the character group sense.  We have also shown that
there are well-defined dual notions of controllability and observability
for codes and behaviors, rather than for state-space realizations of codes
and behaviors as in classical and behavioral linear system theory. 
Finally, we have shown close connections between controllability and finite
generatability, on the one hand, and observability and finite checkability
(completeness), on the other.

It is helpful to keep in mind both the controllability and observability
properties of a code or system.  An uncontrollable (resp.\ unobservable)
system may have simple observability (resp.\ controllability) properties,
as shown in Example 4.  A ``low-rate" code or system is usually more simply
specified in controller form (\eg by a generator matrix, encoder or
image representation), whereas a ``high-rate" code or system is usually
more simply specified in observer form (\eg by a parity-check matrix,
syndrome-former or kernel representation).  Controller memory and observer
memory are both important parameters of a system.

It can also be helpful to characterize a code or system by its dual.  For
example, a complete compact code or system can be characterized by its
finite discrete dual, whose properties are purely algebraic.  Pathologies
in the primal system will be reflected in pathologies in the dual system,
but their nature will usually be quite different (\eg in Examples 2 and
4).

It appears to us that behavioral system theory and symbolic dynamics have
focussed largely on observability structure.  Systems are usually assumed
to be complete and compact, and ``memory" usually means observer memory
(see, \eg \cite{FMST95}).  In automata theory, on the other hand, languages
are usually sets of discrete and finitely supported sequences.  We believe
that each of these fields might benefit from a more balanced viewpoint.
	
There are several clues in this work, as well as in \cite{FT93} and
\cite{FZ99}, that the abelian assumption is inessential.  It is not needed
for the purely algebraic controllability structure discussed in
\cite{FT93}, nor for the more difficult observer-form constructions of
\cite{FZ99}.  The key idea of \cite{FZ99} may be the recognition that even
when a subgroup $H$ (such as a code) is not normal in a group $G$ (such as
its output sequence space), the set $G//H$ of left cosets of $H$ in $G$ is
nonetheless a tractable group-theoretic object upon which $G$ acts
naturally by translation.  Moreover, in this paper we have shown that all
observer granules are isomorphic to ``end-around" controller granules,
which remain well-defined in the nonabelian case.  It may well be useful
therefore to develop an alternative purely algebraic general theory of
observability structure that will apply equally to abelian and nonabelian
group codes.

\section*{Acknowledgments}

This paper has benefited from the comments of many colleagues over its
lengthy development.  We would like to acknowledge particularly Thomas
Ericson, Fabio Fagnani, Andi Loeliger, Brian Marcus, Thomas Mittelholzer,
Sanjoy Mitter, Joachim Rosenthal, Maria Elena Valcher, Jan Willems and
Sandro Zampieri.  We are also grateful to the Associate Editor and referees
for four excellent reviews.  We are particularly indebted to Thomas
Mittelholzer for helpful suggestions and references with regard to the
issues of ``closed projections" and ``completeness," and to Fabio Fagnani
for correction of mathematical errors.



% http://www.ctan.org/tex-archive/macros/latex/contrib/supported/IEEEtran/bibtex
%\bibliographystyle{IEEEtran.bst}

\begin{thebibliography}{10}

\bibitem{AL95} 
H. Andersson and H.-A. Loeliger, ``Codes from iterated
maps," {\em Proc. 1995 IEEE Intl. Symp. Inform. Theory} (Whistler, BC),
p.\ 309, Sept.\ 1995.

\bibitem{D87}  
R. Devaney,  \emph{An Introduction to Chaotic Dynamical Systems}. 
Reading, MA:  Addison-Wesley, 1987.

\bibitem{ES92}  
M. Enock and J.-M. Schwartz, \emph{Kac Algebras and Duality of
Locally Compact Groups}.  New York:  Springer-Verlag, 1992.

\bibitem{F91}  
F. Fagnani,  ``Controllability and observability indices for
linear time-invariant systems:  A new approach,"  \emph{Syst.\ Contr.\
Lett.}, vol.\ 17, pp.\ 243--250, 1991.

\bibitem{F96}  
F. Fagnani,  ``Some results on the classification of
expansive automorphisms of compact Abelian groups,"  \emph{Ergod.\ Theory
Dyn.\ Syst.}, vol.\ 16, pp.\ 45--50, 1996.

\bibitem{F97}  
F. Fagnani, ``Shifts on compact and discrete Lie groups: 
Algebraic-topological invariants and classification problems," 
\emph{Adv.\ Math.}, vol.\ 127, pp.\ 283--306, 1997.

\bibitem{FW92}  
F. Fagnani and J. C. Willems,  ``Controllability of $l_2$
systems,"  \emph{SIAM J. Contr.\ Opt.}, vol.\ 30, pp.\ 1101--1125, 1992.

\bibitem{FZ96}  
F. Fagnani and S. Zampieri,  ``Dynamical systems and
convolutional codes over finite Abelian groups,"  \emph{IEEE Trans.\
Inform.\ Theory}, vol.\ 42, pp.\ 1892--1912, Nov.\ 1996.

\bibitem{FZ97}  
F. Fagnani and S. Zampieri,  ``Classification problems for shifts on
modules over a principal ideal domain,"  \emph{Trans.\ AMS}, vol.\ 349,
pp.\ 1993--2006, 1997.

\bibitem{FZ99}  
F. Fagnani and S. Zampieri,  ``Minimal syndrome formers for
group codes,"  \emph{IEEE Trans.\ Inform.\ Theory}, vol.\ 45, pp.\ 3--31,
Jan.\ 1999.

\bibitem{FV94}  
E. Fornasini and M. E. Valcher,  ``Algebraic aspects of 2D
convolutional codes,"  \emph{IEEE Trans.\ Inform.\ Theory}, vol.\ 40, pp.\
1068--1082, 1994. 

\bibitem{FV96}  
E. Fornasini and M. E. Valcher, 
``Multidimensional systems with finite support behaviors:  Signal
structure, generation and detection,"  \emph{SIAM J. Contr.\ Opt.},
vol.\ 36, pp.\ 760--779, 1998.

\bibitem{F73}  
G. D. Forney, Jr.,  ``Structural analysis of convolutional
codes via dual codes,"  \emph{IEEE Trans.\ Inform.\ Theory}, vol.\ IT-19,
pp.\ 512--518, 1973.

\bibitem{F01}  
G. D. Forney, Jr.,  ``Codes on graphs:  Normal realizations,"  \emph{IEEE
Trans.\ Inform.\ Theory}, vol.\  47, pp.\ 520--548, Feb.\ 2001.

\bibitem{FT93}  
G. D. Forney, Jr.\ and M. D. Trott,  ``The dynamics of group
codes:  State spaces, trellis diagrams and canonical encoders," 
\emph{IEEE Trans.\ Inform.\ Theory}, vol.\ 39, pp.\ 1491--1513, Sept.\
1993.

\bibitem{FT93a}  
G. D. Forney, Jr.\ and M. D. Trott,  ``The dynamics of group
codes:  Syndromes, normal codes and observers,"  \emph{Proc.\ 1993 Intl.\
Symp.\ Inform.\ Theory} (San Antonio, TX), p.\ 177, Jan.\ 1993.

\bibitem{FT93b}  
G. D. Forney, Jr. and M. D. Trott,  ``Duals of abelian group
codes and systems,"  \emph{Proc.\ IEEE Workshop on Coding, System Theory
and Symbolic Dynamics} (Mansfield, MA), pp.\ 36--38, Oct.\ 1993.

\bibitem{FT95} 
G. D. Forney, Jr.\ and M. D. Trott,  ``Controllability, observability and
duality in behavioral group systems,"  \emph{Proc.\ 34th Conf.\ Dec.\
Contr.} (New Orleans, LA), vol.\ 3, pp.\ 3259--3264, Dec.\ 1995.

\bibitem{FMST95}  
G. D. Forney, Jr., B. Marcus, N. T. Sindhushayana and M.
D. Trott,  ``Multilingual dictionary:  System theory, coding theory,
symbolic dynamics and automata theory,"  \emph{Proc.\ Symp.\ Appl.\
Math.}, vol.\ 50, pp.\ 109--137, 1995.

\bibitem{HR79}  
E. Hewitt and K. A. Ross,  \emph{Abstract Harmonic Analysis I}
(second ed.).  New York:  Springer-Verlag, 1979.

\bibitem{K80}  
T. Kailath,  \emph{Linear Systems}.  Englewood Cliffs, NJ: 
Prentice-Hall, 1980.

\bibitem{K48}  
S. Kaplan,  ``Extension of the Pontrjagin duality I: 
Infinite products,"  \emph{Duke Math.\ J.}, vol.\ 15, pp.\ 649--658, 1948.

\bibitem{K87}  
B. Kitchens,  ``Expansive dynamics on zero-dimensional
groups,"  \emph{Ergod.\ Theory Dyn.\ Syst.}, vol.\ 7, pp.\ 249--261, 1987.

\bibitem{KS89}  
B. Kitchens and K. Schmidt,  ``Automorphisms of compact
groups,"  \emph{Ergod.\ Theory Dyn.\ Syst.}, vol.\ 9, pp.\ 691--735, 1989.

\bibitem{KF70}
A. N. Kolmogorov and S. V. Fomin, \emph{Introductory Real Analysis} (R.
A. Silverman, trans./ed.).  New York:  Dover, 1970.

\bibitem{LM95}  
D. Lind and B. Marcus,  \emph{Symbolic Dynamics and
Coding}.  Cambridge, UK:  Cambridge U. Press, 1995.

\bibitem{L93}  
H.-A. Loeliger,  ``Abelian-group convolutional codes need
not be ring codes,"  \emph{Proc.\ 6th Joint Swedish-Russian Intl.\ Wkshp.\
Inform.\ Theory} (M\"{o}lle, Sweden), pp.\ 21--22, Aug.\ 1993.

\bibitem{LFMT94}  
H.-A. Loeliger, G. D. Forney, Jr., T. Mittelholzer and
M D. Trott,  ``Minimality and observability of group systems," 
\emph{Linear Alg.\ Appl.}, vol.\ 205--206, pp.\ 937--963, July 1994.

\bibitem{LM96}  
H.-A. Loeliger and T. Mittelholzer,  ``Convolutional codes
over groups,"  \emph{IEEE Trans.\ Inform.\ Theory}, vol.\ 42, pp.\
1660--1686, Nov.\ 1996.

\bibitem{MT78}  
G. Miles and R. K. Thomas,  ``The breakdown of
automorphisms of compact topological groups,"  in \emph{Studies in
Probability and Ergodic Theory} (Adv. Math. Suppl. Stud.), vol.\ 2, pp.\
207--218.  Academic Press, 1978.

\bibitem{N97} 
H. Narayanan, 
\emph{Submodular Functions and Electrical Networks}
(\emph{Ann.\ Discr.\ Math.\ }{\bf 54}), Chaps.\ 7--8.
Amsterdam:  North Holland, 1997.

\bibitem{N00} 
H. Narayanan, 
``Matroids representable over modules,
electrical network topology and
behavioural systems theory,"
research report, EE Dept., IIT--Bombay,
May 2000.

\bibitem{N000} 
H. Narayanan, 
``On the duality between controllability and
observability in behavioural systems theory,"
{\em Proc.\ CCSP 2000}, Bangalore, pp.\ 183--186, July 2000.

\bibitem{N70}  
N. Noble,  ``$k$-groups and duality,"  \emph{Trans.\ AMS},
vol.\ 151, pp.\ 551--561, Oct.\ 1970.

\bibitem{P46}  
L. Pontryagin,  \emph{Topological Groups}.  Princeton, NJ: 
Princeton U. Press, 1946.

\bibitem{RW91}  
P. Rocha and J. C. Willems,  ``Controllability of 2D
systems,"  \emph{IEEE Trans.\ Automat.\ Contr.}, vol.\ 36, pp.\ 413--423,
1991.

\bibitem{RSV96}  
J. Rosenthal, J. M. Schumacher and E. V. York,  ``On
behaviors and convolutional codes,"  \emph{IEEE Trans.\ Inform.\ Theory},
vol.\ 42, pp.\ 1881--1891, 1996.

\bibitem{R88}
J. J. Rotman, \emph{An Introduction to the Theory of Groups}.  Dubuque,
IA:  Wm.\ C. Brown, 1988.

\bibitem{R90}  
W. Rudin,  \emph{Fourier Analysis on Groups}.  New York:  Wiley,
1990.

\bibitem{S75}
H. Schubert, \emph{Topology} (S. Moran, trans.).  Boston: Allyn and Bacon,
1968.

\bibitem{S90}  
K. Schmidt,  ``Automorphisms of compact abelian groups and
affine varieties," \emph{Proc.\ London Math.\ Soc.}, vol.\ 61, pp.\
480--496, 1990.

\bibitem{T99}
H. Trentelman, ``A truly behavioral approach to the $H_\infty$ control
problem," in \emph{The Mathematics of Systems and Control:  From
Intelligent Control to Behavioral Systems} (J. W. Polderman and H.
Trentelman, eds.), pp.\ 177--190.  Groningen, The Netherlands:  U.
Groningen Press, 1999.

\bibitem{T92}  
M. D. Trott,  ``The algebraic structure of trellis codes," 
Ph.D. dissertation, Dept.\ Elec.\ Engg., Stanford U., Stanford, CA, Aug.\
1992.

%\bibitem{TBGM96}  
%M. D. Trott, S. Benedetto, R. Garello and M. Mondin,  ``Rotational
%invariance of trellis codes--- Part I:  Encoders and precoders," 
%\emph{IEEE Trans.\ Inform.\ Theory}, vol.\ 42, pp.\ 751--765, May 1996.

\bibitem{VF94}  
M. E. Valcher and E. Fornasini,  ``On 2D finite-support
convolutional codes:  An algebraic approach,"  \emph{Multid.\ Syst.\ Sig.\
Proc.}, vol.\ 5, pp.\ 231--243, 1994.

\bibitem{V76}  
R. Venkataraman,  ``A characterization of Pontryagin
duality,"  \emph{Math.\ Z.}, vol.\ 149, pp.\ 109--119, 1976.

\bibitem{W86}  
J. C. Willems,  ``From time series to linear systems, Parts
I--III,"  \emph{Automatica}, vol.\ 22, pp.\ 561--580 and 675--694, 1986; 
vol.\ 23, pp.\ 87--115, 1987.

\bibitem{W89} 
J. C. Willems,  ``Models for dynamics,"  \emph{Dynamics Reported},
vol.\ 2, pp.\ 171--269, 1989.

\bibitem{W91}  
J. C. Willems,  ``Paradigms and puzzles in the theory of
dynamical systems,"  \emph{IEEE Trans.\ Automat.\ Contr.}, vol.\ 36, pp.\
259--294, Mar.\ 1991.

\bibitem{W97}  
J. C. Willems,  ``On interconnections, control and
feedback,"  \emph{IEEE Trans.\ Automat.\ Contr.}, vol.\ 42, pp.\ 326--339,
Mar.\ 1997.

\bibitem{ZM96}  
S. Zampieri and S. Mitter,  ``Linear systems over
Noetherian rings in the behavioral approach,"  \emph{J. Math.\ Syst.\
Estim.\ Contr.}, vol.\ 6, pp.\ 235--238, 1996.

\end{thebibliography}

\newpage

%\begin{biography}[{\includegraphics[width=1in,height=1.25in,clip,keepaspectratio]{mshell}}]{Michael Shell}
% where an .eps filename suffix will be assumed under latex, and a .pdf suffix
% will be assumed for pdflatex; or if you just want to reserve a space for
% a photo:

%\begin{biography}{Michael Shell}
%\end{biography}

\begin{biographynophoto}{G. David Forney, Jr.}
(S'59--M'61--F'73)
received the B.S.E. degree in electrical
engineering from Princeton University, Princeton, NJ, in 1961, and the M.S.\ and
Sc.D.\ degrees in electrical engineering from the Massachusetts Institute of
Technology (MIT), Cambridge, in 1963 and 1965, respectively.

From 1965 to 1999 he was with the Codex Corporation, which was acquired by
Motorola, Inc.\ in 1977, and its successor, the Motorola Information Systems
Group, Mansfield, MA.  Since 1996, he has been Bernard M. Gordon Adjunct
Professor at M.I.T.

Dr. Forney was Editor-in-Chief of the \textsc{IEEE Transactions on Information
Theory} from 1970 to 1973.  He has been a member of the Board of Governors of
the IEEE Information Theory Society during 1970--76, 1986--94, and 2004--, and
was President in 1992.  He has been awarded the 1970 IEEE Information Theory
Group Prize Paper Award, the 1972 IEEE Browder J. Thompson Memorial Prize Paper
Award, the 1990 IEEE Donald G. Fink Prize Paper Award, the 1992 IEEE Edison
Medal, the 1995 IEEE Information Theory Society Claude E. Shannon Award, the 1996
Christopher Columbus International Communications Award, and the 1997 Marconi
International Fellowship.  In 1998 he received an IT Golden Jubilee Award for
Technological Innovation, and two IT Golden Jubilee Paper Awards.  He was
elected a Fellow of the IEEE in 1973, a member of the National Academy of
Engineering (USA) in 1983, a Fellow of the American Association for the
Advancement of Science in 1993, an honorary member of the Popov Society (Russia)
in 1994, a Fellow of the American Academy of Arts and Sciences in 1998, and a
member of the National Academy of Sciences (USA) in 2003. 
\end{biographynophoto}

% insert where needed to balance the two columns on the last page
%\newpage

\begin{biographynophoto}{Mitchell D. Trott}
(S'90--M'92) received the B.S. and M.S. degrees in systems engineering from
Case Western Reserve University, Cleveland, OH, in 1987 and 1988, respectively,
and the Ph.D. degree in electrical engineering from Stanford University,
Stanford, CA, in 1992.

He was an Associate Professor in the Department of Electrical Engineering and 
Computer Science at the Massachusetts Institute of Technology, Cambridge, MA,
from 1992 until 1998, and Director of Research at ArrayComm, Inc., San Jose,
CA, from 1997 through 2002.  He now leads the Streaming Media Technologies
group at Hewlett-Packard Laboratories, Palo Alto, CA.

His research interests include streaming media systems, multi-user and 
wireless communications, and information theory.
\end{biographynophoto}


\vfill


\end{document}